\documentclass[aps,prd,nopacs,floatfix,notitlepage,nofootinbib,superscriptaddress,twocolumn,a4paper]{revtex4-1}
\usepackage{amsfonts,amsmath,units,wasysym,epsfig,graphicx,verbatim,color,subfigure,graphicx,bm,mathrsfs,lipsum,hyperref,cleveref}
\usepackage{booktabs}
\usepackage[normalem]{ulem}  

\newcommand{\ems}{\mathcal{E}_{EM}}
\newcommand{\qsd}{\mathcal{E}_{Q}}
\newcommand{\rsd}{\mathcal{E}_{R}}
\newcommand{\rsat}{\alpha_{\rm S}}

\hypersetup{colorlinks, citecolor =blue}

\begin{document}

\newcommand{\JBCA}{Jodrell Bank Centre for Astrophysics, School of Physics and Astronomy, University of Manchester, Manchester, M13 9PL, UK}
\newcommand{\USAL}{Departamento de F\'isica Fundamental, Universidad de Salamanca and IUFFyM, Plaza de la Merced S/N, E-37008 Salamanca, Spain}
\newcommand{\Uliege}{Space Sciences, Technologies and Astrophysics Research (STAR) Institute, Universit\'e de Li\`ege, B\^at. B5a, 4000 Li\`ege, Belgium}
\newcommand{\SNU}{Department of Physics, School of Natural Sciences,
Shiv Nadar Institution of Eminence, Greater Noida 201314, Uttar Pradesh, India}
\newcommand{\IITD}{Department of Physics, Indian Institute of Technology Delhi, New Delhi 110016, India}

\title{Structural response of neutron stars to rapid rotation \\ and its impact on the braking index}

\author{Avishek Basu}
\email{avishek.basu@manchester.ac.uk}
\affiliation{\JBCA}

\author{Prasanta Char}
\email{prasanta.char@usal.es}
\affiliation{\USAL}\affiliation{\Uliege}

\author{Rana Nandi}
\email{rananandi@iitd.ac.in}
\affiliation{\SNU}\affiliation{\IITD}

\begin{abstract}
Pulsars are rotating neutron stars that are observed to be slowing down, implying a loss of their {  rotational} energy. There can be several different physical mechanisms involved in their spin-down process. The properties of fast-rotating pulsars depend on the nature of the neutron star matter, which can also affect the spin-down mechanisms. In this work, we examine three different physical phenomena contributing to the spin-down: magnetic dipole radiation, gravitational mass quadrupole radiation due to the ``mountain" formation, gravitational mass current quadrupole radiation or the r-modes, and calculate the expressions for the braking indices due to all of them. We have also considered the implications of the uncertainties of the equation of the state of neutron star matter and rapid rotation on the braking indices corresponding to the aforementioned processes and their combinations. In all cases, the rapid rotation results in a departure of the braking index from the standard values when the rotational effects are ignored. If generated with a saturation amplitude within the range of $10^{-4} - 10^{-1}$, the r-mode oscillations dominate the spin-down of millisecond pulsars. Moreover, we explore the braking index in the context of millisecond magnetars. { We also study the effects of different choices of baryon mass on the braking indices.}

\end{abstract}

\maketitle


\section{Introduction}\label{intro}

The majority of neutron stars (NS) are observed as pulsars, which are rapidly rotating and observed to spin down over time, indicating a loss of rotational kinetic energy. The spin-down rate is given as $\dot \Omega = -k \Omega^n$ \cite{MDN+1985}, where $n$ is the braking index and $k$ is a positive constant. The braking index captures information about the mechanism of energy loss. Depending on the emission loss mechanism $k$ can be a function of radius, moment of inertia, magnetic field strength etc (discussed later in the text). In most cases, the pulsar's rotational energy loss is attributed to the magnetic dipolar radiation (MDR) from the pulsar magnetosphere for which the braking index is 3 \cite{GO+1969}. However, pulsars can lose energy via gravitational wave (GW) radiation if there is non-axisymmetric deformation or r-mode instability, for which the braking indices are 5 and 7, respectively \citep{Riles+2023}. 
Therefore, it is natural to expect a rapidly rotating neutron star to lose energy via all three channels if the conditions are favourable. Apart from these three channels of energy loss, pulsars also lose energy due to the particle wind \citep{Harding+1999, Spitkovsky+2006}.

The spin-period of pulsars spans a large range. The slowest pulsar has a rotation period of 76 s \cite{Caleb:2022xyo} to the most rapidly rotating being 1.39 ms  \cite{HRS+2006}. The total population is typically categorized as normal (or ``slow'') pulsars and millisecond pulsars, with the demarcation around 16 ms \cite{Halder:2023rfu}. {The braking index has been measured for a few younger pulsars { (through pulsar timing)} and found {  to} be less than 3 \cite{Livingstone:2007bn, WJE+2011, Livingstone_2011, LPS+1993, BCD+1995, RGL+2012, Espinoza_2011}}. This is often attributed to the evolution of the magnetic field, modified magnetosphere or even the interaction of supernova fallback disk \cite{Chen+1993, Melatos+1997, Harding+1999}. A similar sub-3 braking index can be achieved in rapidly rotating pulsars, which is purely due to structural changes in the star due to rapid rotation. In such a scenario, the $k$ cannot be assumed independent of the spin frequency \cite{Hamil+2015, Lan:2021luk}, which has been discussed in detail in Sec.~\ref{BraIn}.

The effect of structural evolution is prominent over a spin frequency of 200 Hz \cite{Hamil+2015}. Currently, the  ATNF Pulsar Catalogue \footnote{\href{https://www.atnf.csiro.au/research/pulsar/psrcat/}{https://www.atnf.csiro.au/research/pulsar/psrcat/}}  \cite{Manchester:2004bp} records 392 pulsars with spin-frequency more than 200 Hz,  which corresponds to $\sim11\%$ of the total pulsar population discovered till date. These rapidly rotating pulsars are important in probing the physics of dense nuclear matter. The physics of the nuclear interaction and the constituents of the matter determines the dense matter equation of state (EOS), which in turn governs the stellar structure and its response to rapid rotation. There are several ways found in the literature to model the NS EOS. One may use explicit nuclear potentials, or energy density functionals describing phenomenological nucleonic interactions \cite{Dutra:2012mb,Dutra:2014qga,Oertel:2016bki,Nandi:2018ami, Sun:2023xkg}. One may also employ more agnostic and semi-agnostic approaches which do not assume any interactions a priori \cite{Read:2008iy,Lindblom:2010bb,Landry:2018prl, Margueron:2017eqc,Biswas:2020puz}. In this article, we have adopted a semi-agnostic approach to construct the dense matter equation of state following the formalism of \textcite{Gandolfi:2019zpj}, summarized in Sec.~\ref{eos_des}, to study the frequency evolution of the stellar structure and the braking index. 

There are various types of constraints that can be applied to the EOS model. 
Astrophysical observations provide several constraints relevant to the high density part of the EOS. In particular, 
the radio observations of the most massive pulsar, PSR J0740+6620 \cite{Cromartie:2019kug,Fonseca:2021wxt} provide the most stringent constraint on the NS EOS. The tidal deformability measurements from the multimessenger observations of the binary neutron star merger event GW170817 put additional constraints on the EOS \cite{TheLIGOScientific:2017qsa,LIGOScientific:2017ync,LIGOScientific:2018hze}. In recent years, the Neutron star Interior Composition Explorer (NICER) collaboration has reported a few simultaneous mass-radius measurements of PSR J0030+0451, PSR J0740+6620, and PSR J0437-4715, respectively, that also help us to constrain the Mass-radius plane and consequently the EOS parameter space \cite{Miller+2019, Riley+2019,Miller+2021, Riley+2021, Salmi+2022,Vinciguerra+2024,Choudhury:2024xbk}. On the other hand, the latest constraint on the low-density part of the EOS comes from the advancements of the chiral effective field theory ($\chi$-EFT) calculations \cite{Tews:2012fj,Hebeler:2013nza,Lynn:2015jua, Drischler:2017wtt, Huth:2020ozf}.

We have employed a Bayesian analysis framework, supplemented by the measurements from the astrophysical observations to constrain the EOS model parameters and present the results of the analysis in Sec.~\ref{eos_des}. In Sec.~\ref{rot_effect}, we used the constrained EOS to investigate their effects on stellar structure under rapid rotation. In Sec.~\ref{BraIn}, we have presented an ab initio derivation of the braking index under the assumption of energy loss through MDR and GWs due to finite deformation and r-mode oscillation. The analytical expressions have been later combined to study the frequency-dependent evolution of the braking index in context to millisecond pulsars and nascent millisecond magnetars born from NS-NS mergers or other formation scenarios. Finally, we summarise our findings of this work in Sec.~\ref{conclusion}.

\begin{figure*}
    \centering
    \includegraphics[scale=0.4]{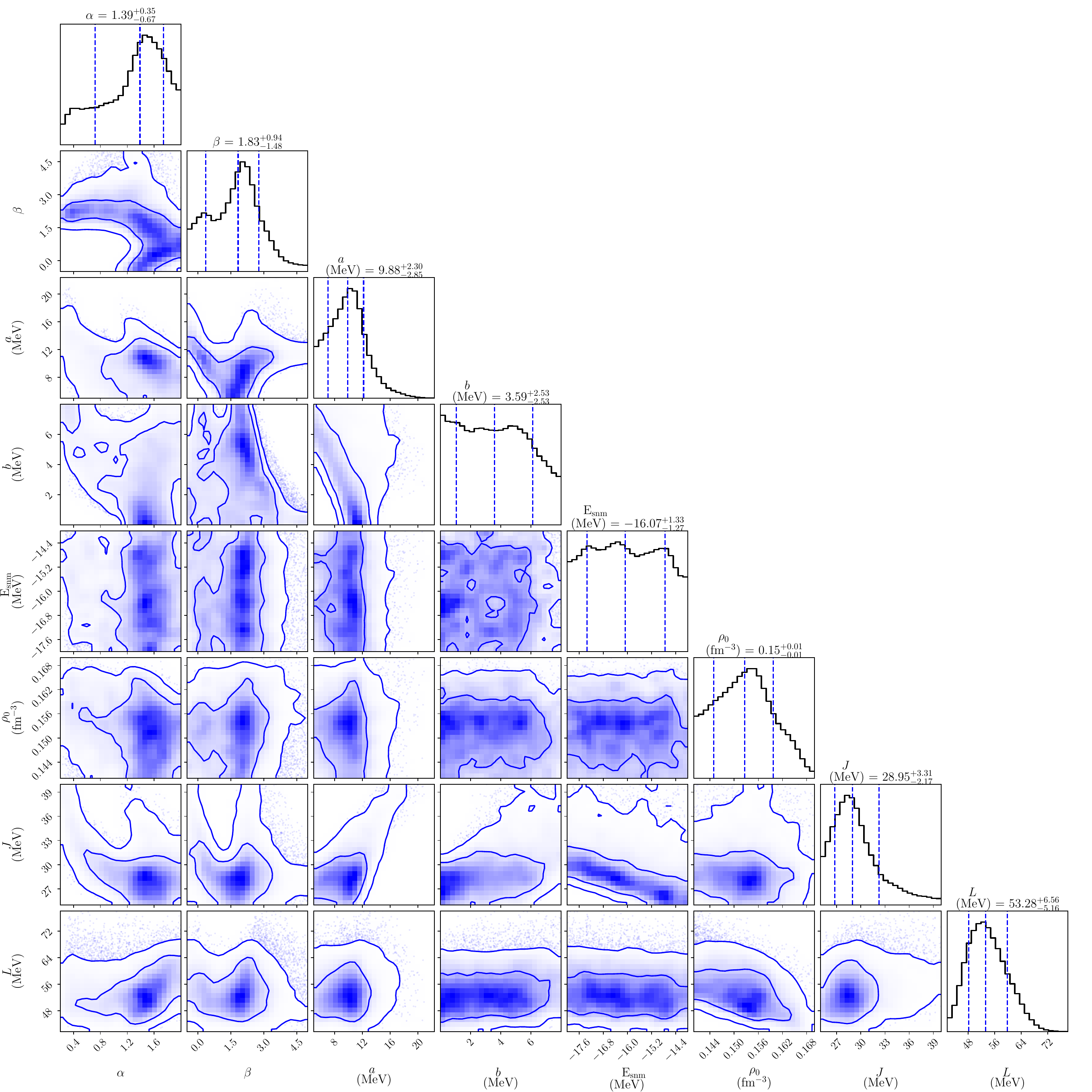}
    \caption{Shows the posterior distribution of the model parameters along with the symmetry energy at the saturation density ($J$) and the slope of symmetry energy at the saturation density ($L$).}
    \label{fig:nuclear}
\end{figure*}

\begin{figure}
    \centering
    \includegraphics[scale=0.40]{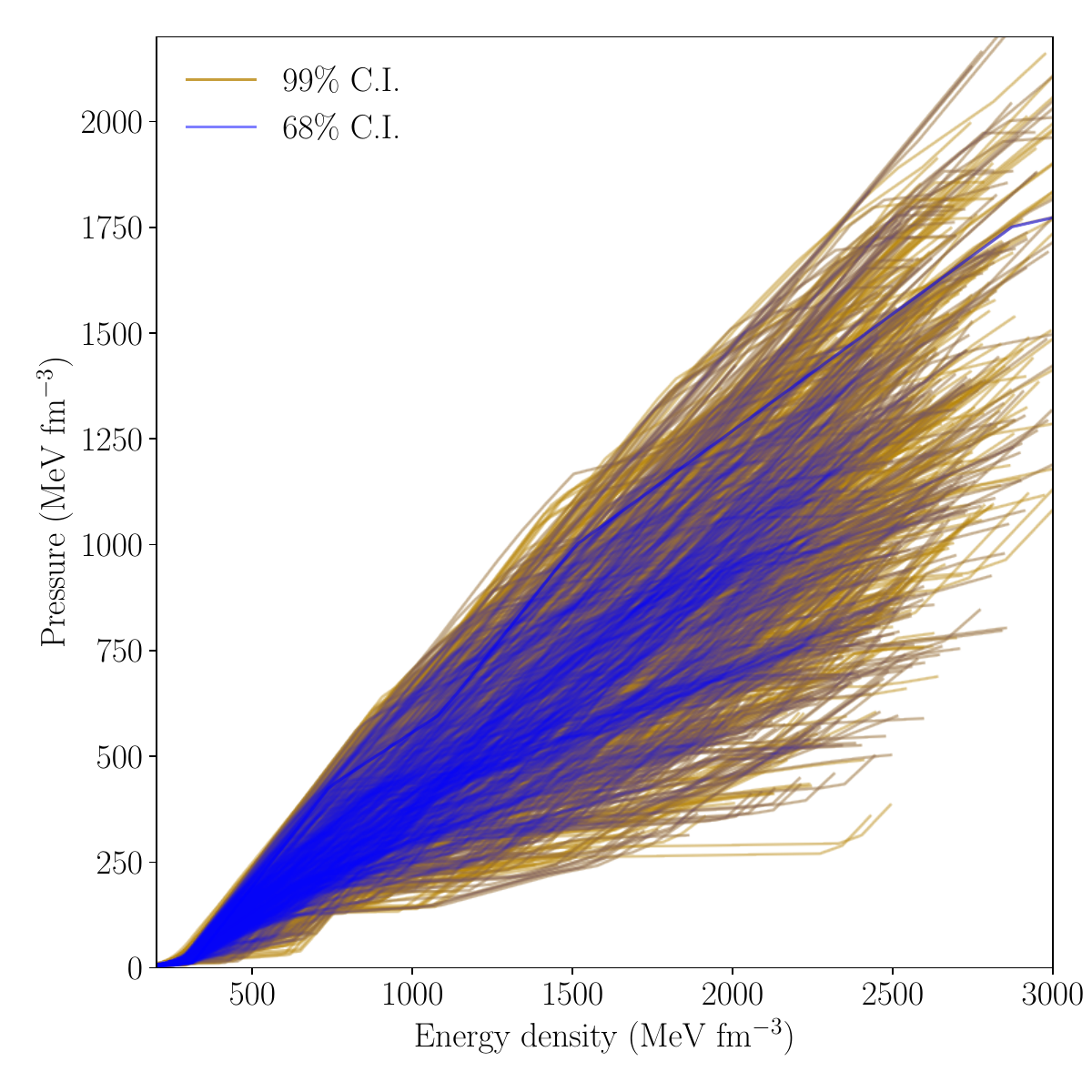}
    \caption{The brown lines indicate the {  99\% confidence interval of the} constrained EOS corresponding to the posterior distributions of figure \ref{fig:nuclear}  and the blue lines indicate the EOS which falls within the 68\% confidence interval of the constrained posterior distribution of the EOS parameters.}
    \label{fig:EOSplot}
\end{figure}

\begin{figure}
    \centering
    \includegraphics[scale=0.40]{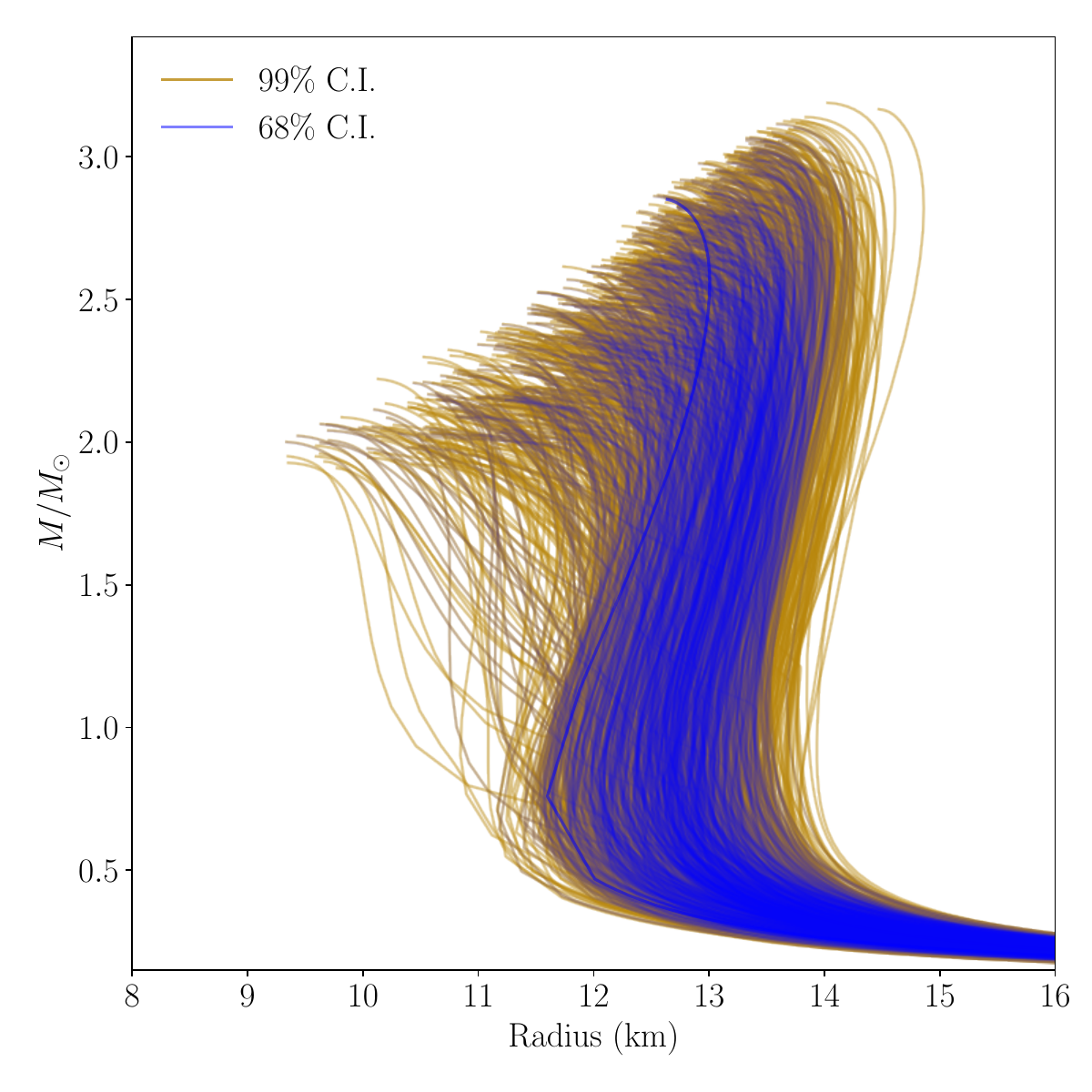}
    \caption{The M-R diagrams for all EOS within the {  99\% confidence interval of the posterior distribution} from figure \ref{fig:EOSplot} have been shown in the brown coloured lines and the blue lines indicate the M-R relations corresponding to the 68\% confidence interval of the posterior distribution of the constrained EOS parameters.}
    \label{fig:mass-radius}
\end{figure}

\section{Description of neutron star matter}\label{eos_des}
In this section, we describe the EOS of dense nuclear matter. We follow the work by  \textcite{Gandolfi:2019zpj} to construct a double polytrope EOS for our purpose. With this particular form of the function, \textcite{Gandolfi:2019zpj} could parametrize the results of the ab-initio microscopic calculations using Quantum Monte Carlo methods. 
We use this parametrization to create the EOS directly from nuclear empirical parameters (NMPs). We summarize the necessary equations in the following. The energy per particle for the pure neutron matter (PNM) is expressed as a function of density,

\begin{equation}
    E_{\rm{PNM}} (\rho) = a \left(\frac{\rho}{\rho_0}\right)^\alpha + b \left(\frac{\rho}{\rho_0}\right)^\beta , 
\end{equation}
where, $\rho_0$ is the nuclear saturation density.

The symmetry energy is defined as, $E_{\rm{sym}} (\rho) =  E_{\rm{PNM}} (\rho) - E_{\rm{SNM}} (\rho) $,  where $E_{\rm{SNM}} (\rho)$ is the energy per particle of the symmetric nuclear matter (SNM). Then, the slope of symmetry energy becomes, $L(\rho) = 3 \rho_0 \partial E_{\rm{sym}} (\rho) / \partial \rho $. At saturation, they can be expressed as,

\begin{eqnarray}
    J = E_{\rm{sym}} (\rho_0) &=& a + b - E_{\rm{SNM}}, \nonumber\\
    L(\rho_0) &=& 3(a \alpha + b \beta).
\label{eqn_sym}    
\end{eqnarray}

The density dependence of the symmetry energy can be parametrized by,
\begin{equation}
    E_{\rm{sym}} (\rho) = C \left( \frac{\rho}{\rho_0}\right)^\gamma.
\end{equation}
Thus, we get, $C = a + b - E_{\rm{SNM}}$ from equation \ref{eqn_sym} at $\rho=\rho_0$. Finally, from the condition, $P = \rho^2 \frac{\partial E_{\rm{SNM}}}{\partial \rho}|_{\rho=\rho_0}$, we get,

\begin{equation}
    \gamma = \frac{a \alpha + b \beta}{a + b - E_{\rm{SNM}}}.
\end{equation}

Using the relations above, the general expression of the EOS as a function of baryon number density ($\rho$) and proton fraction ($x=\rho_p/\rho$,  where $\rho_p$ is the proton density) becomes,
\begin{equation}
    E(\rho,x) = E_{\rm{PNM}} (\rho) +  C\left( \frac{\rho}{\rho_0}\right)^\gamma \left[ \left(1-2x \right)^2 -1 \right].
    \label{enr}
\end{equation}

To describe the NS matter, we also need to include the contribution from the lepton. In a cold NS, the lepton density is regulated by the $\beta$-equilibrium condition, $\mu_n = \mu_p + \mu_e$. Beyond $2\rho_0$, we construct the EOS using the piecewise constant speed of sound EOS model \cite{Tews:2018kmu} up to the density of $12 \rho_0$. 
We randomly sample the density points within this interval of the $c_s - \rho$ plane and the corresponding speed of sound points complying with the constraints $0 \leq c_s < 1$.
For the low-density part of the EOS, we have used the standard Baym-Pethick-Sutherland (BPS) EOS ($ \lesssim 0.001 ~{\rm fm}^{-3}$) for the outer crust and Negele-Vautherin (NV) EOS (up to $0.08 ~{\rm fm}^{-3}$) for the inner crust\cite{Baym:1971pw,Negele:1971vb}. The crust EOS is smoothly joined with the core EOS given by equation \ref{enr}.

After setting up our EOS model, we continue to explore the parameter space of our model using various theoretical, experimental, and observational constraints within a Bayesian analysis. We describe briefly the constraints and their implementation in the following.

First, we consider constraints around saturation density. The EOS generated within our model must satisfy the constraints coming from the chiral effective field theory ($\chi$-EFT) calculations. We used the results from \textcite{Drischler:2017wtt} to implement the ($\chi$-EFT) constraints for the EOSs in $\beta$-equilibrium in the range $(0.5,1.1)\rho_0$. The priors of our model parameters are given in table~\ref{prior_table}. The ranges of $\rho_0$ and $E_{\rm SNM}$ are chosen according to the experimental knowledge. Another condition is on the calculated values of $E_{\rm sym}$ and $L$ at saturation. We have imposed the $E_{\rm sym}$ to be in the range of 25 to 40 MeV and $L$ to be within 30 and 80 MeV.

We have used the mass-measurements of massive pulsars, and combined tidal deformability ($\tilde{\Lambda} $) from GW170817 as our observational constraints on the EOS. In particular, we have used the mass-measurement of J0740+6620  as $2.08 \pm 0.07 M_\odot$ reported by  \cite{Fonseca:2021wxt} as a Gaussian likelihood, 
\begin{equation}
    P(\text{data}_{M_{\text{max}}}|\mathbf{X}) = \frac{1}{2} \left[ 1 + {\rm erf} \left(\frac{M_{\text{max}}(\mathbf{X})/M_\odot - 2.08}{0.07\sqrt{2}} \right) \right],
    \label{mmax}
\end{equation}
 where, ($\mathbf{X} = a, \alpha, b, \beta, E_{\rm{SNM}}, \rho_0$) represents our model parameters. 
Then, we applied the improved constraints on $\tilde{\Lambda}$ from \textcite{LIGOScientific:2018hze}. The value of $\tilde{\Lambda}$ depends on the mass ratio $q=m_1/m_2$ and the chirp mass (${\cal M}_{\rm chirp} = \frac{(m_1 m_2)^{3/5}}{(m_1 + m_2)^{1/5}}$), where $m_1$, $m_2$ are the masses of primary and secondary objects of the binary system, respectively. The chirp mass has been determined rather precisely to be ${\cal M}_{\text{chirp}} = 1.186 \pm 0.001 M_\odot$ for GW170817. We use the publicly available data from LIGO-Virgo-Kagra (LVK) collaboration ~\footnote{LVK collaboration,~\href{https://dcc.ligo.org/LIGO-P1800115/public}{https://dcc.ligo.org/LIGO-P1800115/public}} assuming the low-spin prior and construct the likelihood as,

\begin{align} \label{eq:GW-likelihood}
    P({\rm data}_{\rm LVK}|\mathbf{X}) = \int dm_1 \int dm_2  P(m_1,m_2|\mathbf{X}) \nonumber\\
    \times P({\rm data}_{\rm LVK} | m_1, m_2, \Lambda_1(m_1,\mathbf{X}), \Lambda_2(m_2,\mathbf{X})) \,,
\end{align}

where, $P(m_1,m_2|\mathbf{X})$ is the prior distribution for the component masses of the binary. For simplicity, we have chosen a uniform prior for $m_1$ and $m_2$. We use a Gaussian kernel density estimator to construct the GW likelihood from the discrete data.
We have fixed the chirp mass to its median value because of the high precision of the measurement. Then, we construct binaries corresponding to that chirp mass by varying  $m_1$ and determine the corresponding $m_2$. For each EOS, we compute the tidal deformabilities ($\Lambda_1, \Lambda_2$) for the pairs of ($m_1,m_2$), and find the probability using equation \ref{eq:GW-likelihood}.

The final likelihood function with all the astrophysical constraints is simply the product,
\begin{equation}
    P(\text{data}_{\text{astro}}|\mathbf{X}) = P(\text{data}_{M_{\text{max}}}|\mathbf{X}) \times P(\text{data}_{\text{LVK}}|\mathbf{X}).
\end{equation}

\begin{table}
\begin{tabular}{l l}
\hline
\hline
Parameter & Prior                     \\ \hline \hline
$\alpha$     & $\mathcal{U}(0.2,2)$                   \\ \hline
$\beta$      & $\mathcal{U}(-0.5,5)$                 \\ \hline
$a$ ~~~~~~(MeV)        & $\mathcal{U}(5,25)$                    \\ \hline
$b$ ~~~~~~(MeV)      & $\mathcal{U}(10^{-4},8)$\\ \hline
E$_{\text{SNM}}$ (MeV)  & $\mathcal{U}(-18,-14)$                  \\ \hline
$\rho_0$   ~~~~~(fm$^{-3}$)     &  $\mathcal{U}(0.14,0.17)$           \\ \hline
\end{tabular}
\caption{Summary of the prior range used for our EOS model parameter. \label{prior_table}}
\end{table}
Finally, we sample the posterior using a nested sampling algorithm \cite{Skilling:2006gxv}, in the {\tt dynesty} software package \cite{Speagle:2019ivv}. The posterior distributions of the EOS parameters for nuclear matter are shown in Fig.~\ref{fig:nuclear}. We have also included the derived values for $E_{\rm sym}$ and $L$ at $\rho_0$. We have found $E_{\rm sym} (\rho_0) = 28.95^{+3.31}_{-2.17} $ MeV and $L=53.28^{+6.56}_{-5.16}$ MeV at 68\% confidence interval (C.I.). These values are well within the agreed ranges found in literature \cite{Oertel:2016bki,Margueron:2017eqc}. This affirms the robustness of our EOS samples used to study the rotational properties of pulsars in this work. We have shown the corresponding EOSs and their mass-radius curves in Fig.~\ref{fig:EOSplot} and \ref{fig:mass-radius}, respectively. We can see from Fig.~\ref{fig:mass-radius} that the EOSs generated in our exercise span a large range of radii for a $1.4 M_\odot$ star. Therefore, these curves are able to represent the parameter space of neutron star matter. Note that, we have not used the data from NICER measurements in our analysis as we see that our 68\% mass-radius curves in Fig.~\ref{fig:mass-radius} are consistent with 68\% of those sources.

\section{Rotational effects on the structure of Neutron stars}\label{rot_effect}
To study the effect of rotation on the macroscopic quantities of the star we have used the {\tt RNS} code \footnote{\href{https://github.com/cgca/rns}{https://github.com/cgca/rns}}  \citep{Stergioulas:1994ea,RNS}. For every EOS computed in the Sec~\ref{eos_des}, we calculate the gravitational mass ($M$), the equatorial radius ($R_e$), the ratio of the polar to the equatorial radius ($R_p/R_e$) and the moment of inertia ($I$) as a function of rotation frequency ($f=\Omega/2\pi$, where $\Omega$ is the angular velocity of the star), keeping the baryon mass ($M_b$) fixed at $2M_\odot$. 
\begin{figure*}
    \centering
    \includegraphics[scale=0.8]{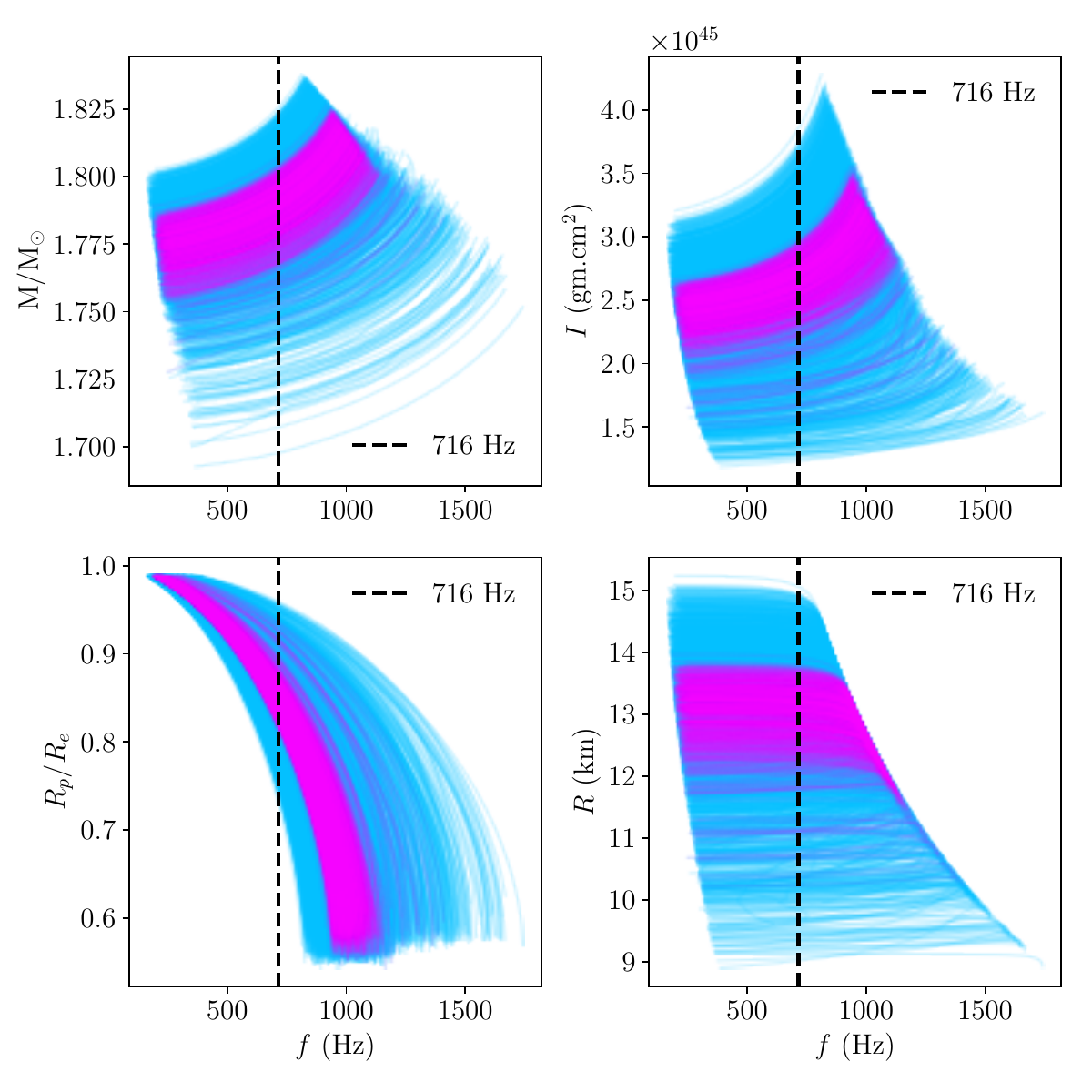}
    \caption{This figure shows the variation of four structural quantities of neutron stars as a function of the rotation frequency for all EOS obtained in Sec~\ref{eos_des}. The blue lines indicate the relation for all $\sim$6K EOS and the purple lines corresponds to the solutions of EOS constructed from the 68\% C.I. of the posterior distribution of parameters shown in Fig.~\ref{fig:nuclear}. The upper left panel shows the variation of the gravitational mass as the function of the rotation frequency, the upper right panel shows the variation of the moment of inertia, the lower left panel shows the variation of the ratio of the polar to {  equatorial } radius and the lower right panel shows the variation of average radius of the star as the function of rotation frequency.  For every EOS the baryon mass of the star is kept fixed at 2$M_\odot$, further details can be found in the Sec.~\ref{rot_effect}.}
    \label{fig:shape_vs_freq}
\end{figure*}

A rapidly rotating star can sustain more mass than a non-rotating star with the same central density. To obtain the sequence of the above-mentioned macroscopic quantities with rotation frequency, we treat central density as the parameter and search for the frequency which a star of baryon mass of 2$M_\odot$ can sustain. However, since a low central density leads to a low-mass star, the frequency needed to meet the $M_b=2M_\odot$ condition may be more than the Kepler frequency (defined as the maximum frequency of rotation that a star can support before mass-shedding) if the central density is too low. Hence, the {\tt RNS} code does not converge. As we increase the central density, the $M_b=2M_\odot$ criteria is fulfilled at increasingly lower frequencies. When the central density is very high, the condition of $M_b=2M_\odot$ is achieved at a frequency ($\lesssim 200$ Hz) that is too low for the {\tt RNS} code to handle. Therefore, we identify the densities corresponding to the highest and the  lowest frequencies for which the {\tt RNS} code converges. Subsequently, we generate the sequence between those two densities for all $\sim$ 6000 EOSs in an automated fashion in our pipeline built around the {\tt RNS} code.

The rapid rotation generates a large amount of centrifugal force, which leads to the departure from the spherical shape of the star that is characterized by the ratio of the polar to the equatorial radius $R_p/R_e$. The dependence of this ratio on frequency is shown in the bottom left panel of Fig.~\ref{fig:shape_vs_freq} for all EOS. Due to the deformation in the shape of the star, the definition of the radius becomes ambiguous in such a case, yet, in many scenarios, for estimating various physical quantities, the radius of the star is required as discussed in Sec.~\ref{BraIn}. Hence, in this work, we define a frequency-dependent radius by averaging over the circumferential radius of a rotating star given as 
\begin{equation}\label{average_radius}
    R \equiv \left <R(\Omega) \right > = \frac{\int_0^{2\pi} R(\theta, \Omega) d\theta}{ \int_0^{2\pi} d\theta}.
\end{equation}
The expression of circumferential radius $R(\theta, \Omega) = R_e[1-{\Bar{\Omega}}^2(0.788 -1.03x)\cos^2 \theta]$ is taken from the previous work by \textcite{AM+2014} and \textcite{SPW+2020}, where $\theta$ is the polar angle, $\Bar{\Omega} = \Omega(R^3_e/GM)^{1/2}$ and $x=GM/c^2R_e$. The EOS and rotation frequency-dependent average radius of the star is shown in the lower right panel of Fig.~\ref{fig:shape_vs_freq}, which can be seen to decrease strongly at higher frequencies ($\gtrsim 850$Hz). Also at the higher frequency, the $R_p$ can be $\sim 25\%$ smaller than the $R_e$, therefore our method of computing a unique value of a radius and attributing a spherical shape to the star may have some shortcomings even if it mathematically renders a valid solution. In the upper left and right panels of Fig.~\ref{fig:shape_vs_freq} we show the variation of the gravitational mass and moment of inertia as a function of the rotation frequency for all EOS. It is evident from Fig.~\ref{fig:shape_vs_freq} that all these physical quantities have significant frequency dependence and will contribute to the braking index of the NS which we will discuss in the next section. We have confined our studies to the rotational frequency above $\gtrsim 200$ Hz following \citet{Hamil+2015}, where the frequency dependence of the macroscopic quantities has been shown to be prominent above  $\gtrsim 200$ Hz only. 

The high-frequency behaviour of the macroscopic quantities observed in Fig.~\ref{fig:shape_vs_freq} is controlled by the stiffness of the EOSs. At a given density, a stiffer EOS produces more pressure than a softer one and, therefore, can defy gravity more easily. Hence, if there is no rotation, the stiffer EOS will make a star of a larger radius and a lower central density if a certain number of baryons (equivalent to a fixed baryon mass) are put together. Consequently, the star with the stiffer EOS will be less bound, resulting in a higher gravitational mass. Note the gravitational binding energy is defined as $M_b-M$. This behaviour remains even if the stars rotate. Therefore, the upper curves in $M$-vs-$f$, $R$-vs-$f$, and $I$-vs-$f$ plots in Fig.~\ref{fig:shape_vs_freq} correspond to stiffer EOSs. With increasing frequency, the equatorial radii of all the stars (irrespective of the underlying EOS) grow, making them less bound and having higher gravitational mass. Furthermore, since at a given rotational frequency, the star with the stiffer EOS is less bound than the softer EOS, it becomes unstable at a lower frequency. In other words, the Kepler frequency is smaller in the case of stiffer EOSs.  It explains the behaviour seen in the $M-f$ plot.
The $R_p/R_e$-vs-$f$ curve captures the dependence of the stellar deformation on the rotation frequency. It is clear from the above discussion that for the stiffer EOSs, the equatorial radius increases much faster with the small change to the rotation frequency compared to the softer EOSs. Consequently, the ratio $R_p/R_e$ falls faster in the case of stiffer EOSs. Therefore, the left curves in the $R_p/R_e$-vs-$f$ plot belong to stiffer EOSs.

\section{Braking Index}\label{BraIn}
In this section, we present the derivation of the braking index and its dependence on the spin frequency. Later in this section, we discuss the results in the context of millisecond pulsars and newly born millisecond magnetars in detail.

\subsection{Calculation of braking index}\label{deriv_brake}
Pulsars spin down at the cost of their rotational kinetic energy.  If the spin-down energy $\dot E$ through a mechanism is proportional to $-\Omega^{n+1}$, the spin-down relation is obtained as
\begin{equation}\label{genspd}
    \frac{d}{dt} \left (\frac{1}{2} I \Omega^2 \right ) = I \Omega \dot \Omega \propto -\Omega^{n+1} \Rightarrow \dot \Omega = -k \Omega^n,
\end{equation}
where $k$ is a constant which depends on various physical parameters of the neutron stars depending on the mechanism of the emission loss discussed later in this section.
The braking index $n$ depends both on the first and the second derivative of the spin frequencies as
\begin{equation}\label{braking_index}
    n = \frac{ \Omega \ddot \Omega}{\dot \Omega^2}.
\end{equation}
The numeric value of the braking index is the signature of the mechanism which dominates the spin-down of a pulsar.

The energy loss by the MDR is given as
\begin{eqnarray}\label{em-spin-down}
|\dot E_{\rm{EM}}| = \frac{2}{3c^3}\mu^2\Omega^4\sin^2\alpha \equiv \ems R^6 \Omega^4,
\end{eqnarray}
where the magnetic moment $|\mu|^2=B_p^2R^6/4$, with the surface magnetic field strength $B_p$, stellar radius $R$, $\alpha$ is the angle of inclination between the rotation and the magnetic axis and $\ems =B_p^2\sin^2 \alpha/6c^3$.
The corresponding spin-down rate using Eq. \ref{genspd} leads to 
\begin{equation}\label{sd-mdr}
    \dot \Omega_{\rm MDR} = -\frac{2B_p^2 R^6 \sin^2\alpha}{3Ic^3} \Omega^3 \Rightarrow -k_{\rm MDR} \Omega^3,
\end{equation}
indicating the MDR leads to braking index 3. However, in Sec.~\ref{rot_effect} we have noticed at higher spin-frequency $R$ and $I$ depend on $f$ making $k_{\rm MDR}$ dependent on $f$ and hence the braking index. On the other hand, the $f$ dependence of $I$, will not allow us to arrive at the expression \ref{sd-mdr}, as $dI/d\Omega \neq 0$. Hence, for a rapidly rotating star, the modification to the spin-down expression is essential.

The expression given in Eq.~\ref{em-spin-down} is valid if the external magnetic field is purely dipole. However, studies show the presence of complex structures of higher order magnetic multipole near the stellar surface \citep{Kalapotharakos+2021}. But, at a larger distance from the surface, the approximation of the dipolar field line is still valid. Therefore in this work, we have assumed that the pulsar magnetosphere is purely dipolar. 
Apart, from the geometrical consideration of the magnetic field, we also ignore any possible evolution of the dipolar field with time. Otherwise, it would contribute by an amount $2 \Omega \dot B/\dot \Omega B$ to the braking index. Similarly, the evolution of $\alpha$ would allow the departure of the braking index by an amount $4 \Omega \dot \alpha / \dot \Omega \tan \alpha$, which has been also ignored for the current study. As we do not consider any evolution in $B$ and $\alpha$ we use them as a parameter in our model.

A long-lived, non-axisymmetric deformation of neutron stars would result in energy loss via the continuous GW (details can be found in a recent review by \textcite{grittin+2024}). These deformations are referred to as mountains and generate a certain amount of ellipticity ($\epsilon$), whose magnitude is still uncertain, depending on the process leading to the formation of the mountains and the crustal EOS. However, a typical value of maximum ellipticity that a neutron star crust can sustain is  $\sim 10^{-6}$ \citep{ushomirsky+2000, haskell+2006, Grittin+2021}. The strong magnetic field can lead to the deformation of a star generating $\epsilon \sim 10^{-12}$ and the misalignment between the magnetic axis and the rotational axis leads to the GW emission \citep{HSGA+2008, KM+2020, SBD+2021}. Similarly, mountains are generated from the accretion process, they can be either thermal mountains \citep{Bilsden+1998} or magnetic mountains. The thermal mountains produced from the temperature-sensitive nuclear reactions on the surface of the neutron stars survive for a much shorter duration $\sim 0.2 $ yrs (this a typical value for more details refer to \textcite{grittin+2024} and the references therein) between the accretion-mediated outburst phase of neutron stars. However, the magnetic field-supported mountains of the accreted material may survive over much longer time scales $\gtrsim 10^8$ yr \citep{Vig_mel+2009} giving rise to $\epsilon \sim 10^{-7}$ to $10^{-8}$. Implies, that the mountain formed at the end of the accretion phase (even after the companion's disappearance) can survive for a decent fraction of a neutron star's lifetime. The amount of energy loss via the GW emission is proportional to the sixth power of the spin frequency and the ellipticity of the star, and is given as
\begin{equation}\label{spin-downquadrupole}
    |\dot E^Q_{\rm{GW}}| = \frac{32}{5} \frac{G}{c^5} I^2\epsilon^2 \Omega^6 \equiv \qsd I^2 \epsilon^2 \Omega^6,
\end{equation}
where $\qsd = 32G/5c^5$. The corresponding spin-down rate, following the Eq.~\ref{genspd} is 
\begin{equation}\label{qgwspd}
\dot \Omega^Q_{\rm GW} = -\qsd I \epsilon^2\Omega^5 \Rightarrow = -k^Q_{\rm GW} \Omega^5,
\end{equation}
indicating the braking index in this case is 5. However, a similar argument like in the case of MDR spin-down rate given in Eq.\ref{sd-mdr}, is valid here, where rotational effects on $I$ would not lead to the expression we have arrived at Eq.~\ref{qgwspd}. 

The all-sky search for the continuous gravitational waves did not lead to the discovery of a signal but has enabled to establish an upper limit of $\epsilon \lesssim 10^{-6}$(depending on the distance and the frequency) \citep{LIGO+2019_ellipticity}. Hence, in our case, we use $\epsilon = 10^{-7}$ which is an order of magnitude lower than the established upper limit. 

The r-modes are quasi-toroidal oscillations in neutron stars where the Coriolis force acts as the restoring force \citep{Owen:1998xg}. Counterrotating r-modes can become unstable to gravitational radiation reaction via the Chandrasekhar-Friedman-Schutz (CFS) mechanism \citep{Chandrasekhar:1970pjp,Friedman:1978hf}. It has been shown that in the absence of any fluid dissipation, CFS instability can arise at any rotational frequency of the star \citep{Andersson:1997xt,Friedman:1997uh,Bildsten:1998ey}. In realistic neutron stars, however, several damping mechanisms are present, such as due to the bulk and shear viscosities, which lead to the r-modes amplitude achieving a saturation value ($\alpha_{\rm S}$), as shown in \textcite{Owen:1998xg}. The microscopic origin of the damping mechanisms relates to the nature and the temperature of the neutron star matter. In the angular velocity-temperature ($\Omega-T$) plane, the interplay between the gravitational wave emission timescale and viscous dissipation timescales curve out an instability region. When the star resides within the unstable region, the gravitational wave emission contributes significantly to the rotational energy loss becoming another mechanism for the spindown of the star. The frequency of the emitted GW is 4/3 the rotation frequency of the star \citep{Riles+2023}. Considering previous studies, the r-mode saturation amplitude can attain a value between $10^{-4}$ to $10^{-1}$. For these values of $\alpha_{\rm S}$, the energy loss due to r-mode is given by \cite{Owen:1998xg, Thorne:1980ru},

\begin{equation}\label{rosbby-mode}
|\dot E^R_{\rm GW}| = \left (\frac{4}{3} \right) ^8\frac{4 \pi G}{25 c^7}\alpha_{\rm S}^2M^2R^6\Omega^8\mathcal{J}^2 \equiv \rsd \alpha_{\rm S}^2M^2R^6\Omega^{8}\mathcal{J}^2
\end{equation}
where, $\rsd = \left (\frac{4}{3} \right) ^8\frac{4 \pi G}{25 c^7} $, and $\mathcal{J}$ is given by,

\begin{equation}
    \mathcal{J} = \frac{1}{MR^4} \int_0^R  \rho(r) r^6 dr.
\end{equation}
The expression \ref{genspd} gives the spin-down rate as 
\begin{equation}\label{sp-rmode}
    \dot \Omega^R_{\rm GW} = -\rsd \frac{\alpha_{\rm S}^2M^2R^6 \mathcal{J}^2}{I} \Omega^7,
\end{equation}
indicating the spin-down due to r-mode leads to a braking index of 7. The above expression \ref{sp-rmode} is under the approximation of the no rotational effect on the stellar structure. However, we have shown that for rapid rotation there exists significant structural evolution, hence a revision to the above expressions is important.

When all the above three mechanisms of energy loss are active, the rotational energy loss is given by 
\begin{equation}\label{energy-balance}
    \frac {d} {dt} \left (\frac{1}{2} I \Omega^2 \right) = -\dot E_{\rm{EM}} - \dot E^{Q}_{\rm{GW}} - \dot E^{R}_{\rm{GW}},
\end{equation}
where $\dot E_{\text{EM}}$, $\dot E^Q_{\text{GW}}$ and $\dot E^R_{\text{GW}}$ are the rate of loss of energy via the electromagnetic radiation, gravitational quadrupolar radiation and the r-mode. The relation between the spin-down rate and the other macroscopic properties of the neutron star's interior and magnetosphere can be obtained by expanding the left-hand side derivative of the Equation \ref{energy-balance}, where we consider a non-zero time evolution of the moment of inertia.
\begin{align}\label{deriv1}
    \dot I + 2 \dot \Omega I = -2\ems R^6 \Omega^3 -2\qsd I^2 \epsilon^2 \Omega^5 - 2\rsd \rsat^2M^2R^6\Omega^7 \mathcal{J}^2
\end{align}
In Equation \ref{deriv1}, we use the chain rule to write $\dot I = \frac{dI}{d\Omega} \frac{d\Omega}{dt} = I^\prime \dot \Omega$ and further simply the equation to obtain $\dot \Omega$ as 
\begin{equation}\label{omegadot}
    \dot \Omega =- \frac{2\ems R^6 \Omega^3 +2\qsd I^2\epsilon^2\Omega^5 +2\rsd\rsat^2 M^2 R^6 \Omega^7 \mathcal{J}^2}{I^\prime \Omega + 2 I}
\end{equation}
Similarly by taking the derivative of Equation \ref{deriv1} and applying the chain rule to write $\ddot I = I^{\prime \prime} \dot \Omega^2 + I^\prime \ddot \Omega$, we obtain
\begin{eqnarray}\label{omegaddot} 
    \ddot \Omega &=& [-6\ems \Omega^2 \dot \Omega R^5 ( 2 R^\prime  \Omega + R) \nonumber \\ 
    &-&2 \qsd I \epsilon \dot \Omega \Omega^4(2I^\prime \epsilon \Omega + 
     5 I \epsilon)  \nonumber\\ 
    &-&  2\rsd\rsat^2 M R^5 \Omega^6 \dot\Omega \mathcal{J} \times \nonumber\\
    && (2M^\prime R \Omega  + 6MR^\prime\Omega + 2MR\Omega \mathcal{J}^\prime  + 7MR)  \nonumber\\ 
    &-&3I^\prime \dot \Omega^2 - I^{\prime \prime} \dot \Omega^2 \Omega]/[2I + I^\prime \Omega].
\end{eqnarray}
All the quantities with ($^{\prime}$) denote the derivative with respect to the angular velocity $\Omega$. Combining the $\dot \Omega$ and $\ddot \Omega$ from the expressions in the given in the Equation \ref{omegadot} and \ref{omegaddot}, we obtain the braking index using the relation \ref{braking_index} as a function of the spin angular velocity $\Omega$.

\begin{widetext}
\begin{eqnarray}
&&n(\Omega) = n_{\rm EM} + n^Q_{\rm{GW}} + n^R_{\rm{GW}}+ n_{I},~~~ \rm {where} \nonumber\\
&&n_{\rm EM} = \frac{6\ems R^5 \Omega^3 (2R^\prime \Omega +R)}{2\ems R^6 \Omega^3 + 2\qsd I^2 \epsilon^2 \Omega^5 + 2\rsd \rsat^2 M^2 R^6 \Omega^7 \mathcal{J}^2}, \nonumber\\
&&n^Q_{\rm{GW}} = \frac{2 \qsd I \Omega^5 \epsilon^2(2I^\prime \Omega 
     + 5 I)}{2\ems R^6 \Omega^3 + 2\qsd I^2 \epsilon^2 \Omega^5 + 2\rsd \rsat^2 M^2 R^6 \Omega^7 \mathcal{J}^2}, \nonumber\\
&&n^R_{\rm{GW}} = 2\rsd \rsat^2R^5M\Omega^7 \mathcal{J} \times 
    \frac{(2M^\prime R \Omega \mathcal{J} + 6 MR^\prime \Omega  \mathcal{J}+2MR\mathcal{J}^\prime \Omega + 7MR\mathcal{J})}{2\ems R^6 \Omega^3 + 2\qsd I^2 \epsilon^2 \Omega^5 + 2\rsd \rsat^2 M^2 R^6 \Omega^7 \mathcal{J}^2},   \nonumber\\
&&    n_{I} = - \frac{3I^\prime \Omega + I^{\prime \prime} \Omega^2}{2I + I^\prime \Omega}.
\label{fullbi}     
\end{eqnarray}
\end{widetext}

The expression of $n(\Omega)$ above shows the dependence of the braking index on the spin frequency, which is further dependent on the EOS through the frequency derivative of the macroscopic quantities such as $M$, $R$ and $I$. The term $n_{\rm EM}$ is the contribution of MDR loss, $n^Q_{\rm{GW}}$ is the contribution from GW radiation due to long-lived deformations, $n^R_{\rm{GW}}$ is the contribution from the emission loss via r-mode oscillations. The term $n_{I}$ depends purely on the moment of inertia and its higher-order derivatives, which has been previously derived in \textcite{GPW:1997}.
The quantities $\ems, \epsilon$ and $\alpha_{\rm S}$ can be considered as switches which by setting to zero help to analyze the braking index in the absence of a particular channel of energy loss. For example, setting $\alpha_{\rm S} =0$ will help to understand the evolution of the braking index purely due to the electric dipolar emission loss and the GW emission loss due to finite $\epsilon$. In the limit of slow rotation where all the derivatives w.r.t $\Omega$ tend to zero, by setting $\epsilon$ and $\alpha_{\rm S}$ to zero we obtain $n=3$, by setting $\ems$ and $\alpha_{\rm S}= 0$, we obtain $n=5$ and finally by setting $\ems=\epsilon=0$, we obtain $n=7$, which are in agreement with the limiting cases known in the literature.  

\begin{figure*}
    \centering
    \includegraphics[scale=0.90]{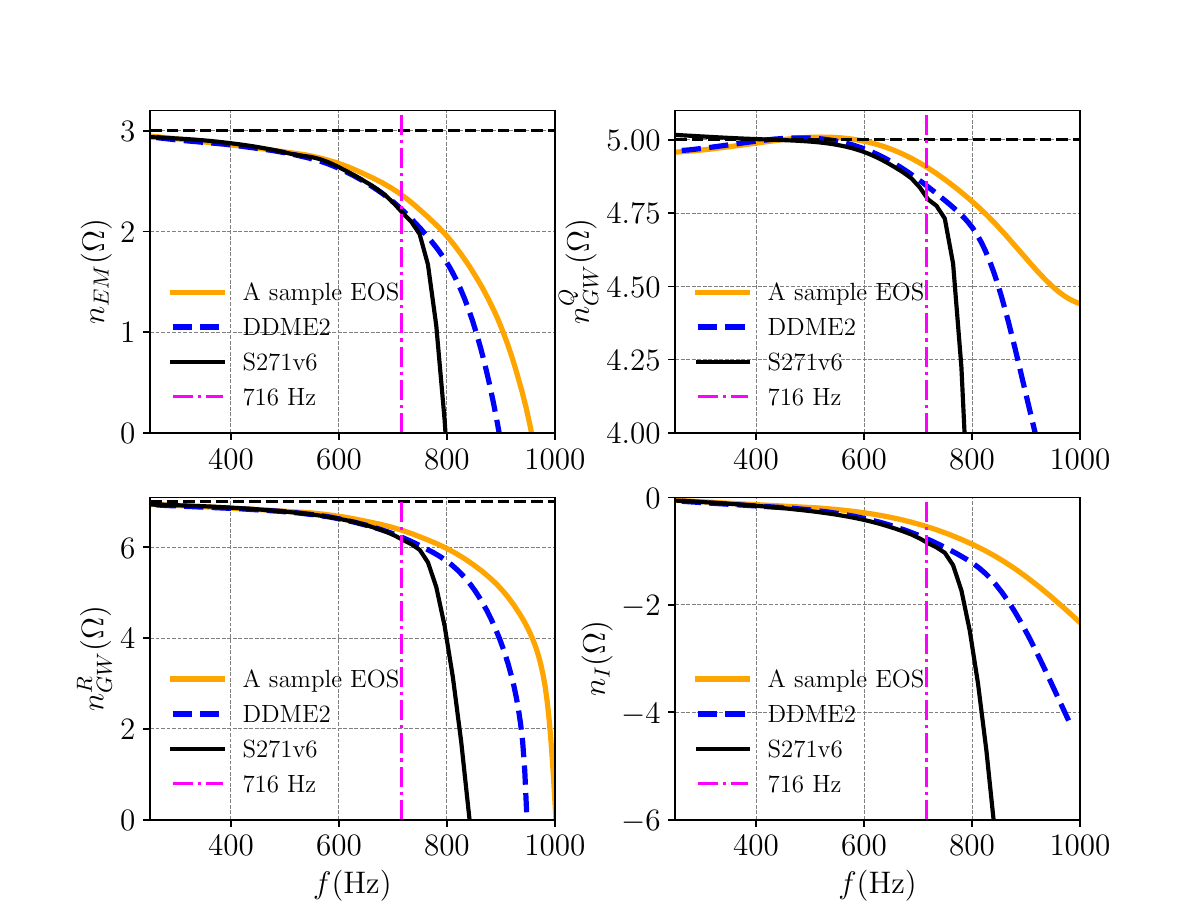}
    \caption{This figure shows the effect of rotation on the braking index of pulsars (see Sec.~\ref{rot_effect} for detailed discussions and references). {  The upper left panel show the frequency dependence of the magnetic braking index, the upper right panel shows the frequency dependence of the braking index if the energy loss happens only through the GW emission in the presence of long-lived deformations, the lower left panel shows the frequency dependence on the braking index due r-mode instability, and the lower right panel shows the variation of $n_I$ as a function of rotation frequency. All these curves are for stars with a baryon mass of 2$M_\odot$.} The solid yellow line represents one of the EOS from our sample (see Sec.~\ref{res1} for the details), the blue dashed line indicates the DDME2 EOS, and the black solid line represents the S271v6 EOS. The vertical magenta line corresponds to $f=716$ Hz, the maximum spin-frequency of a pulsar discovered to date.}
    \label{fig:braking_index_eos}
\end{figure*}

\subsection{Effect of rapid rotation on $n_{\rm EM}, n^Q_{\rm{GW}}$ and $n^R_{\rm{GW}}$}\label{res1}
In Sec~\ref{deriv_brake}, we have shown how the expressions in Eq. \ref{fullbi} reduce to the known values of the braking index when the rotational effects are ignored by setting all derivatives w.r.t. $\Omega$ to zero. However, when the effect of rotation is taken into account, the departure of $n_{\rm EM}, n^Q_{\rm{GW}}$ and $ n^R_{\rm{GW}}$ from 3, 5 and 7 occurs, which depends on the EOS through the variation of the macroscopic quantities and their derivatives as the function of the $\Omega$. 

If the energy loss is considered through the magnetic dipolar radiation only, then 
\begin{equation}\label{MDR-rot}
    n_{\rm EM} = 3\left (1+ \frac{2R^\prime \Omega}{R} \right ) +n_{I},
\end{equation}
 which can be obtained by setting $\epsilon$ and $\alpha_{\rm S} =0$. Similarly by assuming the loss of energy is via the GW emission due to the non-axisymmetric long-lived deformation (that is by setting $B_p=\alpha_{\rm S} =0$) the braking index is given as 
\begin{equation}\label{GWQ-rot}
     n^Q_{\rm{GW}} = 5\left (1+ \frac{2I^\prime \Omega}{5I}\right ) +n_{I}.
\end{equation}
Likewise, for the emission to happen purely from the r-mode oscillation in the absence of the electromagnetic dipolar radiation and the GW emission from deformations the braking index takes the form
\begin{equation}\label{GWR-rot}
    n^R_{\rm{GW}} = 7\left (1 + \frac{2M^\prime \Omega}{7M} + \frac{6R^\prime \Omega}{7R} + \frac{2\mathcal{J}^\prime \Omega}{7\mathcal{J} }\right ) +n_{I},
\end{equation}
which is obtained by setting $B_p=\epsilon=0$. In Fig.~\ref{fig:braking_index_eos} we have shown the dependence of these three braking indices $n_{\rm EM}, n^Q_{\rm{GW}}$ and $ n^R_{\rm{GW}}$ separately as a function of $f$, for three different EOS. The solid yellow line corresponds to one of the EOS described in Sec.~\ref{eos_des} with the highest weight (the parameters of the EOS are: $\alpha=$ 0.699, $\beta =$ 2.151, $a = 9.423$ MeV, $b=5.662$ MeV, $\rho_0=0.152$ fm$^{-3}$ and E$_{\rm SNM}=-17.934$ MeV), the blue dashed line is for the density-dependent relativistic mean field (RMF) EOS, DDME2 \citep{Lalazissis:2005de}  and another RMF EOS with nonlinear meson field couplings, S271v6 \citep{Horowitz:2002mb} is shown by the black solid line. These phenomenological EOSs are chosen as examples because they have been shown to satisfy most of the current constraints \cite{Nandi:2018ami}. The departure of $n_{\rm EM}, n^Q_{\rm{GW}}$ and $ n^R_{\rm{GW}}$ from the known constant values of 3, 5, and 7, respectively, is due to the rotational effect and {  independent of the magnitude of the magnetic field, deformation, and the r-mode saturation amplitude. This implies that if an NS loses energy purely through any of these mechanisms, the deviation of the braking index is due only to structural changes.}

The response to rapid rotation depends on the stiffness of the EOS, which is different for different EOSs (refer to Sec.~\ref{rot_effect} for detailed discussion), giving rise to different braking index curves. 
However, the differences in $n_{\rm EM}$ and $n^R_{\rm{GW}}$ for different EOS (at least for the three representative ones shown in Fig.~\ref{fig:braking_index_eos}) are negligible at frequencies $\lesssim 550$ Hz.
Both $n_{\rm EM}$ and $n^R_{\rm{GW}}$ systematically decrease with increasing spin frequency, which indicates that at frequencies $\gtrsim 200$ Hz the contribution of $n_{I}$ is significant. The importance of $n_{I}$ can be argued from Fig.~\ref{fig:shape_vs_freq}, where $R$ is almost constant at $f\sim 200-750$ Hz, indicating $R^\prime \sim 0$, which would lead to $n_{\rm EM} \simeq 3$ of Eq.~\ref{MDR-rot} in this frequency range. Hence, the deviation seen in $n_{\rm EM}$ from 3 in Fig.~\ref{fig:braking_index_eos} can be attributed to $n_{I}$, which becomes increasingly negative with frequency (see bottom right panel of Fig. \ref{fig:braking_index_eos}). A similar trend in the variation of $n^R_{\rm{GW}}$ can be seen in Fig.~\ref{fig:braking_index_eos}, which is also attributed to $n_{I}$. The finite but small positive value of $M^\prime /M$ and the finite but small negative value of $\mathcal{J}^\prime/\mathcal{J}$ lead to a slope different from $n_{\rm  EM}$ in the initial fall of $n^R_{\rm{GW}}$ from the value 7.

The dependence of $n^Q_{\rm{GW}}$ on spin frequency comes only from $I$ and its derivative, which changes appreciably even at $f \gtrsim 200$ Hz, making $n^Q_{\rm{GW}}$ always $> 5$ (the first term of Eq.~\ref{GWQ-rot}) and an increasing function of $f$ in the absence of $n_{I}$. Whereas $n_{I}$ is always $< 0$, and the nature of its fall depends on the EOS. Therefore, there may exist a range of frequencies for which $n^Q_{\rm{GW}} > 5$ in the presence of the $n_{I}$ term, as shown in Fig.~\ref{fig:braking_index_eos}. However, unlike $n_{\rm EM}$ and $n^R_{\rm{GW}}$ (at least for these EOS), we find distinguishable EOS-dependent variation in $n^Q_{\rm{GW}}$ even at frequencies $\sim 250-300$ Hz. But as the spin frequency increases, the variation of all three braking indices for different EOS can be identified clearly at the $f \sim 716$ Hz, the spin frequency of PSR J1748$-$2446ad (maximum spin frequency of a pulsar known till date), an eclipsing binary millisecond pulsar in the globular cluster Terzan 5 \citep{HRS+2006}. However, at much larger frequencies ($f \gtrsim 850$ Hz), we find the large change in radius (see Fig.~\ref{fig:braking_index_eos}), indicating the size decreases with frequency, which could be an artifact because of forceful fitting of a sphere to a largely deformed star. Hence, the results beyond $f \gtrsim 850$ Hz must be interpreted cautiously.

\begin{figure*}
    \centering
    \includegraphics[scale=0.65]{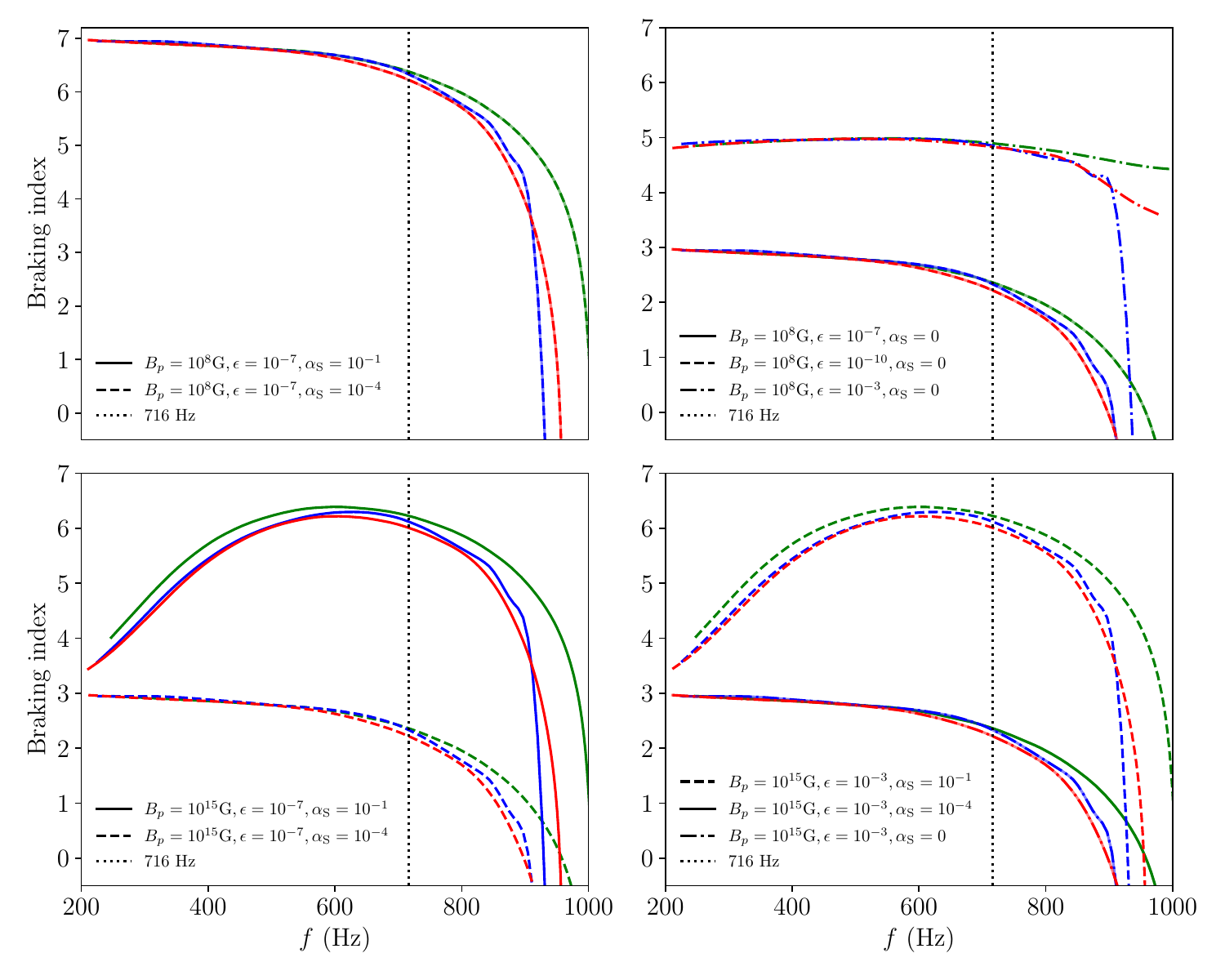}
    \caption{This figure captures the dependence of braking index on the magnetic field strength $B_p$, the ellipticity due to long-lived deformation $\epsilon$, and the r-mode saturation amplitude $\alpha_{\rm S}$ as a function of rotation frequency.  We use a constant inclination angle $\alpha=45^\circ$. The variation has been studied using three representative EOS, the red lines indicate DDME2 EOS, the blue lines indicate S271v6 EOS and the green lines indicate an EOS from our sample (see Sec.~\ref{res2} for details). The results for different combinations of $B_p, \epsilon$ and $\alpha_{\rm S}$ have been shown with different line styles. Refer to Sec.~\ref{res2} for detailed discussion and implications.}
    \label{fig:parameter_explore}
\end{figure*}

\subsection{Effect of $B_p$, $\epsilon$ and $\alpha_{\rm S}$ on the braking index.}\label{res2}
In the previous Sec.~\ref{res1}, we have discussed the rotational effect on all three separate components of the braking index individually in the absence of the other two channels of energy loss. {  There we found that the departure from the expected value of the braking index is purely due to structural change. However, that is not the case when all three energy loss mechanisms are present simultaneously. Rather the variation depends on the strength of the source of energy loss, that is $B_p, \epsilon, \alpha$ and $\alpha_{\rm S}$. In this section,} we present the result of rotational dependence of braking index taking all three energy loss channels into account for the same set of three representative EOSs used in Sec.~\ref{res1}. We further investigate the variation in the braking index for various physically motivated combinations of parameters of $B_p, \epsilon$, and $\alpha_{\rm S}$. The results are shown in Fig.~\ref{fig:parameter_explore}, where the green color indicates the results for the EOS obtained from our agnostic EOS model (see Sec.~\ref{eos_des} for details). The same sample EOS has been used in Sec.~\ref{res1} and has been shown with the solid yellow line in Fig. \ref{fig:braking_index_eos}. The blue and red colors correspond to S271v6 and DDME2 EOS, and different line style indicates the different combinations of  $B_p, \epsilon$, and $\alpha_{\rm S}$.

The upper left panel of the Fig.~\ref{fig:parameter_explore} shows the variation of $n$ for neutron stars with typical values of $B_p=10^8$ G, $\epsilon=10^{-7}$ and for two values of $\alpha_{\rm S}=10^{-1}$ and $10^{-4}$ shown using the solid and dashed line respectively. The justification for using such values of $\epsilon$ and $\alpha_{\rm S}$ has been presented in Sec.~\ref{deriv_brake}. To arrive at the typical value of $B_p$, we find the median value of $B_p$ of pulsars with $f>200$ Hz from the ATNF pulsar catalogue \cite{Manchester:2004bp}. The overlap of the dashed and solid lines for all the EOS implies that  the braking index is independent of the value of $\alpha_{\text{S}}$ in this case. The braking index of $\sim  7$ around $f \sim 200-400$ Hz indicates that it is mostly dominated by the GW emission due to r-mode instability. The MDR loss and GW radiation from the mass quadrupole moment do not contribute significantly towards the braking index. However, the departure from the value of 7 is caused by $n_{I}$ and other frequency-dependent structural quantities (as discussed in Sec.~\ref{res1}), which makes $n<7$. In summary, this figure suggests that if the r-mode instability is generated in a typical millisecond pulsar (which is dependent on the spin frequency and the redshifted stellar temperature, for details see \citet{KGK+2024}) then the spin-down would be dominated by the GW radiation through the r-mode instability, irrespective of the value of $\alpha_{\text{S}}$. 
At higher frequency $f\sim716$ Hz the contribution of $n_{I}$ and rotational effects contributing to $n^R_{\rm{GW}}$ leads to $6 < n < 7$. At a much higher frequency of $f\gtrsim 850$ Hz, the rapid decline in braking index is due to a sharp fall in the radius as discussed previously in Sec.~\ref{res1} and \ref{rot_effect}.

In the upper right panel of Fig.~\ref{fig:parameter_explore}, we show the braking index as a function of frequency in millisecond pulsars in the absence of GW radiation through r-mode instability. We use $B_p =10^8$ G, and two separate values of $\epsilon = 10^{-7} $ and $10^{-10}$ shown in the solid and dashed line respectively. Both lines overlap with braking index $\sim 3$, the deviation from $n=3$ is due to the rotational effect. It indicates that the emission loss through GW is insignificant even with a deformation of $\epsilon = 10^{-7}$. Our findings are consistent with the current observation limit \cite{LIGO+2019_ellipticity}. We find that a large deformation which can generate $\epsilon \gtrsim 10^{-3}$ can result in the emission of GW large enough to impact the rotational evolution of the neutron stars. However, such a high magnitude of ellipticity is unlikely for normal millisecond pulsars \cite{JMcO+2013}.

The lower left panel of Fig.~\ref{fig:parameter_explore} shows the effect of rotation on the braking index if a neutron star has an ultra-strong magnetic field $B_p = 10^{15}$ G, making them fall in the class of magnetars. The primary motivation for using such a strong magnetic field is to study the frequency-dependence of the braking index of a 
newly born millisecond magnetar \cite{DL+1998}, which may form due to the accretion-induced collapse of white dwarfs \citep{Usov+1992} or a merger of two neutron stars \citep{ROM+2013, Lu_2014}. The formation of these objects is believed to manifest as the Gamma Ray bursts (GRBs) \cite{LLL+2019}. From the figure we observe that if $\alpha_{\rm S} =10^{-4}$ and $\epsilon=10^{-7}$ at lower frequency $f\sim 200-600$ Hz the braking index is $\sim 3$, indicating the magnetic dipolar radiation dominates the spin-down rate of the star. At higher frequencies, $f\gtrsim 600$ Hz, the departure from $n\sim3$ is due to the rotational effects and is sensitive to the EOS. However, when the saturation amplitude of the r-mode is large, that is, $\alpha_{\rm S} = 10^{-1}$, we find at low rotation frequency ($f \sim 200$Hz) the braking index is $\sim3$, but finally rises to $\sim $6  close to $f=716$ Hz. The increase in the braking index from $\sim 3.5$ to $\sim 6$ is the manifestation of $\Omega^8$ dependence on the energy loss via the r-mode. The braking index cannot reach the value of 7 because of the rotational effects where $n_{I}$ is also dominant along with the contribution from $R^\prime/R$ and $\mathcal{J}^\prime/ \mathcal{J}$, where all of them are negative quantities. 
The behavior of the braking index found here illustrates the importance of incorporating the rotational effects to interpret the spin-down evolution of the remnants produced from the astrophysical transient events discussed above. For example, if the rotational effect is not considered, one can misinterpret a measurement of $n=5$ for a magnetar as the spin-down dominated by the GW emission due to a finite mass quadrupole moment. 

We have further investigated the frequency dependence of the braking index in the parlance of a newborn millisecond magnetar with a much higher value of ellipticity ($\epsilon = 10^{-3}$), based on the measurement of \citet{XDY+2022}. The results are shown in the lower right panel of Fig.~\ref{fig:parameter_explore}. We use the same colour convention of EOS mentioned earlier in the section. The dashed line style represents the braking index with $B_p=10^{15}$G, $\epsilon=10^{-3}$, and $\alpha_{\rm S}=10^{-1}$. This is the combination with the highest magnetic field, ellipticity and r-mode saturation amplitude. We find that the braking index depends on the rotation frequency exactly in the same way as the set of values $B_p=10^{15}$G, $\epsilon=10^{-7}$ and $\alpha_{\rm S}=10^{-1}$ (shown by the solid lines in lower left panel of Fig.~\ref{fig:parameter_explore}). It implies that even with 4 orders of magnitude larger ellipticity, the frequency dependence of braking index is still dominated by the r-mode instability if the saturation amplitude is $10^{-1}$. Whereas, if the $\alpha_{\rm S} \leq 10^{-4}$, we find the braking index is dominated by the magnetic dipolar radiation, which is shown by the solid line with the $B_p=10^{15}$G, $\epsilon=10^{-3}$ and $\alpha_{\rm S}=10^{-4}.$ The dashed-dotted line represents the variation of braking index with frequency in the absence of r-mode but large deformation ($\epsilon=10^{-3}$) and magnetic field ($B_p=10^{15}$G). The plot again suggests that the magnetic dipole radiation is the dominant mode of energy loss from the neutron star.

\begin{figure*}[htbp]
    \centering
    \subfigure{
        \includegraphics[scale=.42]{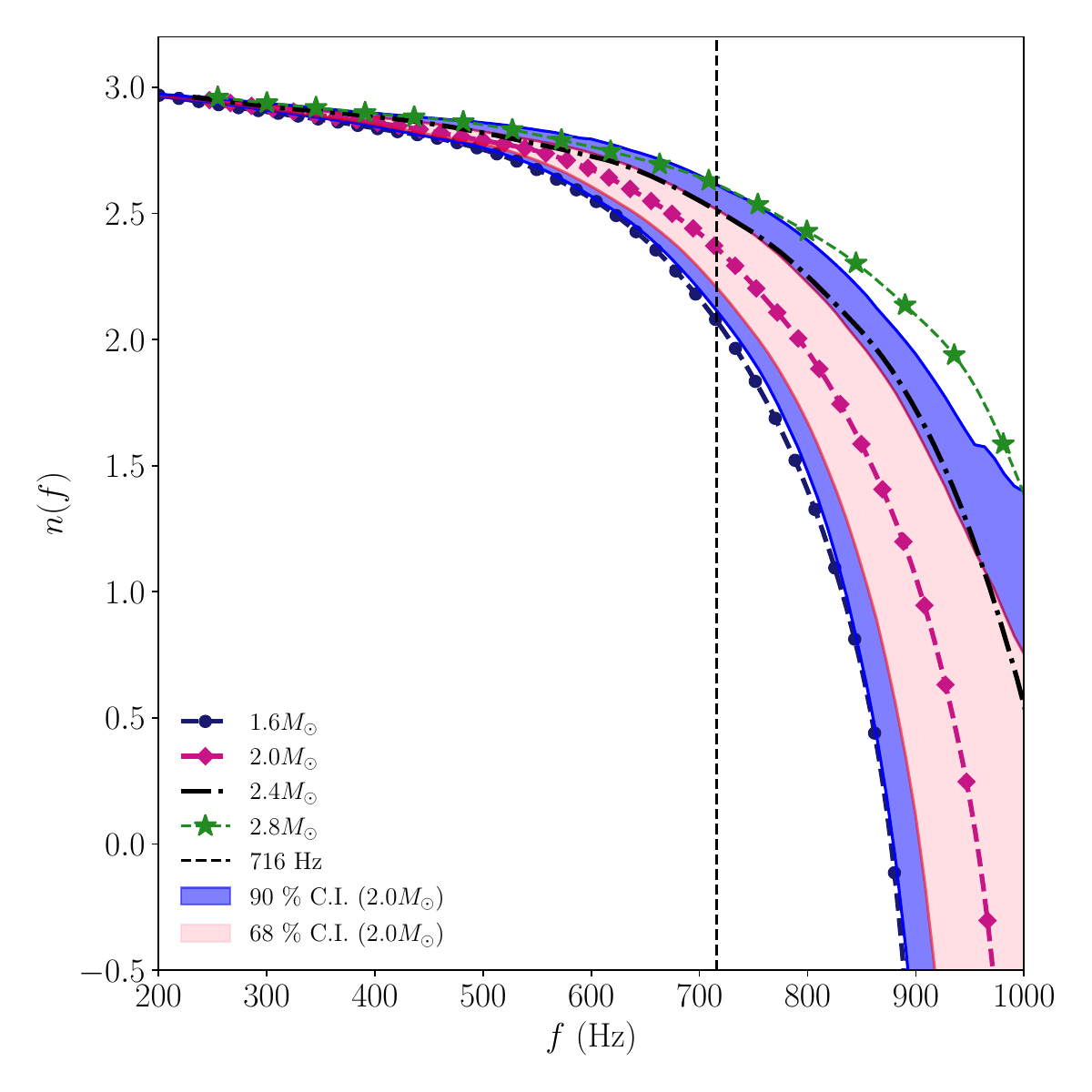}
    }
    \subfigure{
        \includegraphics[scale=.42]{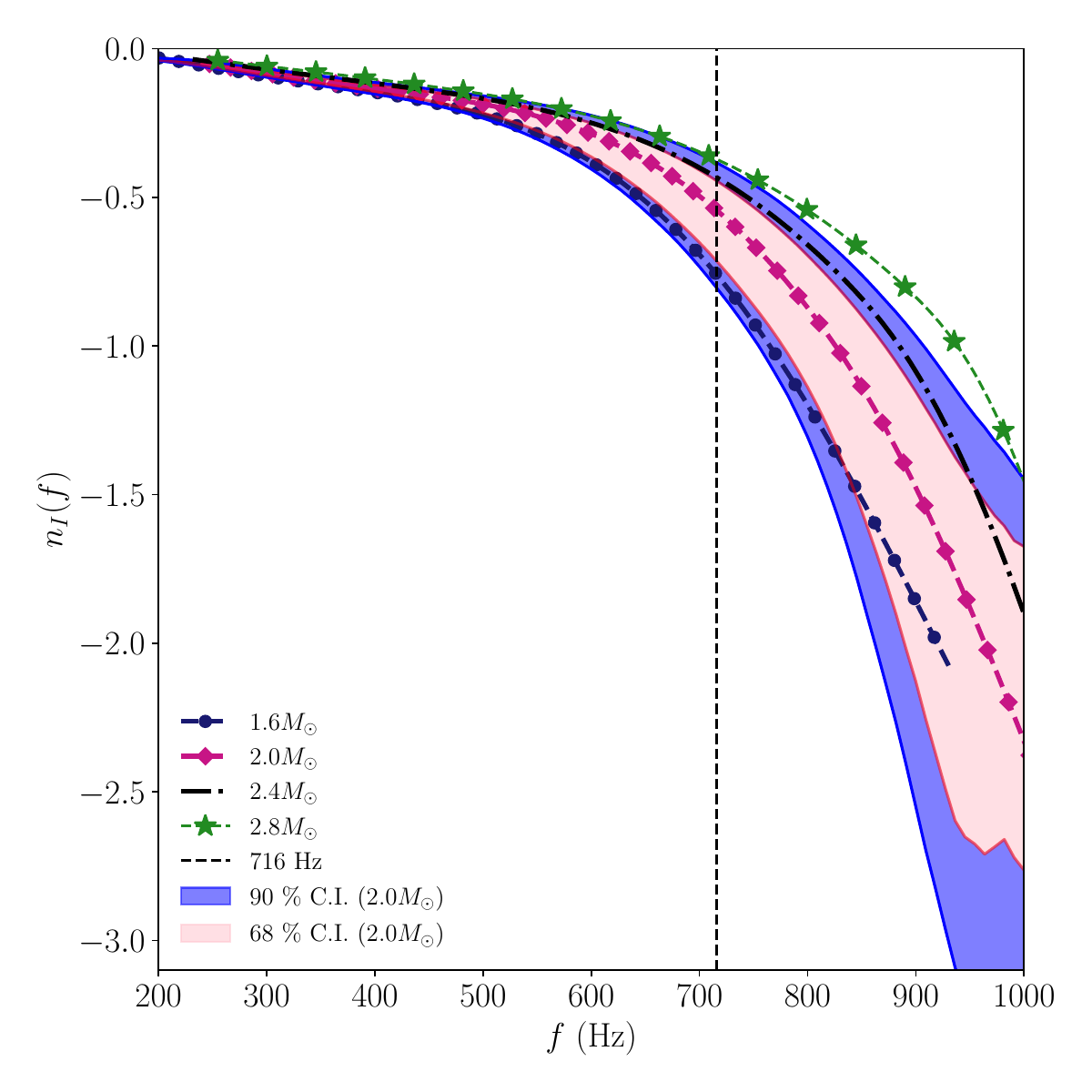}
    }
    \caption{{  {\it Left:} Braking index as a function of rotation frequency assuming a star of 2$M_\odot$ baryon mass. The blue and red bands show the 90\% and 68\% C.I.s of the braking index, respectively, originating from the uncertainty in the EOS. 
The individual curves correspond to different baryon masses but for the same sample EOS used in Fig.~\ref{fig:braking_index_eos}.
The blue dashed-circled line is for 1.6$M_\odot$ baryon mass, the dashed diamond corresponds to 2$M_\odot$ baryon mass, the black dashed-dotted line is for 2.4$M_\odot$ baryon mass,  and the green star dashed line corresponds to 2.8$M_\odot$ baryon mass. {\it Right:} $n_I$ as a function of spin frequency, the color scheme is the same as the left panel. Here we set $B_p=10^{15}$G, $\epsilon=10^{-7}$, and $\alpha_{\rm S}=10^{-4}$.}}
    \label{fig:braking_index}
\end{figure*}

\subsection{EOS uncertainties and the braking index of millisecond magnetar}\label{res3}
The long and short GRBs are observed to show an X-ray plateau phase after the prompt emission, which lasts for a few tens to thousands of seconds, indicating an ongoing energy injection from the central engine \cite{Nousek+2006, OWO+2006, Zhang+2006, ROM+2013, XDY+2022, Lu_2014}. There exists debate on the nature of the central engine, but a millisecond magnetar is one of the plausible models often used in the literature. The spin-down of the magnetar is believed to power the X-ray plateau region of the light curve \citep{ZM+2001, metzger+2011}, which is modeled to obtain the braking index \citep{Lasky+2017, XDY+2022}. 
We have shown above that the braking index of a neutron star is affected by its structural changes due to rotation, especially at higher frequencies. Given the remnant is a newly born millisecond magnetar with spin period $P\sim 1-3$ ms \citep{XDY+2022}, its braking index is expected to be different from that of a slowly rotating neutron star. However, the response of a neutron star to rapid rotation also depends on the underlying EOS. To explore the uncertainty in the braking index due to the EOS, we next calculate it at different frequencies for all the EOS constructed in Sec.~\ref{eos_des}.  We have used Eq.~\ref{fullbi}, supplemented by all the frequency-dependent profiles of $M, I, R$ and $\mathcal{J}$ (as discussed in Sec.~\ref{rot_effect} and shown in Fig.~\ref{fig:shape_vs_freq}) and their derivatives. Here we set $B_p=10^{15}$G, $\epsilon=10^{-7}$, and $\alpha_{\rm S}=10^{-4}$, where the values of $\epsilon$, and $\alpha_{\rm S}$ are conservative estimates based on our results of Sec.~\ref{res2}. The result is shown in Fig.~\ref{fig:braking_index} for a NS of $2M_\odot$ baryon mass. The 90\% C.I. and the 68\% C.I. in the braking index as a function of frequency are shown by the blue and red bands, respectively, which have been computed from the weights on the EOS parameters obtained from the Bayesian analysis in Sec.~\ref{eos_des}. The figure shows that the range of the braking index gets wider as the frequency increases. It happens because the structural changes of a neutron star due to rotation strongly depend on the EOS.

{  While the band of uncertainty obtained in the braking index captures the uncertainty due to the EOS,  uncertainty may arise due to the choice of the baryon mass of the star as well.  Therefore, we calculate the braking indices for stars with baryon masses 1.6, 2.0, 2.4, and 2.8 $M_\odot$ using the same sample EOS used above in the Sec~.\ref{res1}. The results are shown in Fig.~\ref{fig:braking_index} by curves of different colors and line styles as described in the caption. We find that the braking index for a massive star falls less rapidly with frequency than a lower-mass star. As a result, the heavier star has a larger braking index at a given frequency than a lighter one.  A similar mass dependence is reflected in $n_I$. This behavior implies if a massive millisecond magnetar is born from some astrophysical events, it will spin down faster in comparison to a millisecond magnetar born lighter. 
It is also interesting to note that the uncertainty in the braking index from baryon mass completely masks the uncertainty from the EOS in  Fig.~\ref{fig:braking_index}. Therefore, knowing the EOS and rotation frequency is not enough to infer the braking index of a pulsar. We also need to know its gravitational mass. Please note that one can obtain the baryon mass from the gravitational mass by assuming an EOS if the rotation frequency is known. In other words, knowledge of both the gravitational mass and the braking index of a pulsar is necessary to constrain the EOS. Similar conclusions are expected for other combinations of $B_p$, $\epsilon$, and $\alpha_{\rm S}$.}

{  Immediately after the birth of a millisecond magnetar, the object undergoes rapid spin-down, resulting in a significant and time-dependent variation in the braking index. Understanding how the progenitor baryon mass, uncertainties in the EOS, and spin-dependent spin-down evolution influence the energy injection mechanism and the resulting GRB light curve morphology requires a comprehensive and self-consistent numerical treatment. Such an analysis is beyond the scope of the present work.
}
\section{Conclusion}\label{conclusion}
In this paper, we have investigated the effect of stellar rotation on the braking index and its dependence on the EOS {  and baryon mass}. We have generated a large number of EOSs by employing a semi-agnostic model and constrained them using the latest astrophysical observations from the gravitational and radio wave bands within a Bayesian framework.
Our constrained parameter space is also consistent with the observational findings of neutron star's mass and radius from X-ray bands. We then use these constrained EOS to study the effect on the macroscopic quantities of rapidly rotating neutron stars like mass, radius, moment of inertia and the ratio of polar to equatorial radius. Our analysis shows neutron stars with stiffer EOS experience 
larger deformations in the shape at a given frequency, in comparison to the softer EOS. Consequently, neutron stars with stiffer EOS have smaller Kepler frequencies.

Next, we study the effect of rotation on the braking index of neutron stars. {  We present the first calculation of braking index in the presence of the r-mode instability jointly with the other braking mechanism like MDR and GWs due to non-axisymmetric deformations, and incorporating the rotational effect.} We have shown that the rotation of the star can impact its braking index significantly. 
For example, the rotational effects make the braking index smaller than 3 in the case of magnetic dipole radiation when the spin frequency $\gtrsim 200$ Hz.
However, the low braking index in normal pulsars has been explained through various other mechanisms, such as, the evolution of magnetic inclination angle, which has not been considered here. Similarly, we have shown that if the neutron stars spin-down is purely due to GW radiation from non-asymmetric deformations or r-mode oscillations, rapid rotation leads to the significant departure of braking index from their classical values of 5 and 7, respectively.

We further explore the dependence of the braking index on the rotation frequency for various physically motivated combinations of surface magnetic field strength, ellipticity of the star, and the amplitude of r-mode oscillations. We find if there is r-mode instability generated in a typical millisecond pulsar with $\alpha_{\rm S}$ between $10^{-4} - 10^{-1}$, then angular momentum loss through GW r-mode instability will dominate the spin-down of the pulsar. Whereas, extremely large deformation of $\epsilon \gtrsim 10^{-3}$ is required for the spin-down of a typical millisecond pulsar to be dominated by GW emission due to time-dependent mass quadrupolar moment, which has been ruled out under various physical conditions. Our analysis shows that if the $\epsilon \lesssim 10^{-7}$, the spin-down is dominated by the magnetic dipole radiation in the absence of r-mode instability, which is consistent with the current limit of the GW observations.

We extended our study in the light of millisecond magnetars, which are believed to be born from NS-NS mergers, core-collapse of a massive star, or accretion-induced collapse of a white dwarf. The spin-down energy is believed to power the X-ray plateau phase of GRBs. These millisecond magnetars can have large magnetic field strength $\sim 10^{15}$ G. We find that at large spin-frequencies $\gtrsim 600$ Hz, the presence of large amplitude r-mode oscillations dominates the spin-down rate. However, the effect of rapid rotation suppresses the braking index closer to 6 instead of 7. But the impact of non-axisymmetric deformation on the braking index is less pronounced unless $\epsilon > 10^{-3}$. However, if the birth spin frequency is low $\sim 200$ Hz or r-mode amplitude is less than $10^{-1}$, spin-down is always dominated by the magnetic dipole radiation. 

We have used the constrained EOS to compute the permissible range of the braking index as a function of the rotational frequency. The width of the range of the braking index increases with rotation frequency, indicating a wide range of EOS-dependent structural change of neutron stars in response to large centrifugal force. We find a millisecond magnetar with the magnetic field strength $10^{15}$ G, deformation $\epsilon = 10^{-7}$ and r-mode amplitude $\alpha_{\rm S} = 10^{-4}$ has braking index $\sim 3$ at low frequency and $n<3$ at higher frequency\footnote{It is important to emphasize that millisecond magnetar are prone to magnetospheric changes \cite{Contopoulos:2005rs}, where the separatrix radius (distance from center to the edge of the closed magnetosphere, beyond which the magnetic field lines open) can lag behind the radius of light cylinder (distance from the center to the radius of the rigid co-rotating magnetosphere), which naturally lead to $n<3$ scenario. However, this modification in the magnetosphere is beyond the scope of this paper and has been confined to the simple dipolar approximation of the magnetosphere.} as shown in Fig.~\ref{fig:braking_index}. {  We have further investigated the effect of baryon mass on the spin-frequency dependent braking index and found that massive stars have a larger braking index compared to the star with lower baryon mass at a given spin-frequency. We find there exists a degeneracy between the uncertainty due to choice of baryon mass and the uncertainty in the EOS. This work provides a framework for future rigorous analysis for modeling GRB light curves.}

\section*{Acknowledgements}
Pulsar research at Jodrell Bank Centre for Astrophysics and Jodrell Bank Observatory is supported by a consolidated grant from the UK Science and Technology Facilities Council (STFC). This project has received funding from the European Union’s Horizon 2020 research and innovation programme under the Marie Skłodowska-Curie grant agreement No. 101034371. PC acknowledges the support from the European Union's HORIZON MSCA-2022-PF-01-01 Programme under Grant Agreement No. 101109652, project ProMatEx-NS. Part of the computation has been performed in the high-performance computing facility “Magus,” which is available at the Shiv Nadar Institution of Eminence.

\bibliography{refs}

\begin{thebibliography}{93}%
\makeatletter
\providecommand \@ifxundefined [1]{%
 \@ifx{#1\undefined}
}%
\providecommand \@ifnum [1]{%
 \ifnum #1\expandafter \@firstoftwo
 \else \expandafter \@secondoftwo
 \fi
}%
\providecommand \@ifx [1]{%
 \ifx #1\expandafter \@firstoftwo
 \else \expandafter \@secondoftwo
 \fi
}%
\providecommand \natexlab [1]{#1}%
\providecommand \enquote  [1]{``#1''}%
\providecommand \bibnamefont  [1]{#1}%
\providecommand \bibfnamefont [1]{#1}%
\providecommand \citenamefont [1]{#1}%
\providecommand \href@noop [0]{\@secondoftwo}%
\providecommand \href [0]{\begingroup \@sanitize@url \@href}%
\providecommand \@href[1]{\@@startlink{#1}\@@href}%
\providecommand \@@href[1]{\endgroup#1\@@endlink}%
\providecommand \@sanitize@url [0]{\catcode `\\12\catcode `\$12\catcode
  `\&12\catcode `\#12\catcode `\^12\catcode `\_12\catcode `\%12\relax}%
\providecommand \@@startlink[1]{}%
\providecommand \@@endlink[0]{}%
\providecommand \url  [0]{\begingroup\@sanitize@url \@url }%
\providecommand \@url [1]{\endgroup\@href {#1}{\urlprefix }}%
\providecommand \urlprefix  [0]{URL }%
\providecommand \Eprint [0]{\href }%
\providecommand \doibase [0]{http://dx.doi.org/}%
\providecommand \selectlanguage [0]{\@gobble}%
\providecommand \bibinfo  [0]{\@secondoftwo}%
\providecommand \bibfield  [0]{\@secondoftwo}%
\providecommand \translation [1]{[#1]}%
\providecommand \BibitemOpen [0]{}%
\providecommand \bibitemStop [0]{}%
\providecommand \bibitemNoStop [0]{.\EOS\space}%
\providecommand \EOS [0]{\spacefactor3000\relax}%
\providecommand \BibitemShut  [1]{\csname bibitem#1\endcsname}%
\let\auto@bib@innerbib\@empty
\bibitem [{\citenamefont {{Manchester}}\ \emph {et~al.}(1985)\citenamefont
  {{Manchester}}, \citenamefont {{Durdin}},\ and\ \citenamefont
  {{Newton}}}]{MDN+1985}%
  \BibitemOpen
  \bibfield  {author} {\bibinfo {author} {\bibfnamefont {R.~N.}\ \bibnamefont
  {{Manchester}}}, \bibinfo {author} {\bibfnamefont {J.~M.}\ \bibnamefont
  {{Durdin}}}, \ and\ \bibinfo {author} {\bibfnamefont {L.~M.}\ \bibnamefont
  {{Newton}}},\ }\href {\doibase 10.1038/313374a0} {\bibfield  {journal}
  {\bibinfo  {journal} {Nature}\ }\textbf {\bibinfo {volume} {313}},\ \bibinfo
  {pages} {374} (\bibinfo {year} {1985})}\BibitemShut {NoStop}%
\bibitem [{\citenamefont {{Gunn}}\ and\ \citenamefont
  {{Ostriker}}(1969)}]{GO+1969}%
  \BibitemOpen
  \bibfield  {author} {\bibinfo {author} {\bibfnamefont {J.~E.}\ \bibnamefont
  {{Gunn}}}\ and\ \bibinfo {author} {\bibfnamefont {J.~P.}\ \bibnamefont
  {{Ostriker}}},\ }\href {\doibase 10.1038/221454a0} {\bibfield  {journal}
  {\bibinfo  {journal} {Nature}\ }\textbf {\bibinfo {volume} {221}},\ \bibinfo
  {pages} {454} (\bibinfo {year} {1969})}\BibitemShut {NoStop}%
\bibitem [{\citenamefont {{Riles}}(2023)}]{Riles+2023}%
  \BibitemOpen
  \bibfield  {author} {\bibinfo {author} {\bibfnamefont {K.}~\bibnamefont
  {{Riles}}},\ }\href {\doibase 10.1007/s41114-023-00044-3} {\bibfield
  {journal} {\bibinfo  {journal} {Living Reviews in Relativity}\ }\textbf
  {\bibinfo {volume} {26}},\ \bibinfo {eid} {3} (\bibinfo {year} {2023})},\
  \Eprint {http://arxiv.org/abs/2206.06447} {arXiv:2206.06447 [astro-ph.HE]}
  \BibitemShut {NoStop}%
\bibitem [{\citenamefont {{Harding}}\ \emph {et~al.}(1999)\citenamefont
  {{Harding}}, \citenamefont {{Contopoulos}},\ and\ \citenamefont
  {{Kazanas}}}]{Harding+1999}%
  \BibitemOpen
  \bibfield  {author} {\bibinfo {author} {\bibfnamefont {A.~K.}\ \bibnamefont
  {{Harding}}}, \bibinfo {author} {\bibfnamefont {I.}~\bibnamefont
  {{Contopoulos}}}, \ and\ \bibinfo {author} {\bibfnamefont {D.}~\bibnamefont
  {{Kazanas}}},\ }\href {\doibase 10.1086/312339} {\bibfield  {journal}
  {\bibinfo  {journal} {ApJL}\ }\textbf {\bibinfo {volume} {525}},\ \bibinfo
  {pages} {L125} (\bibinfo {year} {1999})},\ \Eprint
  {http://arxiv.org/abs/astro-ph/9908279} {arXiv:astro-ph/9908279 [astro-ph]}
  \BibitemShut {NoStop}%
\bibitem [{\citenamefont {{Spitkovsky}}(2006)}]{Spitkovsky+2006}%
  \BibitemOpen
  \bibfield  {author} {\bibinfo {author} {\bibfnamefont {A.}~\bibnamefont
  {{Spitkovsky}}},\ }\href {\doibase 10.1086/507518} {\bibfield  {journal}
  {\bibinfo  {journal} {ApJL}\ }\textbf {\bibinfo {volume} {648}},\ \bibinfo
  {pages} {L51} (\bibinfo {year} {2006})},\ \Eprint
  {http://arxiv.org/abs/astro-ph/0603147} {arXiv:astro-ph/0603147 [astro-ph]}
  \BibitemShut {NoStop}%
\bibitem [{\citenamefont {Caleb}\ \emph {et~al.}(2022)\citenamefont {Caleb}
  \emph {et~al.}}]{Caleb:2022xyo}%
  \BibitemOpen
  \bibfield  {author} {\bibinfo {author} {\bibfnamefont {M.}~\bibnamefont
  {Caleb}} \emph {et~al.},\ }\href {\doibase 10.1038/s41550-022-01688-x}
  {\bibfield  {journal} {\bibinfo  {journal} {Nature Astron.}\ }\textbf
  {\bibinfo {volume} {6}},\ \bibinfo {pages} {828} (\bibinfo {year} {2022})},\
  \Eprint {http://arxiv.org/abs/2206.01346} {arXiv:2206.01346 [astro-ph.HE]}
  \BibitemShut {NoStop}%
\bibitem [{\citenamefont {{Hessels}}\ \emph {et~al.}(2006)\citenamefont
  {{Hessels}}, \citenamefont {{Ransom}}, \citenamefont {{Stairs}},
  \citenamefont {{Freire}}, \citenamefont {{Kaspi}},\ and\ \citenamefont
  {{Camilo}}}]{HRS+2006}%
  \BibitemOpen
  \bibfield  {author} {\bibinfo {author} {\bibfnamefont {J.~W.~T.}\
  \bibnamefont {{Hessels}}}, \bibinfo {author} {\bibfnamefont {S.~M.}\
  \bibnamefont {{Ransom}}}, \bibinfo {author} {\bibfnamefont {I.~H.}\
  \bibnamefont {{Stairs}}}, \bibinfo {author} {\bibfnamefont {P.~C.~C.}\
  \bibnamefont {{Freire}}}, \bibinfo {author} {\bibfnamefont {V.~M.}\
  \bibnamefont {{Kaspi}}}, \ and\ \bibinfo {author} {\bibfnamefont
  {F.}~\bibnamefont {{Camilo}}},\ }\href {\doibase 10.1126/science.1123430}
  {\bibfield  {journal} {\bibinfo  {journal} {Science}\ }\textbf {\bibinfo
  {volume} {311}},\ \bibinfo {pages} {1901} (\bibinfo {year} {2006})},\ \Eprint
  {http://arxiv.org/abs/astro-ph/0601337} {arXiv:astro-ph/0601337 [astro-ph]}
  \BibitemShut {NoStop}%
\bibitem [{\citenamefont {Halder}\ \emph {et~al.}(2023)\citenamefont {Halder},
  \citenamefont {Goswami}, \citenamefont {Halder}, \citenamefont {Ghosh},\ and\
  \citenamefont {Konar}}]{Halder:2023rfu}%
  \BibitemOpen
  \bibfield  {author} {\bibinfo {author} {\bibfnamefont {P.}~\bibnamefont
  {Halder}}, \bibinfo {author} {\bibfnamefont {S.}~\bibnamefont {Goswami}},
  \bibinfo {author} {\bibfnamefont {P.}~\bibnamefont {Halder}}, \bibinfo
  {author} {\bibfnamefont {U.}~\bibnamefont {Ghosh}}, \ and\ \bibinfo {author}
  {\bibfnamefont {S.}~\bibnamefont {Konar}},\ }\href {\doibase
  10.3847/2515-5172/ad00ac} {\bibfield  {journal} {\bibinfo  {journal} {Res.
  Notes AAS}\ }\textbf {\bibinfo {volume} {7}},\ \bibinfo {pages} {213}
  (\bibinfo {year} {2023})},\ \Eprint {http://arxiv.org/abs/2310.06230}
  {arXiv:2310.06230 [astro-ph.HE]} \BibitemShut {NoStop}%
\bibitem [{\citenamefont {Livingstone}\ \emph {et~al.}(2007)\citenamefont
  {Livingstone}, \citenamefont {Kaspi}, \citenamefont {Gavriil}, \citenamefont
  {Manchester}, \citenamefont {Gotthelf},\ and\ \citenamefont
  {Kuiper}}]{Livingstone:2007bn}%
  \BibitemOpen
  \bibfield  {author} {\bibinfo {author} {\bibfnamefont {M.~A.}\ \bibnamefont
  {Livingstone}}, \bibinfo {author} {\bibfnamefont {V.~M.}\ \bibnamefont
  {Kaspi}}, \bibinfo {author} {\bibfnamefont {F.~P.}\ \bibnamefont {Gavriil}},
  \bibinfo {author} {\bibfnamefont {R.~N.}\ \bibnamefont {Manchester}},
  \bibinfo {author} {\bibfnamefont {E.~V.}\ \bibnamefont {Gotthelf}}, \ and\
  \bibinfo {author} {\bibfnamefont {L.}~\bibnamefont {Kuiper}},\ }\href
  {\doibase 10.1007/s10509-007-9320-3} {\bibfield  {journal} {\bibinfo
  {journal} {Astrophys. Space Sci.}\ }\textbf {\bibinfo {volume} {308}},\
  \bibinfo {pages} {317} (\bibinfo {year} {2007})},\ \Eprint
  {http://arxiv.org/abs/astro-ph/0702196} {arXiv:astro-ph/0702196} \BibitemShut
  {NoStop}%
\bibitem [{\citenamefont {{Weltevrede}}\ \emph {et~al.}(2011)\citenamefont
  {{Weltevrede}}, \citenamefont {{Johnston}},\ and\ \citenamefont
  {{Espinoza}}}]{WJE+2011}%
  \BibitemOpen
  \bibfield  {author} {\bibinfo {author} {\bibfnamefont {P.}~\bibnamefont
  {{Weltevrede}}}, \bibinfo {author} {\bibfnamefont {S.}~\bibnamefont
  {{Johnston}}}, \ and\ \bibinfo {author} {\bibfnamefont {C.~M.}\ \bibnamefont
  {{Espinoza}}},\ }\href {\doibase 10.1111/j.1365-2966.2010.17821.x} {\bibfield
   {journal} {\bibinfo  {journal} {MNRAS}\ }\textbf {\bibinfo {volume} {411}},\
  \bibinfo {pages} {1917} (\bibinfo {year} {2011})},\ \Eprint
  {http://arxiv.org/abs/1010.0857} {arXiv:1010.0857 [astro-ph.SR]} \BibitemShut
  {NoStop}%
\bibitem [{\citenamefont {Livingstone}\ and\ \citenamefont
  {Kaspi}(2011)}]{Livingstone_2011}%
  \BibitemOpen
  \bibfield  {author} {\bibinfo {author} {\bibfnamefont {M.~A.}\ \bibnamefont
  {Livingstone}}\ and\ \bibinfo {author} {\bibfnamefont {V.~M.}\ \bibnamefont
  {Kaspi}},\ }\href {\doibase 10.1088/0004-637X/742/1/31} {\bibfield  {journal}
  {\bibinfo  {journal} {The Astrophysical Journal}\ }\textbf {\bibinfo {volume}
  {742}},\ \bibinfo {pages} {31} (\bibinfo {year} {2011})}\BibitemShut
  {NoStop}%
\bibitem [{\citenamefont {{Lyne}}\ \emph {et~al.}(1993)\citenamefont {{Lyne}},
  \citenamefont {{Pritchard}},\ and\ \citenamefont {{Graham
  Smith}}}]{LPS+1993}%
  \BibitemOpen
  \bibfield  {author} {\bibinfo {author} {\bibfnamefont {A.~G.}\ \bibnamefont
  {{Lyne}}}, \bibinfo {author} {\bibfnamefont {R.~S.}\ \bibnamefont
  {{Pritchard}}}, \ and\ \bibinfo {author} {\bibfnamefont {F.}~\bibnamefont
  {{Graham Smith}}},\ }\href {\doibase 10.1093/mnras/265.4.1003} {\bibfield
  {journal} {\bibinfo  {journal} {MNRAS}\ }\textbf {\bibinfo {volume} {265}},\
  \bibinfo {pages} {1003} (\bibinfo {year} {1993})}\BibitemShut {NoStop}%
\bibitem [{\citenamefont {{Boyd}}\ \emph {et~al.}(1995)\citenamefont {{Boyd}},
  \citenamefont {{van Citters}}, \citenamefont {{Dolan}}, \citenamefont
  {{Wolinski}}, \citenamefont {{Percival}}, \citenamefont {{Bless}},
  \citenamefont {{Elliot}}, \citenamefont {{Nelson}},\ and\ \citenamefont
  {{Taylor}}}]{BCD+1995}%
  \BibitemOpen
  \bibfield  {author} {\bibinfo {author} {\bibfnamefont {P.~T.}\ \bibnamefont
  {{Boyd}}}, \bibinfo {author} {\bibfnamefont {G.~W.}\ \bibnamefont {{van
  Citters}}}, \bibinfo {author} {\bibfnamefont {J.~F.}\ \bibnamefont
  {{Dolan}}}, \bibinfo {author} {\bibfnamefont {K.~G.}\ \bibnamefont
  {{Wolinski}}}, \bibinfo {author} {\bibfnamefont {J.~W.}\ \bibnamefont
  {{Percival}}}, \bibinfo {author} {\bibfnamefont {R.~C.}\ \bibnamefont
  {{Bless}}}, \bibinfo {author} {\bibfnamefont {J.~L.}\ \bibnamefont
  {{Elliot}}}, \bibinfo {author} {\bibfnamefont {M.~J.}\ \bibnamefont
  {{Nelson}}}, \ and\ \bibinfo {author} {\bibfnamefont {M.~J.}\ \bibnamefont
  {{Taylor}}},\ }\href {\doibase 10.1086/175967} {\bibfield  {journal}
  {\bibinfo  {journal} {ApJ}\ }\textbf {\bibinfo {volume} {448}},\ \bibinfo
  {pages} {365} (\bibinfo {year} {1995})}\BibitemShut {NoStop}%
\bibitem [{\citenamefont {{Roy}}\ \emph {et~al.}(2012)\citenamefont {{Roy}},
  \citenamefont {{Gupta}},\ and\ \citenamefont {{Lewandowski}}}]{RGL+2012}%
  \BibitemOpen
  \bibfield  {author} {\bibinfo {author} {\bibfnamefont {J.}~\bibnamefont
  {{Roy}}}, \bibinfo {author} {\bibfnamefont {Y.}~\bibnamefont {{Gupta}}}, \
  and\ \bibinfo {author} {\bibfnamefont {W.}~\bibnamefont {{Lewandowski}}},\
  }\href {\doibase 10.1111/j.1365-2966.2012.21380.x} {\bibfield  {journal}
  {\bibinfo  {journal} {MNRAS}\ }\textbf {\bibinfo {volume} {424}},\ \bibinfo
  {pages} {2213} (\bibinfo {year} {2012})},\ \Eprint
  {http://arxiv.org/abs/1205.6264} {arXiv:1205.6264 [astro-ph.SR]} \BibitemShut
  {NoStop}%
\bibitem [{\citenamefont {Espinoza}\ \emph {et~al.}(2011)\citenamefont
  {Espinoza}, \citenamefont {Lyne}, \citenamefont {Kramer}, \citenamefont
  {Manchester},\ and\ \citenamefont {Kaspi}}]{Espinoza_2011}%
  \BibitemOpen
  \bibfield  {author} {\bibinfo {author} {\bibfnamefont {C.~M.}\ \bibnamefont
  {Espinoza}}, \bibinfo {author} {\bibfnamefont {A.~G.}\ \bibnamefont {Lyne}},
  \bibinfo {author} {\bibfnamefont {M.}~\bibnamefont {Kramer}}, \bibinfo
  {author} {\bibfnamefont {R.~N.}\ \bibnamefont {Manchester}}, \ and\ \bibinfo
  {author} {\bibfnamefont {V.~M.}\ \bibnamefont {Kaspi}},\ }\href {\doibase
  10.1088/2041-8205/741/1/L13} {\bibfield  {journal} {\bibinfo  {journal} {The
  Astrophysical Journal Letters}\ }\textbf {\bibinfo {volume} {741}},\ \bibinfo
  {pages} {L13} (\bibinfo {year} {2011})}\BibitemShut {NoStop}%
\bibitem [{\citenamefont {{Chen}}\ and\ \citenamefont
  {{Ruderman}}(1993)}]{Chen+1993}%
  \BibitemOpen
  \bibfield  {author} {\bibinfo {author} {\bibfnamefont {K.}~\bibnamefont
  {{Chen}}}\ and\ \bibinfo {author} {\bibfnamefont {M.}~\bibnamefont
  {{Ruderman}}},\ }\href {\doibase 10.1086/172129} {\bibfield  {journal}
  {\bibinfo  {journal} {ApJ}\ }\textbf {\bibinfo {volume} {402}},\ \bibinfo
  {pages} {264} (\bibinfo {year} {1993})}\BibitemShut {NoStop}%
\bibitem [{\citenamefont {{Melatos}}(1997)}]{Melatos+1997}%
  \BibitemOpen
  \bibfield  {author} {\bibinfo {author} {\bibfnamefont {A.}~\bibnamefont
  {{Melatos}}},\ }\href {\doibase 10.1093/mnras/288.4.1049} {\bibfield
  {journal} {\bibinfo  {journal} {MNRAS}\ }\textbf {\bibinfo {volume} {288}},\
  \bibinfo {pages} {1049} (\bibinfo {year} {1997})}\BibitemShut {NoStop}%
\bibitem [{\citenamefont {{Hamil}}\ \emph {et~al.}(2015)\citenamefont
  {{Hamil}}, \citenamefont {{Stone}}, \citenamefont {{Urbanec}},\ and\
  \citenamefont {{Urbancov{\'a}}}}]{Hamil+2015}%
  \BibitemOpen
  \bibfield  {author} {\bibinfo {author} {\bibfnamefont {O.}~\bibnamefont
  {{Hamil}}}, \bibinfo {author} {\bibfnamefont {J.~R.}\ \bibnamefont
  {{Stone}}}, \bibinfo {author} {\bibfnamefont {M.}~\bibnamefont {{Urbanec}}},
  \ and\ \bibinfo {author} {\bibfnamefont {G.}~\bibnamefont
  {{Urbancov{\'a}}}},\ }\href {\doibase 10.1103/PhysRevD.91.063007} {\bibfield
  {journal} {\bibinfo  {journal} {Phys. Rev. D}\ }\textbf {\bibinfo {volume}
  {91}},\ \bibinfo {eid} {063007} (\bibinfo {year} {2015})},\ \Eprint
  {http://arxiv.org/abs/1608.01383} {arXiv:1608.01383 [astro-ph.HE]}
  \BibitemShut {NoStop}%
\bibitem [{\citenamefont {Lan}\ \emph {et~al.}(2021)\citenamefont {Lan},
  \citenamefont {Gao}, \citenamefont {Ai},\ and\ \citenamefont
  {Li}}]{Lan:2021luk}%
  \BibitemOpen
  \bibfield  {author} {\bibinfo {author} {\bibfnamefont {L.}~\bibnamefont
  {Lan}}, \bibinfo {author} {\bibfnamefont {H.}~\bibnamefont {Gao}}, \bibinfo
  {author} {\bibfnamefont {S.}~\bibnamefont {Ai}}, \ and\ \bibinfo {author}
  {\bibfnamefont {S.-Z.}\ \bibnamefont {Li}},\ }\href {\doibase
  10.3847/1538-4357/ac167d} {\bibfield  {journal} {\bibinfo  {journal}
  {Astrophys. J.}\ }\textbf {\bibinfo {volume} {919}},\ \bibinfo {pages} {14}
  (\bibinfo {year} {2021})},\ \Eprint {http://arxiv.org/abs/2111.05542}
  {arXiv:2111.05542 [astro-ph.HE]} \BibitemShut {NoStop}%
\bibitem [{\citenamefont {Manchester}\ \emph {et~al.}(2005)\citenamefont
  {Manchester}, \citenamefont {Hobbs}, \citenamefont {Teoh},\ and\
  \citenamefont {Hobbs}}]{Manchester:2004bp}%
  \BibitemOpen
  \bibfield  {author} {\bibinfo {author} {\bibfnamefont {R.~N.}\ \bibnamefont
  {Manchester}}, \bibinfo {author} {\bibfnamefont {G.~B.}\ \bibnamefont
  {Hobbs}}, \bibinfo {author} {\bibfnamefont {A.}~\bibnamefont {Teoh}}, \ and\
  \bibinfo {author} {\bibfnamefont {M.}~\bibnamefont {Hobbs}},\ }\href
  {\doibase 10.1086/428488} {\bibfield  {journal} {\bibinfo  {journal} {Astron.
  J.}\ }\textbf {\bibinfo {volume} {129}},\ \bibinfo {pages} {1993} (\bibinfo
  {year} {2005})},\ \Eprint {http://arxiv.org/abs/astro-ph/0412641}
  {arXiv:astro-ph/0412641} \BibitemShut {NoStop}%
\bibitem [{\citenamefont {Dutra}\ \emph {et~al.}(2012)\citenamefont {Dutra},
  \citenamefont {Lourenco}, \citenamefont {Sa~Martins}, \citenamefont
  {Delfino}, \citenamefont {Stone},\ and\ \citenamefont
  {Stevenson}}]{Dutra:2012mb}%
  \BibitemOpen
  \bibfield  {author} {\bibinfo {author} {\bibfnamefont {M.}~\bibnamefont
  {Dutra}}, \bibinfo {author} {\bibfnamefont {O.}~\bibnamefont {Lourenco}},
  \bibinfo {author} {\bibfnamefont {J.~S.}\ \bibnamefont {Sa~Martins}},
  \bibinfo {author} {\bibfnamefont {A.}~\bibnamefont {Delfino}}, \bibinfo
  {author} {\bibfnamefont {J.~R.}\ \bibnamefont {Stone}}, \ and\ \bibinfo
  {author} {\bibfnamefont {P.~D.}\ \bibnamefont {Stevenson}},\ }\href {\doibase
  10.1103/PhysRevC.85.035201} {\bibfield  {journal} {\bibinfo  {journal} {Phys.
  Rev. C}\ }\textbf {\bibinfo {volume} {85}},\ \bibinfo {pages} {035201}
  (\bibinfo {year} {2012})},\ \Eprint {http://arxiv.org/abs/1202.3902}
  {arXiv:1202.3902 [nucl-th]} \BibitemShut {NoStop}%
\bibitem [{\citenamefont {Dutra}\ \emph {et~al.}(2014)\citenamefont {Dutra},
  \citenamefont {Louren\c{c}o}, \citenamefont {Avancini}, \citenamefont
  {Carlson}, \citenamefont {Delfino}, \citenamefont {Menezes}, \citenamefont
  {Provid\^encia}, \citenamefont {Typel},\ and\ \citenamefont
  {Stone}}]{Dutra:2014qga}%
  \BibitemOpen
  \bibfield  {author} {\bibinfo {author} {\bibfnamefont {M.}~\bibnamefont
  {Dutra}}, \bibinfo {author} {\bibfnamefont {O.}~\bibnamefont {Louren\c{c}o}},
  \bibinfo {author} {\bibfnamefont {S.~S.}\ \bibnamefont {Avancini}}, \bibinfo
  {author} {\bibfnamefont {B.~V.}\ \bibnamefont {Carlson}}, \bibinfo {author}
  {\bibfnamefont {A.}~\bibnamefont {Delfino}}, \bibinfo {author} {\bibfnamefont
  {D.~P.}\ \bibnamefont {Menezes}}, \bibinfo {author} {\bibfnamefont
  {C.}~\bibnamefont {Provid\^encia}}, \bibinfo {author} {\bibfnamefont
  {S.}~\bibnamefont {Typel}}, \ and\ \bibinfo {author} {\bibfnamefont {J.~R.}\
  \bibnamefont {Stone}},\ }\href {\doibase 10.1103/PhysRevC.90.055203}
  {\bibfield  {journal} {\bibinfo  {journal} {Phys. Rev. C}\ }\textbf {\bibinfo
  {volume} {90}},\ \bibinfo {pages} {055203} (\bibinfo {year} {2014})},\
  \Eprint {http://arxiv.org/abs/1405.3633} {arXiv:1405.3633 [nucl-th]}
  \BibitemShut {NoStop}%
\bibitem [{\citenamefont {Oertel}\ \emph {et~al.}(2017)\citenamefont {Oertel},
  \citenamefont {Hempel}, \citenamefont {Klähn},\ and\ \citenamefont
  {Typel}}]{Oertel:2016bki}%
  \BibitemOpen
  \bibfield  {author} {\bibinfo {author} {\bibfnamefont {M.}~\bibnamefont
  {Oertel}}, \bibinfo {author} {\bibfnamefont {M.}~\bibnamefont {Hempel}},
  \bibinfo {author} {\bibfnamefont {T.}~\bibnamefont {Klähn}}, \ and\ \bibinfo
  {author} {\bibfnamefont {S.}~\bibnamefont {Typel}},\ }\href {\doibase
  10.1103/RevModPhys.89.015007} {\bibfield  {journal} {\bibinfo  {journal}
  {Rev. Mod. Phys.}\ }\textbf {\bibinfo {volume} {89}},\ \bibinfo {pages}
  {015007} (\bibinfo {year} {2017})},\ \Eprint
  {http://arxiv.org/abs/1610.03361} {arXiv:1610.03361 [astro-ph.HE]}
  \BibitemShut {NoStop}%
\bibitem [{\citenamefont {Nandi}\ \emph {et~al.}(2019)\citenamefont {Nandi},
  \citenamefont {Char},\ and\ \citenamefont {Pal}}]{Nandi:2018ami}%
  \BibitemOpen
  \bibfield  {author} {\bibinfo {author} {\bibfnamefont {R.}~\bibnamefont
  {Nandi}}, \bibinfo {author} {\bibfnamefont {P.}~\bibnamefont {Char}}, \ and\
  \bibinfo {author} {\bibfnamefont {S.}~\bibnamefont {Pal}},\ }\href {\doibase
  10.1103/PhysRevC.99.052802} {\bibfield  {journal} {\bibinfo  {journal} {Phys.
  Rev. C}\ }\textbf {\bibinfo {volume} {99}},\ \bibinfo {pages} {052802}
  (\bibinfo {year} {2019})},\ \Eprint {http://arxiv.org/abs/1809.07108}
  {arXiv:1809.07108 [astro-ph.HE]} \BibitemShut {NoStop}%
\bibitem [{\citenamefont {Sun}\ \emph {et~al.}(2024)\citenamefont {Sun},
  \citenamefont {Bhattiprolu},\ and\ \citenamefont {Lattimer}}]{Sun:2023xkg}%
  \BibitemOpen
  \bibfield  {author} {\bibinfo {author} {\bibfnamefont {B.}~\bibnamefont
  {Sun}}, \bibinfo {author} {\bibfnamefont {S.}~\bibnamefont {Bhattiprolu}}, \
  and\ \bibinfo {author} {\bibfnamefont {J.~M.}\ \bibnamefont {Lattimer}},\
  }\href {\doibase 10.1103/PhysRevC.109.055801} {\bibfield  {journal} {\bibinfo
   {journal} {Phys. Rev. C}\ }\textbf {\bibinfo {volume} {109}},\ \bibinfo
  {pages} {055801} (\bibinfo {year} {2024})},\ \Eprint
  {http://arxiv.org/abs/2311.00843} {arXiv:2311.00843 [nucl-th]} \BibitemShut
  {NoStop}%
\bibitem [{\citenamefont {Read}\ \emph {et~al.}(2009)\citenamefont {Read},
  \citenamefont {Lackey}, \citenamefont {Owen},\ and\ \citenamefont
  {Friedman}}]{Read:2008iy}%
  \BibitemOpen
  \bibfield  {author} {\bibinfo {author} {\bibfnamefont {J.~S.}\ \bibnamefont
  {Read}}, \bibinfo {author} {\bibfnamefont {B.~D.}\ \bibnamefont {Lackey}},
  \bibinfo {author} {\bibfnamefont {B.~J.}\ \bibnamefont {Owen}}, \ and\
  \bibinfo {author} {\bibfnamefont {J.~L.}\ \bibnamefont {Friedman}},\ }\href
  {\doibase 10.1103/PhysRevD.79.124032} {\bibfield  {journal} {\bibinfo
  {journal} {Phys. Rev.}\ }\textbf {\bibinfo {volume} {D79}},\ \bibinfo {pages}
  {124032} (\bibinfo {year} {2009})},\ \Eprint {http://arxiv.org/abs/0812.2163}
  {arXiv:0812.2163 [astro-ph]} \BibitemShut {NoStop}%
\bibitem [{\citenamefont {Lindblom}(2010)}]{Lindblom:2010bb}%
  \BibitemOpen
  \bibfield  {author} {\bibinfo {author} {\bibfnamefont {L.}~\bibnamefont
  {Lindblom}},\ }\href {\doibase 10.1103/PhysRevD.82.103011} {\bibfield
  {journal} {\bibinfo  {journal} {Phys. Rev.}\ }\textbf {\bibinfo {volume}
  {D82}},\ \bibinfo {pages} {103011} (\bibinfo {year} {2010})},\ \Eprint
  {http://arxiv.org/abs/1009.0738} {arXiv:1009.0738 [astro-ph.HE]} \BibitemShut
  {NoStop}%
\bibitem [{\citenamefont {Landry}\ and\ \citenamefont
  {Essick}(2019)}]{Landry:2018prl}%
  \BibitemOpen
  \bibfield  {author} {\bibinfo {author} {\bibfnamefont {P.}~\bibnamefont
  {Landry}}\ and\ \bibinfo {author} {\bibfnamefont {R.}~\bibnamefont
  {Essick}},\ }\href {\doibase 10.1103/PhysRevD.99.084049} {\bibfield
  {journal} {\bibinfo  {journal} {Phys. Rev. D}\ }\textbf {\bibinfo {volume}
  {99}},\ \bibinfo {pages} {084049} (\bibinfo {year} {2019})},\ \Eprint
  {http://arxiv.org/abs/1811.12529} {arXiv:1811.12529 [gr-qc]} \BibitemShut
  {NoStop}%
\bibitem [{\citenamefont {Margueron}\ \emph {et~al.}(2018)\citenamefont
  {Margueron}, \citenamefont {Hoffmann~Casali},\ and\ \citenamefont
  {Gulminelli}}]{Margueron:2017eqc}%
  \BibitemOpen
  \bibfield  {author} {\bibinfo {author} {\bibfnamefont {J.}~\bibnamefont
  {Margueron}}, \bibinfo {author} {\bibfnamefont {R.}~\bibnamefont
  {Hoffmann~Casali}}, \ and\ \bibinfo {author} {\bibfnamefont {F.}~\bibnamefont
  {Gulminelli}},\ }\href {\doibase 10.1103/PhysRevC.97.025805} {\bibfield
  {journal} {\bibinfo  {journal} {Phys. Rev.}\ }\textbf {\bibinfo {volume}
  {C97}},\ \bibinfo {pages} {025805} (\bibinfo {year} {2018})},\ \Eprint
  {http://arxiv.org/abs/1708.06894} {arXiv:1708.06894 [nucl-th]} \BibitemShut
  {NoStop}%
\bibitem [{\citenamefont {Biswas}\ \emph {et~al.}(2021)\citenamefont {Biswas},
  \citenamefont {Char}, \citenamefont {Nandi},\ and\ \citenamefont
  {Bose}}]{Biswas:2020puz}%
  \BibitemOpen
  \bibfield  {author} {\bibinfo {author} {\bibfnamefont {B.}~\bibnamefont
  {Biswas}}, \bibinfo {author} {\bibfnamefont {P.}~\bibnamefont {Char}},
  \bibinfo {author} {\bibfnamefont {R.}~\bibnamefont {Nandi}}, \ and\ \bibinfo
  {author} {\bibfnamefont {S.}~\bibnamefont {Bose}},\ }\href {\doibase
  10.1103/PhysRevD.103.103015} {\bibfield  {journal} {\bibinfo  {journal}
  {Phys. Rev. D}\ }\textbf {\bibinfo {volume} {103}},\ \bibinfo {pages}
  {103015} (\bibinfo {year} {2021})},\ \Eprint
  {http://arxiv.org/abs/2008.01582} {arXiv:2008.01582 [astro-ph.HE]}
  \BibitemShut {NoStop}%
\bibitem [{\citenamefont {Gandolfi}\ \emph {et~al.}(2019)\citenamefont
  {Gandolfi}, \citenamefont {Lippuner}, \citenamefont {Steiner}, \citenamefont
  {Tews}, \citenamefont {Du},\ and\ \citenamefont
  {Al-Mamun}}]{Gandolfi:2019zpj}%
  \BibitemOpen
  \bibfield  {author} {\bibinfo {author} {\bibfnamefont {S.}~\bibnamefont
  {Gandolfi}}, \bibinfo {author} {\bibfnamefont {J.}~\bibnamefont {Lippuner}},
  \bibinfo {author} {\bibfnamefont {A.~W.}\ \bibnamefont {Steiner}}, \bibinfo
  {author} {\bibfnamefont {I.}~\bibnamefont {Tews}}, \bibinfo {author}
  {\bibfnamefont {X.}~\bibnamefont {Du}}, \ and\ \bibinfo {author}
  {\bibfnamefont {M.}~\bibnamefont {Al-Mamun}},\ }\href {\doibase
  10.1088/1361-6471/ab29b3} {\bibfield  {journal} {\bibinfo  {journal} {J.
  Phys. G}\ }\textbf {\bibinfo {volume} {46}},\ \bibinfo {pages} {103001}
  (\bibinfo {year} {2019})},\ \Eprint {http://arxiv.org/abs/1903.06730}
  {arXiv:1903.06730 [nucl-th]} \BibitemShut {NoStop}%
\bibitem [{\citenamefont {Cromartie}\ \emph {et~al.}(2019)\citenamefont
  {Cromartie} \emph {et~al.}}]{Cromartie:2019kug}%
  \BibitemOpen
  \bibfield  {author} {\bibinfo {author} {\bibfnamefont {H.~T.}\ \bibnamefont
  {Cromartie}} \emph {et~al.},\ }\href {\doibase 10.1038/s41550-019-0880-2}
  {\bibfield  {journal} {\bibinfo  {journal} {Nature Astron.}\ }\textbf
  {\bibinfo {volume} {4}},\ \bibinfo {pages} {72} (\bibinfo {year} {2019})},\
  \Eprint {http://arxiv.org/abs/1904.06759} {arXiv:1904.06759 [astro-ph.HE]}
  \BibitemShut {NoStop}%
\bibitem [{\citenamefont {Fonseca}\ \emph {et~al.}(2021)\citenamefont {Fonseca}
  \emph {et~al.}}]{Fonseca:2021wxt}%
  \BibitemOpen
  \bibfield  {author} {\bibinfo {author} {\bibfnamefont {E.}~\bibnamefont
  {Fonseca}} \emph {et~al.},\ }\href {\doibase 10.3847/2041-8213/ac03b8}
  {\bibfield  {journal} {\bibinfo  {journal} {Astrophys. J. Lett.}\ }\textbf
  {\bibinfo {volume} {915}},\ \bibinfo {pages} {L12} (\bibinfo {year}
  {2021})},\ \Eprint {http://arxiv.org/abs/2104.00880} {arXiv:2104.00880
  [astro-ph.HE]} \BibitemShut {NoStop}%
\bibitem [{\citenamefont {Abbott}\ \emph
  {et~al.}(2017{\natexlab{a}})\citenamefont {Abbott} \emph
  {et~al.}}]{TheLIGOScientific:2017qsa}%
  \BibitemOpen
  \bibfield  {author} {\bibinfo {author} {\bibfnamefont {B.~P.}\ \bibnamefont
  {Abbott}} \emph {et~al.} (\bibinfo {collaboration} {LIGO Scientific,
  Virgo}),\ }\href {\doibase 10.1103/PhysRevLett.119.161101} {\bibfield
  {journal} {\bibinfo  {journal} {Phys. Rev. Lett.}\ }\textbf {\bibinfo
  {volume} {119}},\ \bibinfo {pages} {161101} (\bibinfo {year}
  {2017}{\natexlab{a}})},\ \Eprint {http://arxiv.org/abs/1710.05832}
  {arXiv:1710.05832 [gr-qc]} \BibitemShut {NoStop}%
\bibitem [{\citenamefont {Abbott}\ \emph
  {et~al.}(2017{\natexlab{b}})\citenamefont {Abbott} \emph
  {et~al.}}]{LIGOScientific:2017ync}%
  \BibitemOpen
  \bibfield  {author} {\bibinfo {author} {\bibfnamefont {B.~P.}\ \bibnamefont
  {Abbott}} \emph {et~al.} (\bibinfo {collaboration} {LIGO Scientific, Virgo,
  Fermi GBM, INTEGRAL, IceCube, AstroSat Cadmium Zinc Telluride Imager Team,
  IPN, Insight-Hxmt, ANTARES, Swift, AGILE Team, 1M2H Team, Dark Energy Camera
  GW-EM, DES, DLT40, GRAWITA, Fermi-LAT, ATCA, ASKAP, Las Cumbres Observatory
  Group, OzGrav, DWF (Deeper Wider Faster Program), AST3, CAASTRO, VINROUGE,
  MASTER, J-GEM, GROWTH, JAGWAR, CaltechNRAO, TTU-NRAO, NuSTAR, Pan-STARRS,
  MAXI Team, TZAC Consortium, KU, Nordic Optical Telescope, ePESSTO, GROND,
  Texas Tech University, SALT Group, TOROS, BOOTES, MWA, CALET, IKI-GW
  Follow-up, H.E.S.S., LOFAR, LWA, HAWC, Pierre Auger, ALMA, Euro VLBI Team, Pi
  of Sky, Chandra Team at McGill University, DFN, ATLAS Telescopes, High Time
  Resolution Universe Survey, RIMAS, RATIR, SKA South Africa/MeerKAT}),\ }\href
  {\doibase 10.3847/2041-8213/aa91c9} {\bibfield  {journal} {\bibinfo
  {journal} {Astrophys. J. Lett.}\ }\textbf {\bibinfo {volume} {848}},\
  \bibinfo {pages} {L12} (\bibinfo {year} {2017}{\natexlab{b}})},\ \Eprint
  {http://arxiv.org/abs/1710.05833} {arXiv:1710.05833 [astro-ph.HE]}
  \BibitemShut {NoStop}%
\bibitem [{\citenamefont {Abbott}\ \emph
  {et~al.}(2019{\natexlab{a}})\citenamefont {Abbott} \emph
  {et~al.}}]{LIGOScientific:2018hze}%
  \BibitemOpen
  \bibfield  {author} {\bibinfo {author} {\bibfnamefont {B.~P.}\ \bibnamefont
  {Abbott}} \emph {et~al.} (\bibinfo {collaboration} {LIGO Scientific,
  Virgo}),\ }\href {\doibase 10.1103/PhysRevX.9.011001} {\bibfield  {journal}
  {\bibinfo  {journal} {Phys. Rev. X}\ }\textbf {\bibinfo {volume} {9}},\
  \bibinfo {pages} {011001} (\bibinfo {year} {2019}{\natexlab{a}})},\ \Eprint
  {http://arxiv.org/abs/1805.11579} {arXiv:1805.11579 [gr-qc]} \BibitemShut
  {NoStop}%
\bibitem [{\citenamefont {{Miller}}\ \emph {et~al.}(2019)\citenamefont
  {{Miller}}, \citenamefont {{Lamb}}, \citenamefont {{Dittmann}}, \citenamefont
  {{Bogdanov}}, \citenamefont {{Arzoumanian}}, \citenamefont {{Gendreau}},
  \citenamefont {{Guillot}}, \citenamefont {{Harding}}, \citenamefont {{Ho}},
  \citenamefont {{Lattimer}}, \citenamefont {{Ludlam}}, \citenamefont
  {{Mahmoodifar}}, \citenamefont {{Morsink}}, \citenamefont {{Ray}},
  \citenamefont {{Strohmayer}}, \citenamefont {{Wood}}, \citenamefont
  {{Enoto}}, \citenamefont {{Foster}}, \citenamefont {{Okajima}}, \citenamefont
  {{Prigozhin}},\ and\ \citenamefont {{Soong}}}]{Miller+2019}%
  \BibitemOpen
  \bibfield  {author} {\bibinfo {author} {\bibfnamefont {M.~C.}\ \bibnamefont
  {{Miller}}}, \bibinfo {author} {\bibfnamefont {F.~K.}\ \bibnamefont
  {{Lamb}}}, \bibinfo {author} {\bibfnamefont {A.~J.}\ \bibnamefont
  {{Dittmann}}}, \bibinfo {author} {\bibfnamefont {S.}~\bibnamefont
  {{Bogdanov}}}, \bibinfo {author} {\bibfnamefont {Z.}~\bibnamefont
  {{Arzoumanian}}}, \bibinfo {author} {\bibfnamefont {K.~C.}\ \bibnamefont
  {{Gendreau}}}, \bibinfo {author} {\bibfnamefont {S.}~\bibnamefont
  {{Guillot}}}, \bibinfo {author} {\bibfnamefont {A.~K.}\ \bibnamefont
  {{Harding}}}, \bibinfo {author} {\bibfnamefont {W.~C.~G.}\ \bibnamefont
  {{Ho}}}, \bibinfo {author} {\bibfnamefont {J.~M.}\ \bibnamefont
  {{Lattimer}}}, \bibinfo {author} {\bibfnamefont {R.~M.}\ \bibnamefont
  {{Ludlam}}}, \bibinfo {author} {\bibfnamefont {S.}~\bibnamefont
  {{Mahmoodifar}}}, \bibinfo {author} {\bibfnamefont {S.~M.}\ \bibnamefont
  {{Morsink}}}, \bibinfo {author} {\bibfnamefont {P.~S.}\ \bibnamefont
  {{Ray}}}, \bibinfo {author} {\bibfnamefont {T.~E.}\ \bibnamefont
  {{Strohmayer}}}, \bibinfo {author} {\bibfnamefont {K.~S.}\ \bibnamefont
  {{Wood}}}, \bibinfo {author} {\bibfnamefont {T.}~\bibnamefont {{Enoto}}},
  \bibinfo {author} {\bibfnamefont {R.}~\bibnamefont {{Foster}}}, \bibinfo
  {author} {\bibfnamefont {T.}~\bibnamefont {{Okajima}}}, \bibinfo {author}
  {\bibfnamefont {G.}~\bibnamefont {{Prigozhin}}}, \ and\ \bibinfo {author}
  {\bibfnamefont {Y.}~\bibnamefont {{Soong}}},\ }\href {\doibase
  10.3847/2041-8213/ab50c5} {\bibfield  {journal} {\bibinfo  {journal} {ApJL}\
  }\textbf {\bibinfo {volume} {887}},\ \bibinfo {eid} {L24} (\bibinfo {year}
  {2019})},\ \Eprint {http://arxiv.org/abs/1912.05705} {arXiv:1912.05705
  [astro-ph.HE]} \BibitemShut {NoStop}%
\bibitem [{\citenamefont {{Riley}}\ \emph {et~al.}(2019)\citenamefont
  {{Riley}}, \citenamefont {{Watts}}, \citenamefont {{Bogdanov}}, \citenamefont
  {{Ray}}, \citenamefont {{Ludlam}}, \citenamefont {{Guillot}}, \citenamefont
  {{Arzoumanian}}, \citenamefont {{Baker}}, \citenamefont {{Bilous}},
  \citenamefont {{Chakrabarty}}, \citenamefont {{Gendreau}}, \citenamefont
  {{Harding}}, \citenamefont {{Ho}}, \citenamefont {{Lattimer}}, \citenamefont
  {{Morsink}},\ and\ \citenamefont {{Strohmayer}}}]{Riley+2019}%
  \BibitemOpen
  \bibfield  {author} {\bibinfo {author} {\bibfnamefont {T.~E.}\ \bibnamefont
  {{Riley}}}, \bibinfo {author} {\bibfnamefont {A.~L.}\ \bibnamefont
  {{Watts}}}, \bibinfo {author} {\bibfnamefont {S.}~\bibnamefont {{Bogdanov}}},
  \bibinfo {author} {\bibfnamefont {P.~S.}\ \bibnamefont {{Ray}}}, \bibinfo
  {author} {\bibfnamefont {R.~M.}\ \bibnamefont {{Ludlam}}}, \bibinfo {author}
  {\bibfnamefont {S.}~\bibnamefont {{Guillot}}}, \bibinfo {author}
  {\bibfnamefont {Z.}~\bibnamefont {{Arzoumanian}}}, \bibinfo {author}
  {\bibfnamefont {C.~L.}\ \bibnamefont {{Baker}}}, \bibinfo {author}
  {\bibfnamefont {A.~V.}\ \bibnamefont {{Bilous}}}, \bibinfo {author}
  {\bibfnamefont {D.}~\bibnamefont {{Chakrabarty}}}, \bibinfo {author}
  {\bibfnamefont {K.~C.}\ \bibnamefont {{Gendreau}}}, \bibinfo {author}
  {\bibfnamefont {A.~K.}\ \bibnamefont {{Harding}}}, \bibinfo {author}
  {\bibfnamefont {W.~C.~G.}\ \bibnamefont {{Ho}}}, \bibinfo {author}
  {\bibfnamefont {J.~M.}\ \bibnamefont {{Lattimer}}}, \bibinfo {author}
  {\bibfnamefont {S.~M.}\ \bibnamefont {{Morsink}}}, \ and\ \bibinfo {author}
  {\bibfnamefont {T.~E.}\ \bibnamefont {{Strohmayer}}},\ }\href {\doibase
  10.3847/2041-8213/ab481c} {\bibfield  {journal} {\bibinfo  {journal} {ApJL}\
  }\textbf {\bibinfo {volume} {887}},\ \bibinfo {eid} {L21} (\bibinfo {year}
  {2019})},\ \Eprint {http://arxiv.org/abs/1912.05702} {arXiv:1912.05702
  [astro-ph.HE]} \BibitemShut {NoStop}%
\bibitem [{\citenamefont {{Miller}}\ \emph {et~al.}(2021)\citenamefont
  {{Miller}}, \citenamefont {{Lamb}}, \citenamefont {{Dittmann}}, \citenamefont
  {{Bogdanov}}, \citenamefont {{Arzoumanian}}, \citenamefont {{Gendreau}},
  \citenamefont {{Guillot}}, \citenamefont {{Ho}}, \citenamefont {{Lattimer}},
  \citenamefont {{Loewenstein}}, \citenamefont {{Morsink}}, \citenamefont
  {{Ray}}, \citenamefont {{Wolff}}, \citenamefont {{Baker}}, \citenamefont
  {{Cazeau}}, \citenamefont {{Manthripragada}}, \citenamefont {{Markwardt}},
  \citenamefont {{Okajima}}, \citenamefont {{Pollard}}, \citenamefont
  {{Cognard}}, \citenamefont {{Cromartie}}, \citenamefont {{Fonseca}},
  \citenamefont {{Guillemot}}, \citenamefont {{Kerr}}, \citenamefont
  {{Parthasarathy}}, \citenamefont {{Pennucci}}, \citenamefont {{Ransom}},\
  and\ \citenamefont {{Stairs}}}]{Miller+2021}%
  \BibitemOpen
  \bibfield  {author} {\bibinfo {author} {\bibfnamefont {M.~C.}\ \bibnamefont
  {{Miller}}}, \bibinfo {author} {\bibfnamefont {F.~K.}\ \bibnamefont
  {{Lamb}}}, \bibinfo {author} {\bibfnamefont {A.~J.}\ \bibnamefont
  {{Dittmann}}}, \bibinfo {author} {\bibfnamefont {S.}~\bibnamefont
  {{Bogdanov}}}, \bibinfo {author} {\bibfnamefont {Z.}~\bibnamefont
  {{Arzoumanian}}}, \bibinfo {author} {\bibfnamefont {K.~C.}\ \bibnamefont
  {{Gendreau}}}, \bibinfo {author} {\bibfnamefont {S.}~\bibnamefont
  {{Guillot}}}, \bibinfo {author} {\bibfnamefont {W.~C.~G.}\ \bibnamefont
  {{Ho}}}, \bibinfo {author} {\bibfnamefont {J.~M.}\ \bibnamefont
  {{Lattimer}}}, \bibinfo {author} {\bibfnamefont {M.}~\bibnamefont
  {{Loewenstein}}}, \bibinfo {author} {\bibfnamefont {S.~M.}\ \bibnamefont
  {{Morsink}}}, \bibinfo {author} {\bibfnamefont {P.~S.}\ \bibnamefont
  {{Ray}}}, \bibinfo {author} {\bibfnamefont {M.~T.}\ \bibnamefont {{Wolff}}},
  \bibinfo {author} {\bibfnamefont {C.~L.}\ \bibnamefont {{Baker}}}, \bibinfo
  {author} {\bibfnamefont {T.}~\bibnamefont {{Cazeau}}}, \bibinfo {author}
  {\bibfnamefont {S.}~\bibnamefont {{Manthripragada}}}, \bibinfo {author}
  {\bibfnamefont {C.~B.}\ \bibnamefont {{Markwardt}}}, \bibinfo {author}
  {\bibfnamefont {T.}~\bibnamefont {{Okajima}}}, \bibinfo {author}
  {\bibfnamefont {S.}~\bibnamefont {{Pollard}}}, \bibinfo {author}
  {\bibfnamefont {I.}~\bibnamefont {{Cognard}}}, \bibinfo {author}
  {\bibfnamefont {H.~T.}\ \bibnamefont {{Cromartie}}}, \bibinfo {author}
  {\bibfnamefont {E.}~\bibnamefont {{Fonseca}}}, \bibinfo {author}
  {\bibfnamefont {L.}~\bibnamefont {{Guillemot}}}, \bibinfo {author}
  {\bibfnamefont {M.}~\bibnamefont {{Kerr}}}, \bibinfo {author} {\bibfnamefont
  {A.}~\bibnamefont {{Parthasarathy}}}, \bibinfo {author} {\bibfnamefont
  {T.~T.}\ \bibnamefont {{Pennucci}}}, \bibinfo {author} {\bibfnamefont
  {S.}~\bibnamefont {{Ransom}}}, \ and\ \bibinfo {author} {\bibfnamefont
  {I.}~\bibnamefont {{Stairs}}},\ }\href {\doibase 10.3847/2041-8213/ac089b}
  {\bibfield  {journal} {\bibinfo  {journal} {ApJL}\ }\textbf {\bibinfo
  {volume} {918}},\ \bibinfo {eid} {L28} (\bibinfo {year} {2021})},\ \Eprint
  {http://arxiv.org/abs/2105.06979} {arXiv:2105.06979 [astro-ph.HE]}
  \BibitemShut {NoStop}%
\bibitem [{\citenamefont {{Riley}}\ \emph {et~al.}(2021)\citenamefont
  {{Riley}}, \citenamefont {{Watts}}, \citenamefont {{Ray}}, \citenamefont
  {{Bogdanov}}, \citenamefont {{Guillot}}, \citenamefont {{Morsink}},
  \citenamefont {{Bilous}}, \citenamefont {{Arzoumanian}}, \citenamefont
  {{Choudhury}}, \citenamefont {{Deneva}}, \citenamefont {{Gendreau}},
  \citenamefont {{Harding}}, \citenamefont {{Ho}}, \citenamefont {{Lattimer}},
  \citenamefont {{Loewenstein}}, \citenamefont {{Ludlam}}, \citenamefont
  {{Markwardt}}, \citenamefont {{Okajima}}, \citenamefont
  {{Prescod-Weinstein}}, \citenamefont {{Remillard}}, \citenamefont {{Wolff}},
  \citenamefont {{Fonseca}}, \citenamefont {{Cromartie}}, \citenamefont
  {{Kerr}}, \citenamefont {{Pennucci}}, \citenamefont {{Parthasarathy}},
  \citenamefont {{Ransom}}, \citenamefont {{Stairs}}, \citenamefont
  {{Guillemot}},\ and\ \citenamefont {{Cognard}}}]{Riley+2021}%
  \BibitemOpen
  \bibfield  {author} {\bibinfo {author} {\bibfnamefont {T.~E.}\ \bibnamefont
  {{Riley}}}, \bibinfo {author} {\bibfnamefont {A.~L.}\ \bibnamefont
  {{Watts}}}, \bibinfo {author} {\bibfnamefont {P.~S.}\ \bibnamefont {{Ray}}},
  \bibinfo {author} {\bibfnamefont {S.}~\bibnamefont {{Bogdanov}}}, \bibinfo
  {author} {\bibfnamefont {S.}~\bibnamefont {{Guillot}}}, \bibinfo {author}
  {\bibfnamefont {S.~M.}\ \bibnamefont {{Morsink}}}, \bibinfo {author}
  {\bibfnamefont {A.~V.}\ \bibnamefont {{Bilous}}}, \bibinfo {author}
  {\bibfnamefont {Z.}~\bibnamefont {{Arzoumanian}}}, \bibinfo {author}
  {\bibfnamefont {D.}~\bibnamefont {{Choudhury}}}, \bibinfo {author}
  {\bibfnamefont {J.~S.}\ \bibnamefont {{Deneva}}}, \bibinfo {author}
  {\bibfnamefont {K.~C.}\ \bibnamefont {{Gendreau}}}, \bibinfo {author}
  {\bibfnamefont {A.~K.}\ \bibnamefont {{Harding}}}, \bibinfo {author}
  {\bibfnamefont {W.~C.~G.}\ \bibnamefont {{Ho}}}, \bibinfo {author}
  {\bibfnamefont {J.~M.}\ \bibnamefont {{Lattimer}}}, \bibinfo {author}
  {\bibfnamefont {M.}~\bibnamefont {{Loewenstein}}}, \bibinfo {author}
  {\bibfnamefont {R.~M.}\ \bibnamefont {{Ludlam}}}, \bibinfo {author}
  {\bibfnamefont {C.~B.}\ \bibnamefont {{Markwardt}}}, \bibinfo {author}
  {\bibfnamefont {T.}~\bibnamefont {{Okajima}}}, \bibinfo {author}
  {\bibfnamefont {C.}~\bibnamefont {{Prescod-Weinstein}}}, \bibinfo {author}
  {\bibfnamefont {R.~A.}\ \bibnamefont {{Remillard}}}, \bibinfo {author}
  {\bibfnamefont {M.~T.}\ \bibnamefont {{Wolff}}}, \bibinfo {author}
  {\bibfnamefont {E.}~\bibnamefont {{Fonseca}}}, \bibinfo {author}
  {\bibfnamefont {H.~T.}\ \bibnamefont {{Cromartie}}}, \bibinfo {author}
  {\bibfnamefont {M.}~\bibnamefont {{Kerr}}}, \bibinfo {author} {\bibfnamefont
  {T.~T.}\ \bibnamefont {{Pennucci}}}, \bibinfo {author} {\bibfnamefont
  {A.}~\bibnamefont {{Parthasarathy}}}, \bibinfo {author} {\bibfnamefont
  {S.}~\bibnamefont {{Ransom}}}, \bibinfo {author} {\bibfnamefont
  {I.}~\bibnamefont {{Stairs}}}, \bibinfo {author} {\bibfnamefont
  {L.}~\bibnamefont {{Guillemot}}}, \ and\ \bibinfo {author} {\bibfnamefont
  {I.}~\bibnamefont {{Cognard}}},\ }\href {\doibase 10.3847/2041-8213/ac0a81}
  {\bibfield  {journal} {\bibinfo  {journal} {ApJL}\ }\textbf {\bibinfo
  {volume} {918}},\ \bibinfo {eid} {L27} (\bibinfo {year} {2021})},\ \Eprint
  {http://arxiv.org/abs/2105.06980} {arXiv:2105.06980 [astro-ph.HE]}
  \BibitemShut {NoStop}%
\bibitem [{\citenamefont {{Salmi}}\ \emph {et~al.}(2022)\citenamefont
  {{Salmi}}, \citenamefont {{Vinciguerra}}, \citenamefont {{Choudhury}},
  \citenamefont {{Riley}}, \citenamefont {{Watts}}, \citenamefont
  {{Remillard}}, \citenamefont {{Ray}}, \citenamefont {{Bogdanov}},
  \citenamefont {{Guillot}}, \citenamefont {{Arzoumanian}}, \citenamefont
  {{Chirenti}}, \citenamefont {{Dittmann}}, \citenamefont {{Gendreau}},
  \citenamefont {{Ho}}, \citenamefont {{Miller}}, \citenamefont {{Morsink}},
  \citenamefont {{Wadiasingh}},\ and\ \citenamefont {{Wolff}}}]{Salmi+2022}%
  \BibitemOpen
  \bibfield  {author} {\bibinfo {author} {\bibfnamefont {T.}~\bibnamefont
  {{Salmi}}}, \bibinfo {author} {\bibfnamefont {S.}~\bibnamefont
  {{Vinciguerra}}}, \bibinfo {author} {\bibfnamefont {D.}~\bibnamefont
  {{Choudhury}}}, \bibinfo {author} {\bibfnamefont {T.~E.}\ \bibnamefont
  {{Riley}}}, \bibinfo {author} {\bibfnamefont {A.~L.}\ \bibnamefont
  {{Watts}}}, \bibinfo {author} {\bibfnamefont {R.~A.}\ \bibnamefont
  {{Remillard}}}, \bibinfo {author} {\bibfnamefont {P.~S.}\ \bibnamefont
  {{Ray}}}, \bibinfo {author} {\bibfnamefont {S.}~\bibnamefont {{Bogdanov}}},
  \bibinfo {author} {\bibfnamefont {S.}~\bibnamefont {{Guillot}}}, \bibinfo
  {author} {\bibfnamefont {Z.}~\bibnamefont {{Arzoumanian}}}, \bibinfo {author}
  {\bibfnamefont {C.}~\bibnamefont {{Chirenti}}}, \bibinfo {author}
  {\bibfnamefont {A.~J.}\ \bibnamefont {{Dittmann}}}, \bibinfo {author}
  {\bibfnamefont {K.~C.}\ \bibnamefont {{Gendreau}}}, \bibinfo {author}
  {\bibfnamefont {W.~C.~G.}\ \bibnamefont {{Ho}}}, \bibinfo {author}
  {\bibfnamefont {M.~C.}\ \bibnamefont {{Miller}}}, \bibinfo {author}
  {\bibfnamefont {S.~M.}\ \bibnamefont {{Morsink}}}, \bibinfo {author}
  {\bibfnamefont {Z.}~\bibnamefont {{Wadiasingh}}}, \ and\ \bibinfo {author}
  {\bibfnamefont {M.~T.}\ \bibnamefont {{Wolff}}},\ }\href {\doibase
  10.3847/1538-4357/ac983d} {\bibfield  {journal} {\bibinfo  {journal} {ApJ}\
  }\textbf {\bibinfo {volume} {941}},\ \bibinfo {eid} {150} (\bibinfo {year}
  {2022})},\ \Eprint {http://arxiv.org/abs/2209.12840} {arXiv:2209.12840
  [astro-ph.HE]} \BibitemShut {NoStop}%
\bibitem [{\citenamefont {{Vinciguerra}}\ \emph {et~al.}(2024)\citenamefont
  {{Vinciguerra}}, \citenamefont {{Salmi}}, \citenamefont {{Watts}},
  \citenamefont {{Choudhury}}, \citenamefont {{Riley}}, \citenamefont {{Ray}},
  \citenamefont {{Bogdanov}}, \citenamefont {{Kini}}, \citenamefont
  {{Guillot}}, \citenamefont {{Chakrabarty}}, \citenamefont {{Ho}},
  \citenamefont {{Huppenkothen}}, \citenamefont {{Morsink}}, \citenamefont
  {{Wadiasingh}},\ and\ \citenamefont {{Wolff}}}]{Vinciguerra+2024}%
  \BibitemOpen
  \bibfield  {author} {\bibinfo {author} {\bibfnamefont {S.}~\bibnamefont
  {{Vinciguerra}}}, \bibinfo {author} {\bibfnamefont {T.}~\bibnamefont
  {{Salmi}}}, \bibinfo {author} {\bibfnamefont {A.~L.}\ \bibnamefont
  {{Watts}}}, \bibinfo {author} {\bibfnamefont {D.}~\bibnamefont
  {{Choudhury}}}, \bibinfo {author} {\bibfnamefont {T.~E.}\ \bibnamefont
  {{Riley}}}, \bibinfo {author} {\bibfnamefont {P.~S.}\ \bibnamefont {{Ray}}},
  \bibinfo {author} {\bibfnamefont {S.}~\bibnamefont {{Bogdanov}}}, \bibinfo
  {author} {\bibfnamefont {Y.}~\bibnamefont {{Kini}}}, \bibinfo {author}
  {\bibfnamefont {S.}~\bibnamefont {{Guillot}}}, \bibinfo {author}
  {\bibfnamefont {D.}~\bibnamefont {{Chakrabarty}}}, \bibinfo {author}
  {\bibfnamefont {W.~C.~G.}\ \bibnamefont {{Ho}}}, \bibinfo {author}
  {\bibfnamefont {D.}~\bibnamefont {{Huppenkothen}}}, \bibinfo {author}
  {\bibfnamefont {S.~M.}\ \bibnamefont {{Morsink}}}, \bibinfo {author}
  {\bibfnamefont {Z.}~\bibnamefont {{Wadiasingh}}}, \ and\ \bibinfo {author}
  {\bibfnamefont {M.~T.}\ \bibnamefont {{Wolff}}},\ }\href {\doibase
  10.3847/1538-4357/acfb83} {\bibfield  {journal} {\bibinfo  {journal} {ApJ}\
  }\textbf {\bibinfo {volume} {961}},\ \bibinfo {eid} {62} (\bibinfo {year}
  {2024})},\ \Eprint {http://arxiv.org/abs/2308.09469} {arXiv:2308.09469
  [astro-ph.HE]} \BibitemShut {NoStop}%
\bibitem [{\citenamefont {Choudhury}\ \emph {et~al.}(2024)\citenamefont
  {Choudhury} \emph {et~al.}}]{Choudhury:2024xbk}%
  \BibitemOpen
  \bibfield  {author} {\bibinfo {author} {\bibfnamefont {D.}~\bibnamefont
  {Choudhury}} \emph {et~al.},\ }\href {\doibase 10.3847/2041-8213/ad5a6f}
  {\bibfield  {journal} {\bibinfo  {journal} {Astrophys. J. Lett.}\ }\textbf
  {\bibinfo {volume} {971}},\ \bibinfo {pages} {L20} (\bibinfo {year}
  {2024})},\ \Eprint {http://arxiv.org/abs/2407.06789} {arXiv:2407.06789
  [astro-ph.HE]} \BibitemShut {NoStop}%
\bibitem [{\citenamefont {Tews}\ \emph {et~al.}(2013)\citenamefont {Tews},
  \citenamefont {Kr\"uger}, \citenamefont {Hebeler},\ and\ \citenamefont
  {Schwenk}}]{Tews:2012fj}%
  \BibitemOpen
  \bibfield  {author} {\bibinfo {author} {\bibfnamefont {I.}~\bibnamefont
  {Tews}}, \bibinfo {author} {\bibfnamefont {T.}~\bibnamefont {Kr\"uger}},
  \bibinfo {author} {\bibfnamefont {K.}~\bibnamefont {Hebeler}}, \ and\
  \bibinfo {author} {\bibfnamefont {A.}~\bibnamefont {Schwenk}},\ }\href
  {\doibase 10.1103/PhysRevLett.110.032504} {\bibfield  {journal} {\bibinfo
  {journal} {Phys. Rev. Lett.}\ }\textbf {\bibinfo {volume} {110}},\ \bibinfo
  {pages} {032504} (\bibinfo {year} {2013})},\ \Eprint
  {http://arxiv.org/abs/1206.0025} {arXiv:1206.0025 [nucl-th]} \BibitemShut
  {NoStop}%
\bibitem [{\citenamefont {Hebeler}\ \emph {et~al.}(2013)\citenamefont
  {Hebeler}, \citenamefont {Lattimer}, \citenamefont {Pethick},\ and\
  \citenamefont {Schwenk}}]{Hebeler:2013nza}%
  \BibitemOpen
  \bibfield  {author} {\bibinfo {author} {\bibfnamefont {K.}~\bibnamefont
  {Hebeler}}, \bibinfo {author} {\bibfnamefont {J.~M.}\ \bibnamefont
  {Lattimer}}, \bibinfo {author} {\bibfnamefont {C.~J.}\ \bibnamefont
  {Pethick}}, \ and\ \bibinfo {author} {\bibfnamefont {A.}~\bibnamefont
  {Schwenk}},\ }\href {\doibase 10.1088/0004-637X/773/1/11} {\bibfield
  {journal} {\bibinfo  {journal} {Astrophys. J.}\ }\textbf {\bibinfo {volume}
  {773}},\ \bibinfo {pages} {11} (\bibinfo {year} {2013})},\ \Eprint
  {http://arxiv.org/abs/1303.4662} {arXiv:1303.4662 [astro-ph.SR]} \BibitemShut
  {NoStop}%
\bibitem [{\citenamefont {Lynn}\ \emph {et~al.}(2016)\citenamefont {Lynn},
  \citenamefont {Tews}, \citenamefont {Carlson}, \citenamefont {Gandolfi},
  \citenamefont {Gezerlis}, \citenamefont {Schmidt},\ and\ \citenamefont
  {Schwenk}}]{Lynn:2015jua}%
  \BibitemOpen
  \bibfield  {author} {\bibinfo {author} {\bibfnamefont {J.~E.}\ \bibnamefont
  {Lynn}}, \bibinfo {author} {\bibfnamefont {I.}~\bibnamefont {Tews}}, \bibinfo
  {author} {\bibfnamefont {J.}~\bibnamefont {Carlson}}, \bibinfo {author}
  {\bibfnamefont {S.}~\bibnamefont {Gandolfi}}, \bibinfo {author}
  {\bibfnamefont {A.}~\bibnamefont {Gezerlis}}, \bibinfo {author}
  {\bibfnamefont {K.~E.}\ \bibnamefont {Schmidt}}, \ and\ \bibinfo {author}
  {\bibfnamefont {A.}~\bibnamefont {Schwenk}},\ }\href {\doibase
  10.1103/PhysRevLett.116.062501} {\bibfield  {journal} {\bibinfo  {journal}
  {Phys. Rev. Lett.}\ }\textbf {\bibinfo {volume} {116}},\ \bibinfo {pages}
  {062501} (\bibinfo {year} {2016})},\ \Eprint
  {http://arxiv.org/abs/1509.03470} {arXiv:1509.03470 [nucl-th]} \BibitemShut
  {NoStop}%
\bibitem [{\citenamefont {Drischler}\ \emph {et~al.}(2019)\citenamefont
  {Drischler}, \citenamefont {Hebeler},\ and\ \citenamefont
  {Schwenk}}]{Drischler:2017wtt}%
  \BibitemOpen
  \bibfield  {author} {\bibinfo {author} {\bibfnamefont {C.}~\bibnamefont
  {Drischler}}, \bibinfo {author} {\bibfnamefont {K.}~\bibnamefont {Hebeler}},
  \ and\ \bibinfo {author} {\bibfnamefont {A.}~\bibnamefont {Schwenk}},\ }\href
  {\doibase 10.1103/PhysRevLett.122.042501} {\bibfield  {journal} {\bibinfo
  {journal} {Phys. Rev. Lett.}\ }\textbf {\bibinfo {volume} {122}},\ \bibinfo
  {pages} {042501} (\bibinfo {year} {2019})},\ \Eprint
  {http://arxiv.org/abs/1710.08220} {arXiv:1710.08220 [nucl-th]} \BibitemShut
  {NoStop}%
\bibitem [{\citenamefont {Huth}\ \emph {et~al.}(2021)\citenamefont {Huth},
  \citenamefont {Wellenhofer},\ and\ \citenamefont {Schwenk}}]{Huth:2020ozf}%
  \BibitemOpen
  \bibfield  {author} {\bibinfo {author} {\bibfnamefont {S.}~\bibnamefont
  {Huth}}, \bibinfo {author} {\bibfnamefont {C.}~\bibnamefont {Wellenhofer}}, \
  and\ \bibinfo {author} {\bibfnamefont {A.}~\bibnamefont {Schwenk}},\ }\href
  {\doibase 10.1103/PhysRevC.103.025803} {\bibfield  {journal} {\bibinfo
  {journal} {Phys. Rev. C}\ }\textbf {\bibinfo {volume} {103}},\ \bibinfo
  {pages} {025803} (\bibinfo {year} {2021})},\ \Eprint
  {http://arxiv.org/abs/2009.08885} {arXiv:2009.08885 [nucl-th]} \BibitemShut
  {NoStop}%
\bibitem [{\citenamefont {Tews}\ \emph {et~al.}(2018)\citenamefont {Tews},
  \citenamefont {Carlson}, \citenamefont {Gandolfi},\ and\ \citenamefont
  {Reddy}}]{Tews:2018kmu}%
  \BibitemOpen
  \bibfield  {author} {\bibinfo {author} {\bibfnamefont {I.}~\bibnamefont
  {Tews}}, \bibinfo {author} {\bibfnamefont {J.}~\bibnamefont {Carlson}},
  \bibinfo {author} {\bibfnamefont {S.}~\bibnamefont {Gandolfi}}, \ and\
  \bibinfo {author} {\bibfnamefont {S.}~\bibnamefont {Reddy}},\ }\href
  {\doibase 10.3847/1538-4357/aac267} {\bibfield  {journal} {\bibinfo
  {journal} {Astrophys. J.}\ }\textbf {\bibinfo {volume} {860}},\ \bibinfo
  {pages} {149} (\bibinfo {year} {2018})},\ \Eprint
  {http://arxiv.org/abs/1801.01923} {arXiv:1801.01923 [nucl-th]} \BibitemShut
  {NoStop}%
\bibitem [{\citenamefont {Baym}\ \emph {et~al.}(1971)\citenamefont {Baym},
  \citenamefont {Pethick},\ and\ \citenamefont {Sutherland}}]{Baym:1971pw}%
  \BibitemOpen
  \bibfield  {author} {\bibinfo {author} {\bibfnamefont {G.}~\bibnamefont
  {Baym}}, \bibinfo {author} {\bibfnamefont {C.}~\bibnamefont {Pethick}}, \
  and\ \bibinfo {author} {\bibfnamefont {P.}~\bibnamefont {Sutherland}},\
  }\href {\doibase 10.1086/151216} {\bibfield  {journal} {\bibinfo  {journal}
  {Astrophys. J.}\ }\textbf {\bibinfo {volume} {170}},\ \bibinfo {pages} {299}
  (\bibinfo {year} {1971})}\BibitemShut {NoStop}%
\bibitem [{\citenamefont {Negele}\ and\ \citenamefont
  {Vautherin}(1973)}]{Negele:1971vb}%
  \BibitemOpen
  \bibfield  {author} {\bibinfo {author} {\bibfnamefont {J.~W.}\ \bibnamefont
  {Negele}}\ and\ \bibinfo {author} {\bibfnamefont {D.}~\bibnamefont
  {Vautherin}},\ }\href {\doibase 10.1016/0375-9474(73)90349-7} {\bibfield
  {journal} {\bibinfo  {journal} {Nucl. Phys. A}\ }\textbf {\bibinfo {volume}
  {207}},\ \bibinfo {pages} {298} (\bibinfo {year} {1973})}\BibitemShut
  {NoStop}%
\bibitem [{\citenamefont {Skilling}(2006)}]{Skilling:2006gxv}%
  \BibitemOpen
  \bibfield  {author} {\bibinfo {author} {\bibfnamefont {J.}~\bibnamefont
  {Skilling}},\ }\href {\doibase 10.1214/06-BA127} {\bibfield  {journal}
  {\bibinfo  {journal} {Bayesian Analysis}\ }\textbf {\bibinfo {volume} {1}},\
  \bibinfo {pages} {833} (\bibinfo {year} {2006})}\BibitemShut {NoStop}%
\bibitem [{\citenamefont {Speagle}(2020)}]{Speagle:2019ivv}%
  \BibitemOpen
  \bibfield  {author} {\bibinfo {author} {\bibfnamefont {J.~S.}\ \bibnamefont
  {Speagle}},\ }\href {\doibase 10.1093/mnras/staa278} {\bibfield  {journal}
  {\bibinfo  {journal} {Mon. Not. Roy. Astron. Soc.}\ }\textbf {\bibinfo
  {volume} {493}},\ \bibinfo {pages} {3132} (\bibinfo {year} {2020})},\ \Eprint
  {http://arxiv.org/abs/1904.02180} {arXiv:1904.02180 [astro-ph.IM]}
  \BibitemShut {NoStop}%
\bibitem [{\citenamefont {Stergioulas}\ and\ \citenamefont
  {Friedman}(1995)}]{Stergioulas:1994ea}%
  \BibitemOpen
  \bibfield  {author} {\bibinfo {author} {\bibfnamefont {N.}~\bibnamefont
  {Stergioulas}}\ and\ \bibinfo {author} {\bibfnamefont {J.~L.}\ \bibnamefont
  {Friedman}},\ }\href {\doibase 10.1086/175605} {\bibfield  {journal}
  {\bibinfo  {journal} {Astrophys. J.}\ }\textbf {\bibinfo {volume} {444}},\
  \bibinfo {pages} {306} (\bibinfo {year} {1995})},\ \Eprint
  {http://arxiv.org/abs/astro-ph/9411032} {arXiv:astro-ph/9411032} \BibitemShut
  {NoStop}%
\bibitem [{\citenamefont {Nozawa}\ \emph {et~al.}(1998)\citenamefont {Nozawa},
  \citenamefont {Stergioulas}, \citenamefont {Gourgoulhon},\ and\ \citenamefont
  {Eriguchi}}]{RNS}%
  \BibitemOpen
  \bibfield  {author} {\bibinfo {author} {\bibfnamefont {T.}~\bibnamefont
  {Nozawa}}, \bibinfo {author} {\bibfnamefont {N.}~\bibnamefont {Stergioulas}},
  \bibinfo {author} {\bibfnamefont {E.}~\bibnamefont {Gourgoulhon}}, \ and\
  \bibinfo {author} {\bibfnamefont {Y.}~\bibnamefont {Eriguchi}},\ }\href
  {\doibase 10.1051/aas:1998304} {\bibfield  {journal} {\bibinfo  {journal}
  {Astron. Astrophys. Suppl. Ser.}\ }\textbf {\bibinfo {volume} {132}},\
  \bibinfo {pages} {431} (\bibinfo {year} {1998})},\ \Eprint
  {http://arxiv.org/abs/gr-qc/9804048} {arXiv:gr-qc/9804048} \BibitemShut
  {NoStop}%
\bibitem [{\citenamefont {{AlGendy}}\ and\ \citenamefont
  {{Morsink}}(2014)}]{AM+2014}%
  \BibitemOpen
  \bibfield  {author} {\bibinfo {author} {\bibfnamefont {M.}~\bibnamefont
  {{AlGendy}}}\ and\ \bibinfo {author} {\bibfnamefont {S.~M.}\ \bibnamefont
  {{Morsink}}},\ }\href {\doibase 10.1088/0004-637X/791/2/78} {\bibfield
  {journal} {\bibinfo  {journal} {ApJ}\ }\textbf {\bibinfo {volume} {791}},\
  \bibinfo {eid} {78} (\bibinfo {year} {2014})},\ \Eprint
  {http://arxiv.org/abs/1404.0609} {arXiv:1404.0609 [astro-ph.HE]} \BibitemShut
  {NoStop}%
\bibitem [{\citenamefont {{Suleimanov}}\ \emph {et~al.}(2020)\citenamefont
  {{Suleimanov}}, \citenamefont {{Poutanen}},\ and\ \citenamefont
  {{Werner}}}]{SPW+2020}%
  \BibitemOpen
  \bibfield  {author} {\bibinfo {author} {\bibfnamefont {V.~F.}\ \bibnamefont
  {{Suleimanov}}}, \bibinfo {author} {\bibfnamefont {J.}~\bibnamefont
  {{Poutanen}}}, \ and\ \bibinfo {author} {\bibfnamefont {K.}~\bibnamefont
  {{Werner}}},\ }\href {\doibase 10.1051/0004-6361/202037502} {\bibfield
  {journal} {\bibinfo  {journal} {A\&A}\ }\textbf {\bibinfo {volume} {639}},\
  \bibinfo {eid} {A33} (\bibinfo {year} {2020})},\ \Eprint
  {http://arxiv.org/abs/2005.09759} {arXiv:2005.09759 [astro-ph.HE]}
  \BibitemShut {NoStop}%
\bibitem [{\citenamefont {{Kalapotharakos}}\ \emph {et~al.}(2021)\citenamefont
  {{Kalapotharakos}}, \citenamefont {{Wadiasingh}}, \citenamefont {{Harding}},\
  and\ \citenamefont {{Kazanas}}}]{Kalapotharakos+2021}%
  \BibitemOpen
  \bibfield  {author} {\bibinfo {author} {\bibfnamefont {C.}~\bibnamefont
  {{Kalapotharakos}}}, \bibinfo {author} {\bibfnamefont {Z.}~\bibnamefont
  {{Wadiasingh}}}, \bibinfo {author} {\bibfnamefont {A.~K.}\ \bibnamefont
  {{Harding}}}, \ and\ \bibinfo {author} {\bibfnamefont {D.}~\bibnamefont
  {{Kazanas}}},\ }\href {\doibase 10.3847/1538-4357/abcec0} {\bibfield
  {journal} {\bibinfo  {journal} {ApJ}\ }\textbf {\bibinfo {volume} {907}},\
  \bibinfo {eid} {63} (\bibinfo {year} {2021})},\ \Eprint
  {http://arxiv.org/abs/2009.08567} {arXiv:2009.08567 [astro-ph.HE]}
  \BibitemShut {NoStop}%
\bibitem [{\citenamefont {{Gittins}}(2024)}]{grittin+2024}%
  \BibitemOpen
  \bibfield  {author} {\bibinfo {author} {\bibfnamefont {F.}~\bibnamefont
  {{Gittins}}},\ }\href {\doibase 10.1088/1361-6382/ad1c35} {\bibfield
  {journal} {\bibinfo  {journal} {Classical and Quantum Gravity}\ }\textbf
  {\bibinfo {volume} {41}},\ \bibinfo {eid} {043001} (\bibinfo {year}
  {2024})},\ \Eprint {http://arxiv.org/abs/2401.01670} {arXiv:2401.01670
  [gr-qc]} \BibitemShut {NoStop}%
\bibitem [{\citenamefont {{Ushomirsky}}\ \emph {et~al.}(2000)\citenamefont
  {{Ushomirsky}}, \citenamefont {{Cutler}},\ and\ \citenamefont
  {{Bildsten}}}]{ushomirsky+2000}%
  \BibitemOpen
  \bibfield  {author} {\bibinfo {author} {\bibfnamefont {G.}~\bibnamefont
  {{Ushomirsky}}}, \bibinfo {author} {\bibfnamefont {C.}~\bibnamefont
  {{Cutler}}}, \ and\ \bibinfo {author} {\bibfnamefont {L.}~\bibnamefont
  {{Bildsten}}},\ }\href {\doibase 10.1046/j.1365-8711.2000.03938.x} {\bibfield
   {journal} {\bibinfo  {journal} {MNRAS}\ }\textbf {\bibinfo {volume} {319}},\
  \bibinfo {pages} {902} (\bibinfo {year} {2000})},\ \Eprint
  {http://arxiv.org/abs/astro-ph/0001136} {arXiv:astro-ph/0001136 [astro-ph]}
  \BibitemShut {NoStop}%
\bibitem [{\citenamefont {{Haskell}}\ \emph {et~al.}(2006)\citenamefont
  {{Haskell}}, \citenamefont {{Jones}},\ and\ \citenamefont
  {{Andersson}}}]{haskell+2006}%
  \BibitemOpen
  \bibfield  {author} {\bibinfo {author} {\bibfnamefont {B.}~\bibnamefont
  {{Haskell}}}, \bibinfo {author} {\bibfnamefont {D.~I.}\ \bibnamefont
  {{Jones}}}, \ and\ \bibinfo {author} {\bibfnamefont {N.}~\bibnamefont
  {{Andersson}}},\ }\href {\doibase 10.1111/j.1365-2966.2006.10998.x}
  {\bibfield  {journal} {\bibinfo  {journal} {MNRAS}\ }\textbf {\bibinfo
  {volume} {373}},\ \bibinfo {pages} {1423} (\bibinfo {year} {2006})},\ \Eprint
  {http://arxiv.org/abs/astro-ph/0609438} {arXiv:astro-ph/0609438 [astro-ph]}
  \BibitemShut {NoStop}%
\bibitem [{\citenamefont {{Gittins}}\ \emph {et~al.}(2021)\citenamefont
  {{Gittins}}, \citenamefont {{Andersson}},\ and\ \citenamefont
  {{Jones}}}]{Grittin+2021}%
  \BibitemOpen
  \bibfield  {author} {\bibinfo {author} {\bibfnamefont {F.}~\bibnamefont
  {{Gittins}}}, \bibinfo {author} {\bibfnamefont {N.}~\bibnamefont
  {{Andersson}}}, \ and\ \bibinfo {author} {\bibfnamefont {D.~I.}\ \bibnamefont
  {{Jones}}},\ }\href {\doibase 10.1093/mnras/staa3635} {\bibfield  {journal}
  {\bibinfo  {journal} {MNRAS}\ }\textbf {\bibinfo {volume} {500}},\ \bibinfo
  {pages} {5570} (\bibinfo {year} {2021})},\ \Eprint
  {http://arxiv.org/abs/2009.12794} {arXiv:2009.12794 [astro-ph.HE]}
  \BibitemShut {NoStop}%
\bibitem [{\citenamefont {{Haskell}}\ \emph {et~al.}(2008)\citenamefont
  {{Haskell}}, \citenamefont {{Samuelsson}}, \citenamefont {{Glampedakis}},\
  and\ \citenamefont {{Andersson}}}]{HSGA+2008}%
  \BibitemOpen
  \bibfield  {author} {\bibinfo {author} {\bibfnamefont {B.}~\bibnamefont
  {{Haskell}}}, \bibinfo {author} {\bibfnamefont {L.}~\bibnamefont
  {{Samuelsson}}}, \bibinfo {author} {\bibfnamefont {K.}~\bibnamefont
  {{Glampedakis}}}, \ and\ \bibinfo {author} {\bibfnamefont {N.}~\bibnamefont
  {{Andersson}}},\ }\href {\doibase 10.1111/j.1365-2966.2008.12861.x}
  {\bibfield  {journal} {\bibinfo  {journal} {MNRAS}\ }\textbf {\bibinfo
  {volume} {385}},\ \bibinfo {pages} {531} (\bibinfo {year} {2008})},\ \Eprint
  {http://arxiv.org/abs/0705.1780} {arXiv:0705.1780 [astro-ph]} \BibitemShut
  {NoStop}%
\bibitem [{\citenamefont {{Kalita}}\ and\ \citenamefont
  {{Mukhopadhyay}}(2020)}]{KM+2020}%
  \BibitemOpen
  \bibfield  {author} {\bibinfo {author} {\bibfnamefont {S.}~\bibnamefont
  {{Kalita}}}\ and\ \bibinfo {author} {\bibfnamefont {B.}~\bibnamefont
  {{Mukhopadhyay}}},\ }\href {\doibase 10.1093/mnras/stz3383} {\bibfield
  {journal} {\bibinfo  {journal} {MNRAS}\ }\textbf {\bibinfo {volume} {491}},\
  \bibinfo {pages} {4396} (\bibinfo {year} {2020})}\BibitemShut {NoStop}%
\bibitem [{\citenamefont {{Soldateschi}}\ \emph {et~al.}(2021)\citenamefont
  {{Soldateschi}}, \citenamefont {{Bucciantini}},\ and\ \citenamefont {{Del
  Zanna}}}]{SBD+2021}%
  \BibitemOpen
  \bibfield  {author} {\bibinfo {author} {\bibfnamefont {J.}~\bibnamefont
  {{Soldateschi}}}, \bibinfo {author} {\bibfnamefont {N.}~\bibnamefont
  {{Bucciantini}}}, \ and\ \bibinfo {author} {\bibfnamefont {L.}~\bibnamefont
  {{Del Zanna}}},\ }\href {\doibase 10.1051/0004-6361/202141448} {\bibfield
  {journal} {\bibinfo  {journal} {A\&A}\ }\textbf {\bibinfo {volume} {654}},\
  \bibinfo {eid} {A162} (\bibinfo {year} {2021})},\ \Eprint
  {http://arxiv.org/abs/2106.00603} {arXiv:2106.00603 [astro-ph.HE]}
  \BibitemShut {NoStop}%
\bibitem [{\citenamefont {{Bildsten}}(1998)}]{Bilsden+1998}%
  \BibitemOpen
  \bibfield  {author} {\bibinfo {author} {\bibfnamefont {L.}~\bibnamefont
  {{Bildsten}}},\ }\href {\doibase 10.1086/311440} {\bibfield  {journal}
  {\bibinfo  {journal} {ApJL}\ }\textbf {\bibinfo {volume} {501}},\ \bibinfo
  {pages} {L89} (\bibinfo {year} {1998})},\ \Eprint
  {http://arxiv.org/abs/astro-ph/9804325} {arXiv:astro-ph/9804325 [astro-ph]}
  \BibitemShut {NoStop}%
\bibitem [{\citenamefont {{Vigelius}}\ and\ \citenamefont
  {{Melatos}}(2009)}]{Vig_mel+2009}%
  \BibitemOpen
  \bibfield  {author} {\bibinfo {author} {\bibfnamefont {M.}~\bibnamefont
  {{Vigelius}}}\ and\ \bibinfo {author} {\bibfnamefont {A.}~\bibnamefont
  {{Melatos}}},\ }\href {\doibase 10.1111/j.1365-2966.2009.14698.x} {\bibfield
  {journal} {\bibinfo  {journal} {MNRAS}\ }\textbf {\bibinfo {volume} {395}},\
  \bibinfo {pages} {1985} (\bibinfo {year} {2009})},\ \Eprint
  {http://arxiv.org/abs/0902.4484} {arXiv:0902.4484 [astro-ph.HE]} \BibitemShut
  {NoStop}%
\bibitem [{\citenamefont {Abbott}\ \emph
  {et~al.}(2019{\natexlab{b}})\citenamefont {Abbott} \emph
  {et~al.}}]{LIGO+2019_ellipticity}%
  \BibitemOpen
  \bibfield  {author} {\bibinfo {author} {\bibfnamefont {B.~P.}\ \bibnamefont
  {Abbott}} \emph {et~al.} (\bibinfo {collaboration} {LIGO Scientific,
  Virgo}),\ }\href {\doibase 10.1103/PhysRevD.100.024004} {\bibfield  {journal}
  {\bibinfo  {journal} {Phys. Rev. D}\ }\textbf {\bibinfo {volume} {100}},\
  \bibinfo {pages} {024004} (\bibinfo {year} {2019}{\natexlab{b}})},\ \Eprint
  {http://arxiv.org/abs/1903.01901} {arXiv:1903.01901 [astro-ph.HE]}
  \BibitemShut {NoStop}%
\bibitem [{\citenamefont {Owen}\ \emph {et~al.}(1998)\citenamefont {Owen},
  \citenamefont {Lindblom}, \citenamefont {Cutler}, \citenamefont {Schutz},
  \citenamefont {Vecchio},\ and\ \citenamefont {Andersson}}]{Owen:1998xg}%
  \BibitemOpen
  \bibfield  {author} {\bibinfo {author} {\bibfnamefont {B.~J.}\ \bibnamefont
  {Owen}}, \bibinfo {author} {\bibfnamefont {L.}~\bibnamefont {Lindblom}},
  \bibinfo {author} {\bibfnamefont {C.}~\bibnamefont {Cutler}}, \bibinfo
  {author} {\bibfnamefont {B.~F.}\ \bibnamefont {Schutz}}, \bibinfo {author}
  {\bibfnamefont {A.}~\bibnamefont {Vecchio}}, \ and\ \bibinfo {author}
  {\bibfnamefont {N.}~\bibnamefont {Andersson}},\ }\href {\doibase
  10.1103/PhysRevD.58.084020} {\bibfield  {journal} {\bibinfo  {journal} {Phys.
  Rev. D}\ }\textbf {\bibinfo {volume} {58}},\ \bibinfo {pages} {084020}
  (\bibinfo {year} {1998})},\ \Eprint {http://arxiv.org/abs/gr-qc/9804044}
  {arXiv:gr-qc/9804044} \BibitemShut {NoStop}%
\bibitem [{\citenamefont {Chandrasekhar}(1970)}]{Chandrasekhar:1970pjp}%
  \BibitemOpen
  \bibfield  {author} {\bibinfo {author} {\bibfnamefont {S.}~\bibnamefont
  {Chandrasekhar}},\ }\href {\doibase 10.1103/PhysRevLett.24.611} {\bibfield
  {journal} {\bibinfo  {journal} {Phys. Rev. Lett.}\ }\textbf {\bibinfo
  {volume} {24}},\ \bibinfo {pages} {611} (\bibinfo {year} {1970})}\BibitemShut
  {NoStop}%
\bibitem [{\citenamefont {Friedman}\ and\ \citenamefont
  {Schutz}(1978)}]{Friedman:1978hf}%
  \BibitemOpen
  \bibfield  {author} {\bibinfo {author} {\bibfnamefont {J.~L.}\ \bibnamefont
  {Friedman}}\ and\ \bibinfo {author} {\bibfnamefont {B.~F.}\ \bibnamefont
  {Schutz}},\ }\href {\doibase 10.1086/156143} {\bibfield  {journal} {\bibinfo
  {journal} {Astrophys. J.}\ }\textbf {\bibinfo {volume} {222}},\ \bibinfo
  {pages} {281} (\bibinfo {year} {1978})}\BibitemShut {NoStop}%
\bibitem [{\citenamefont {Andersson}(1998)}]{Andersson:1997xt}%
  \BibitemOpen
  \bibfield  {author} {\bibinfo {author} {\bibfnamefont {N.}~\bibnamefont
  {Andersson}},\ }\href {\doibase 10.1086/305919} {\bibfield  {journal}
  {\bibinfo  {journal} {Astrophys. J.}\ }\textbf {\bibinfo {volume} {502}},\
  \bibinfo {pages} {708} (\bibinfo {year} {1998})},\ \Eprint
  {http://arxiv.org/abs/gr-qc/9706075} {arXiv:gr-qc/9706075} \BibitemShut
  {NoStop}%
\bibitem [{\citenamefont {Friedman}\ and\ \citenamefont
  {Morsink}(1998)}]{Friedman:1997uh}%
  \BibitemOpen
  \bibfield  {author} {\bibinfo {author} {\bibfnamefont {J.~L.}\ \bibnamefont
  {Friedman}}\ and\ \bibinfo {author} {\bibfnamefont {S.~M.}\ \bibnamefont
  {Morsink}},\ }\href {\doibase 10.1086/305920} {\bibfield  {journal} {\bibinfo
   {journal} {Astrophys. J.}\ }\textbf {\bibinfo {volume} {502}},\ \bibinfo
  {pages} {714} (\bibinfo {year} {1998})},\ \Eprint
  {http://arxiv.org/abs/gr-qc/9706073} {arXiv:gr-qc/9706073} \BibitemShut
  {NoStop}%
\bibitem [{\citenamefont {Bildsten}(1998)}]{Bildsten:1998ey}%
  \BibitemOpen
  \bibfield  {author} {\bibinfo {author} {\bibfnamefont {L.}~\bibnamefont
  {Bildsten}},\ }\href {\doibase 10.1086/311440} {\bibfield  {journal}
  {\bibinfo  {journal} {Astrophys. J. Lett.}\ }\textbf {\bibinfo {volume}
  {501}},\ \bibinfo {pages} {L89} (\bibinfo {year} {1998})},\ \Eprint
  {http://arxiv.org/abs/astro-ph/9804325} {arXiv:astro-ph/9804325} \BibitemShut
  {NoStop}%
\bibitem [{\citenamefont {Thorne}(1980)}]{Thorne:1980ru}%
  \BibitemOpen
  \bibfield  {author} {\bibinfo {author} {\bibfnamefont {K.~S.}\ \bibnamefont
  {Thorne}},\ }\href {\doibase 10.1103/RevModPhys.52.299} {\bibfield  {journal}
  {\bibinfo  {journal} {Rev. Mod. Phys.}\ }\textbf {\bibinfo {volume} {52}},\
  \bibinfo {pages} {299} (\bibinfo {year} {1980})}\BibitemShut {NoStop}%
\bibitem [{\citenamefont {Glendenning}\ \emph {et~al.}(1997)\citenamefont
  {Glendenning}, \citenamefont {Pei},\ and\ \citenamefont {Weber}}]{GPW:1997}%
  \BibitemOpen
  \bibfield  {author} {\bibinfo {author} {\bibfnamefont {N.~K.}\ \bibnamefont
  {Glendenning}}, \bibinfo {author} {\bibfnamefont {S.}~\bibnamefont {Pei}}, \
  and\ \bibinfo {author} {\bibfnamefont {F.}~\bibnamefont {Weber}},\ }\href
  {\doibase 10.1103/PhysRevLett.79.1603} {\bibfield  {journal} {\bibinfo
  {journal} {Phys. Rev. Lett.}\ }\textbf {\bibinfo {volume} {79}},\ \bibinfo
  {pages} {1603} (\bibinfo {year} {1997})}\BibitemShut {NoStop}%
\bibitem [{\citenamefont {Lalazissis}\ \emph {et~al.}(2005)\citenamefont
  {Lalazissis}, \citenamefont {Niksic}, \citenamefont {Vretenar},\ and\
  \citenamefont {Ring}}]{Lalazissis:2005de}%
  \BibitemOpen
  \bibfield  {author} {\bibinfo {author} {\bibfnamefont {G.~A.}\ \bibnamefont
  {Lalazissis}}, \bibinfo {author} {\bibfnamefont {T.}~\bibnamefont {Niksic}},
  \bibinfo {author} {\bibfnamefont {D.}~\bibnamefont {Vretenar}}, \ and\
  \bibinfo {author} {\bibfnamefont {P.}~\bibnamefont {Ring}},\ }\href {\doibase
  10.1103/PhysRevC.71.024312} {\bibfield  {journal} {\bibinfo  {journal} {Phys.
  Rev. C}\ }\textbf {\bibinfo {volume} {71}},\ \bibinfo {pages} {024312}
  (\bibinfo {year} {2005})}\BibitemShut {NoStop}%
\bibitem [{\citenamefont {Horowitz}\ and\ \citenamefont
  {Piekarewicz}(2002)}]{Horowitz:2002mb}%
  \BibitemOpen
  \bibfield  {author} {\bibinfo {author} {\bibfnamefont {C.~J.}\ \bibnamefont
  {Horowitz}}\ and\ \bibinfo {author} {\bibfnamefont {J.}~\bibnamefont
  {Piekarewicz}},\ }\href {\doibase 10.1103/PhysRevC.66.055803} {\bibfield
  {journal} {\bibinfo  {journal} {Phys. Rev. C}\ }\textbf {\bibinfo {volume}
  {66}},\ \bibinfo {pages} {055803} (\bibinfo {year} {2002})},\ \Eprint
  {http://arxiv.org/abs/nucl-th/0207067} {arXiv:nucl-th/0207067} \BibitemShut
  {NoStop}%
\bibitem [{\citenamefont {Kraav}\ \emph {et~al.}(2024)\citenamefont {Kraav},
  \citenamefont {Gusakov},\ and\ \citenamefont {Kantor}}]{KGK+2024}%
  \BibitemOpen
  \bibfield  {author} {\bibinfo {author} {\bibfnamefont {K.~Y.}\ \bibnamefont
  {Kraav}}, \bibinfo {author} {\bibfnamefont {M.~E.}\ \bibnamefont {Gusakov}},
  \ and\ \bibinfo {author} {\bibfnamefont {E.~M.}\ \bibnamefont {Kantor}},\
  }\href {\doibase 10.1103/PhysRevD.109.043012} {\bibfield  {journal} {\bibinfo
   {journal} {Phys. Rev. D}\ }\textbf {\bibinfo {volume} {109}},\ \bibinfo
  {pages} {043012} (\bibinfo {year} {2024})}\BibitemShut {NoStop}%
\bibitem [{\citenamefont {Johnson-McDaniel}\ and\ \citenamefont
  {Owen}(2013)}]{JMcO+2013}%
  \BibitemOpen
  \bibfield  {author} {\bibinfo {author} {\bibfnamefont {N.~K.}\ \bibnamefont
  {Johnson-McDaniel}}\ and\ \bibinfo {author} {\bibfnamefont {B.~J.}\
  \bibnamefont {Owen}},\ }\href {\doibase 10.1103/PhysRevD.88.044004}
  {\bibfield  {journal} {\bibinfo  {journal} {Phys. Rev. D}\ }\textbf {\bibinfo
  {volume} {88}},\ \bibinfo {pages} {044004} (\bibinfo {year} {2013})},\
  \Eprint {http://arxiv.org/abs/1208.5227} {arXiv:1208.5227 [astro-ph.SR]}
  \BibitemShut {NoStop}%
\bibitem [{\citenamefont {{Dai}}\ and\ \citenamefont {{Lu}}(1998)}]{DL+1998}%
  \BibitemOpen
  \bibfield  {author} {\bibinfo {author} {\bibfnamefont {Z.~G.}\ \bibnamefont
  {{Dai}}}\ and\ \bibinfo {author} {\bibfnamefont {T.}~\bibnamefont {{Lu}}},\
  }\href {\doibase 10.48550/arXiv.astro-ph/9810402} {\bibfield  {journal}
  {\bibinfo  {journal} {A\&A}\ }\textbf {\bibinfo {volume} {333}},\ \bibinfo
  {pages} {L87} (\bibinfo {year} {1998})},\ \Eprint
  {http://arxiv.org/abs/astro-ph/9810402} {arXiv:astro-ph/9810402 [astro-ph]}
  \BibitemShut {NoStop}%
\bibitem [{\citenamefont {{Usov}}(1992)}]{Usov+1992}%
  \BibitemOpen
  \bibfield  {author} {\bibinfo {author} {\bibfnamefont {V.~V.}\ \bibnamefont
  {{Usov}}},\ }\href {\doibase 10.1038/357472a0} {\bibfield  {journal}
  {\bibinfo  {journal} {Nature}\ }\textbf {\bibinfo {volume} {357}},\ \bibinfo
  {pages} {472} (\bibinfo {year} {1992})}\BibitemShut {NoStop}%
\bibitem [{\citenamefont {{Rowlinson}}\ \emph {et~al.}(2013)\citenamefont
  {{Rowlinson}}, \citenamefont {{O'Brien}}, \citenamefont {{Metzger}},
  \citenamefont {{Tanvir}},\ and\ \citenamefont {{Levan}}}]{ROM+2013}%
  \BibitemOpen
  \bibfield  {author} {\bibinfo {author} {\bibfnamefont {A.}~\bibnamefont
  {{Rowlinson}}}, \bibinfo {author} {\bibfnamefont {P.~T.}\ \bibnamefont
  {{O'Brien}}}, \bibinfo {author} {\bibfnamefont {B.~D.}\ \bibnamefont
  {{Metzger}}}, \bibinfo {author} {\bibfnamefont {N.~R.}\ \bibnamefont
  {{Tanvir}}}, \ and\ \bibinfo {author} {\bibfnamefont {A.~J.}\ \bibnamefont
  {{Levan}}},\ }\href {\doibase 10.1093/mnras/sts683} {\bibfield  {journal}
  {\bibinfo  {journal} {MNRAS}\ }\textbf {\bibinfo {volume} {430}},\ \bibinfo
  {pages} {1061} (\bibinfo {year} {2013})},\ \Eprint
  {http://arxiv.org/abs/1301.0629} {arXiv:1301.0629 [astro-ph.HE]} \BibitemShut
  {NoStop}%
\bibitem [{\citenamefont {Lü}\ and\ \citenamefont {Zhang}(2014)}]{Lu_2014}%
  \BibitemOpen
  \bibfield  {author} {\bibinfo {author} {\bibfnamefont {H.-J.}\ \bibnamefont
  {Lü}}\ and\ \bibinfo {author} {\bibfnamefont {B.}~\bibnamefont {Zhang}},\
  }\href {\doibase 10.1088/0004-637X/785/1/74} {\bibfield  {journal} {\bibinfo
  {journal} {ApJ}\ }\textbf {\bibinfo {volume} {785}},\ \bibinfo {pages} {74}
  (\bibinfo {year} {2014})}\BibitemShut {NoStop}%
\bibitem [{\citenamefont {{L{\"u}}}\ \emph {et~al.}(2019)\citenamefont
  {{L{\"u}}}, \citenamefont {{Lan}},\ and\ \citenamefont {{Liang}}}]{LLL+2019}%
  \BibitemOpen
  \bibfield  {author} {\bibinfo {author} {\bibfnamefont {H.-J.}\ \bibnamefont
  {{L{\"u}}}}, \bibinfo {author} {\bibfnamefont {L.}~\bibnamefont {{Lan}}}, \
  and\ \bibinfo {author} {\bibfnamefont {E.-W.}\ \bibnamefont {{Liang}}},\
  }\href {\doibase 10.3847/1538-4357/aaf71d} {\bibfield  {journal} {\bibinfo
  {journal} {ApJ}\ }\textbf {\bibinfo {volume} {871}},\ \bibinfo {eid} {54}
  (\bibinfo {year} {2019})},\ \Eprint {http://arxiv.org/abs/1812.03465}
  {arXiv:1812.03465 [astro-ph.HE]} \BibitemShut {NoStop}%
\bibitem [{\citenamefont {Xie}\ \emph {et~al.}(2022)\citenamefont {Xie},
  \citenamefont {Wei}, \citenamefont {Wang},\ and\ \citenamefont
  {Jin}}]{XDY+2022}%
  \BibitemOpen
  \bibfield  {author} {\bibinfo {author} {\bibfnamefont {L.}~\bibnamefont
  {Xie}}, \bibinfo {author} {\bibfnamefont {D.-M.}\ \bibnamefont {Wei}},
  \bibinfo {author} {\bibfnamefont {Y.}~\bibnamefont {Wang}}, \ and\ \bibinfo
  {author} {\bibfnamefont {Z.-P.}\ \bibnamefont {Jin}},\ }\href {\doibase
  10.3847/1538-4357/ac7c13} {\bibfield  {journal} {\bibinfo  {journal} {The
  Astrophysical Journal}\ }\textbf {\bibinfo {volume} {934}},\ \bibinfo {pages}
  {125} (\bibinfo {year} {2022})}\BibitemShut {NoStop}%
\bibitem [{\citenamefont {{Nousek}}\ \emph {et~al.}(2006)\citenamefont
  {{Nousek}}, \citenamefont {{Kouveliotou}}, \citenamefont {{Grupe}},
  \citenamefont {{Page}}, \citenamefont {{Granot}}, \citenamefont
  {{Ramirez-Ruiz}}, \citenamefont {{Patel}}, \citenamefont {{Burrows}},
  \citenamefont {{Mangano}}, \citenamefont {{Barthelmy}}, \citenamefont
  {{Beardmore}}, \citenamefont {{Campana}}, \citenamefont {{Capalbi}},
  \citenamefont {{Chincarini}}, \citenamefont {{Cusumano}}, \citenamefont
  {{Falcone}}, \citenamefont {{Gehrels}}, \citenamefont {{Giommi}},
  \citenamefont {{Goad}}, \citenamefont {{Godet}}, \citenamefont {{Hurkett}},
  \citenamefont {{Kennea}}, \citenamefont {{Moretti}}, \citenamefont
  {{O'Brien}}, \citenamefont {{Osborne}}, \citenamefont {{Romano}},
  \citenamefont {{Tagliaferri}},\ and\ \citenamefont {{Wells}}}]{Nousek+2006}%
  \BibitemOpen
  \bibfield  {author} {\bibinfo {author} {\bibfnamefont {J.~A.}\ \bibnamefont
  {{Nousek}}}, \bibinfo {author} {\bibfnamefont {C.}~\bibnamefont
  {{Kouveliotou}}}, \bibinfo {author} {\bibfnamefont {D.}~\bibnamefont
  {{Grupe}}}, \bibinfo {author} {\bibfnamefont {K.~L.}\ \bibnamefont {{Page}}},
  \bibinfo {author} {\bibfnamefont {J.}~\bibnamefont {{Granot}}}, \bibinfo
  {author} {\bibfnamefont {E.}~\bibnamefont {{Ramirez-Ruiz}}}, \bibinfo
  {author} {\bibfnamefont {S.~K.}\ \bibnamefont {{Patel}}}, \bibinfo {author}
  {\bibfnamefont {D.~N.}\ \bibnamefont {{Burrows}}}, \bibinfo {author}
  {\bibfnamefont {V.}~\bibnamefont {{Mangano}}}, \bibinfo {author}
  {\bibfnamefont {S.}~\bibnamefont {{Barthelmy}}}, \bibinfo {author}
  {\bibfnamefont {A.~P.}\ \bibnamefont {{Beardmore}}}, \bibinfo {author}
  {\bibfnamefont {S.}~\bibnamefont {{Campana}}}, \bibinfo {author}
  {\bibfnamefont {M.}~\bibnamefont {{Capalbi}}}, \bibinfo {author}
  {\bibfnamefont {G.}~\bibnamefont {{Chincarini}}}, \bibinfo {author}
  {\bibfnamefont {G.}~\bibnamefont {{Cusumano}}}, \bibinfo {author}
  {\bibfnamefont {A.~D.}\ \bibnamefont {{Falcone}}}, \bibinfo {author}
  {\bibfnamefont {N.}~\bibnamefont {{Gehrels}}}, \bibinfo {author}
  {\bibfnamefont {P.}~\bibnamefont {{Giommi}}}, \bibinfo {author}
  {\bibfnamefont {M.~R.}\ \bibnamefont {{Goad}}}, \bibinfo {author}
  {\bibfnamefont {O.}~\bibnamefont {{Godet}}}, \bibinfo {author} {\bibfnamefont
  {C.~P.}\ \bibnamefont {{Hurkett}}}, \bibinfo {author} {\bibfnamefont {J.~A.}\
  \bibnamefont {{Kennea}}}, \bibinfo {author} {\bibfnamefont {A.}~\bibnamefont
  {{Moretti}}}, \bibinfo {author} {\bibfnamefont {P.~T.}\ \bibnamefont
  {{O'Brien}}}, \bibinfo {author} {\bibfnamefont {J.~P.}\ \bibnamefont
  {{Osborne}}}, \bibinfo {author} {\bibfnamefont {P.}~\bibnamefont {{Romano}}},
  \bibinfo {author} {\bibfnamefont {G.}~\bibnamefont {{Tagliaferri}}}, \ and\
  \bibinfo {author} {\bibfnamefont {A.~A.}\ \bibnamefont {{Wells}}},\ }\href
  {\doibase 10.1086/500724} {\bibfield  {journal} {\bibinfo  {journal} {ApJ}\
  }\textbf {\bibinfo {volume} {642}},\ \bibinfo {pages} {389} (\bibinfo {year}
  {2006})},\ \Eprint {http://arxiv.org/abs/astro-ph/0508332}
  {arXiv:astro-ph/0508332 [astro-ph]} \BibitemShut {NoStop}%
\bibitem [{\citenamefont {{O'Brien}}\ \emph {et~al.}(2006)\citenamefont
  {{O'Brien}}, \citenamefont {{Willingale}}, \citenamefont {{Osborne}},
  \citenamefont {{Goad}}, \citenamefont {{Page}}, \citenamefont {{Vaughan}},
  \citenamefont {{Rol}}, \citenamefont {{Beardmore}}, \citenamefont {{Godet}},
  \citenamefont {{Hurkett}}, \citenamefont {{Wells}}, \citenamefont {{Zhang}},
  \citenamefont {{Kobayashi}}, \citenamefont {{Burrows}}, \citenamefont
  {{Nousek}}, \citenamefont {{Kennea}}, \citenamefont {{Falcone}},
  \citenamefont {{Grupe}}, \citenamefont {{Gehrels}}, \citenamefont
  {{Barthelmy}}, \citenamefont {{Cannizzo}}, \citenamefont {{Cummings}},
  \citenamefont {{Hill}}, \citenamefont {{Krimm}}, \citenamefont
  {{Chincarini}}, \citenamefont {{Tagliaferri}}, \citenamefont {{Campana}},
  \citenamefont {{Moretti}}, \citenamefont {{Giommi}}, \citenamefont {{Perri}},
  \citenamefont {{Mangano}},\ and\ \citenamefont {{LaParola}}}]{OWO+2006}%
  \BibitemOpen
  \bibfield  {author} {\bibinfo {author} {\bibfnamefont {P.~T.}\ \bibnamefont
  {{O'Brien}}}, \bibinfo {author} {\bibfnamefont {R.}~\bibnamefont
  {{Willingale}}}, \bibinfo {author} {\bibfnamefont {J.}~\bibnamefont
  {{Osborne}}}, \bibinfo {author} {\bibfnamefont {M.~R.}\ \bibnamefont
  {{Goad}}}, \bibinfo {author} {\bibfnamefont {K.~L.}\ \bibnamefont {{Page}}},
  \bibinfo {author} {\bibfnamefont {S.}~\bibnamefont {{Vaughan}}}, \bibinfo
  {author} {\bibfnamefont {E.}~\bibnamefont {{Rol}}}, \bibinfo {author}
  {\bibfnamefont {A.}~\bibnamefont {{Beardmore}}}, \bibinfo {author}
  {\bibfnamefont {O.}~\bibnamefont {{Godet}}}, \bibinfo {author} {\bibfnamefont
  {C.~P.}\ \bibnamefont {{Hurkett}}}, \bibinfo {author} {\bibfnamefont
  {A.}~\bibnamefont {{Wells}}}, \bibinfo {author} {\bibfnamefont
  {B.}~\bibnamefont {{Zhang}}}, \bibinfo {author} {\bibfnamefont
  {S.}~\bibnamefont {{Kobayashi}}}, \bibinfo {author} {\bibfnamefont {D.~N.}\
  \bibnamefont {{Burrows}}}, \bibinfo {author} {\bibfnamefont {J.~A.}\
  \bibnamefont {{Nousek}}}, \bibinfo {author} {\bibfnamefont {J.~A.}\
  \bibnamefont {{Kennea}}}, \bibinfo {author} {\bibfnamefont {A.}~\bibnamefont
  {{Falcone}}}, \bibinfo {author} {\bibfnamefont {D.}~\bibnamefont {{Grupe}}},
  \bibinfo {author} {\bibfnamefont {N.}~\bibnamefont {{Gehrels}}}, \bibinfo
  {author} {\bibfnamefont {S.}~\bibnamefont {{Barthelmy}}}, \bibinfo {author}
  {\bibfnamefont {J.}~\bibnamefont {{Cannizzo}}}, \bibinfo {author}
  {\bibfnamefont {J.}~\bibnamefont {{Cummings}}}, \bibinfo {author}
  {\bibfnamefont {J.~E.}\ \bibnamefont {{Hill}}}, \bibinfo {author}
  {\bibfnamefont {H.}~\bibnamefont {{Krimm}}}, \bibinfo {author} {\bibfnamefont
  {G.}~\bibnamefont {{Chincarini}}}, \bibinfo {author} {\bibfnamefont
  {G.}~\bibnamefont {{Tagliaferri}}}, \bibinfo {author} {\bibfnamefont
  {S.}~\bibnamefont {{Campana}}}, \bibinfo {author} {\bibfnamefont
  {A.}~\bibnamefont {{Moretti}}}, \bibinfo {author} {\bibfnamefont
  {P.}~\bibnamefont {{Giommi}}}, \bibinfo {author} {\bibfnamefont
  {M.}~\bibnamefont {{Perri}}}, \bibinfo {author} {\bibfnamefont
  {V.}~\bibnamefont {{Mangano}}}, \ and\ \bibinfo {author} {\bibfnamefont
  {V.}~\bibnamefont {{LaParola}}},\ }\href {\doibase 10.1086/505457} {\bibfield
   {journal} {\bibinfo  {journal} {ApJ}\ }\textbf {\bibinfo {volume} {647}},\
  \bibinfo {pages} {1213} (\bibinfo {year} {2006})},\ \Eprint
  {http://arxiv.org/abs/astro-ph/0601125} {arXiv:astro-ph/0601125 [astro-ph]}
  \BibitemShut {NoStop}%
\bibitem [{\citenamefont {{Zhang}}\ \emph {et~al.}(2006)\citenamefont
  {{Zhang}}, \citenamefont {{Fan}}, \citenamefont {{Dyks}}, \citenamefont
  {{Kobayashi}}, \citenamefont {{M{\'e}sz{\'a}ros}}, \citenamefont {{Burrows}},
  \citenamefont {{Nousek}},\ and\ \citenamefont {{Gehrels}}}]{Zhang+2006}%
  \BibitemOpen
  \bibfield  {author} {\bibinfo {author} {\bibfnamefont {B.}~\bibnamefont
  {{Zhang}}}, \bibinfo {author} {\bibfnamefont {Y.~Z.}\ \bibnamefont {{Fan}}},
  \bibinfo {author} {\bibfnamefont {J.}~\bibnamefont {{Dyks}}}, \bibinfo
  {author} {\bibfnamefont {S.}~\bibnamefont {{Kobayashi}}}, \bibinfo {author}
  {\bibfnamefont {P.}~\bibnamefont {{M{\'e}sz{\'a}ros}}}, \bibinfo {author}
  {\bibfnamefont {D.~N.}\ \bibnamefont {{Burrows}}}, \bibinfo {author}
  {\bibfnamefont {J.~A.}\ \bibnamefont {{Nousek}}}, \ and\ \bibinfo {author}
  {\bibfnamefont {N.}~\bibnamefont {{Gehrels}}},\ }\href {\doibase
  10.1086/500723} {\bibfield  {journal} {\bibinfo  {journal} {ApJ}\ }\textbf
  {\bibinfo {volume} {642}},\ \bibinfo {pages} {354} (\bibinfo {year}
  {2006})},\ \Eprint {http://arxiv.org/abs/astro-ph/0508321}
  {arXiv:astro-ph/0508321 [astro-ph]} \BibitemShut {NoStop}%
\bibitem [{\citenamefont {{Zhang}}\ and\ \citenamefont
  {{M{\'e}sz{\'a}ros}}(2001)}]{ZM+2001}%
  \BibitemOpen
  \bibfield  {author} {\bibinfo {author} {\bibfnamefont {B.}~\bibnamefont
  {{Zhang}}}\ and\ \bibinfo {author} {\bibfnamefont {P.}~\bibnamefont
  {{M{\'e}sz{\'a}ros}}},\ }\href {\doibase 10.1086/320255} {\bibfield
  {journal} {\bibinfo  {journal} {ApJL}\ }\textbf {\bibinfo {volume} {552}},\
  \bibinfo {pages} {L35} (\bibinfo {year} {2001})},\ \Eprint
  {http://arxiv.org/abs/astro-ph/0011133} {arXiv:astro-ph/0011133 [astro-ph]}
  \BibitemShut {NoStop}%
\bibitem [{\citenamefont {{Metzger}}\ \emph {et~al.}(2011)\citenamefont
  {{Metzger}}, \citenamefont {{Giannios}}, \citenamefont {{Thompson}},
  \citenamefont {{Bucciantini}},\ and\ \citenamefont
  {{Quataert}}}]{metzger+2011}%
  \BibitemOpen
  \bibfield  {author} {\bibinfo {author} {\bibfnamefont {B.~D.}\ \bibnamefont
  {{Metzger}}}, \bibinfo {author} {\bibfnamefont {D.}~\bibnamefont
  {{Giannios}}}, \bibinfo {author} {\bibfnamefont {T.~A.}\ \bibnamefont
  {{Thompson}}}, \bibinfo {author} {\bibfnamefont {N.}~\bibnamefont
  {{Bucciantini}}}, \ and\ \bibinfo {author} {\bibfnamefont {E.}~\bibnamefont
  {{Quataert}}},\ }\href {\doibase 10.1111/j.1365-2966.2011.18280.x} {\bibfield
   {journal} {\bibinfo  {journal} {MNRAS}\ }\textbf {\bibinfo {volume} {413}},\
  \bibinfo {pages} {2031} (\bibinfo {year} {2011})},\ \Eprint
  {http://arxiv.org/abs/1012.0001} {arXiv:1012.0001 [astro-ph.HE]} \BibitemShut
  {NoStop}%
\bibitem [{\citenamefont {{Lasky}}\ \emph {et~al.}(2017)\citenamefont
  {{Lasky}}, \citenamefont {{Leris}}, \citenamefont {{Rowlinson}},\ and\
  \citenamefont {{Glampedakis}}}]{Lasky+2017}%
  \BibitemOpen
  \bibfield  {author} {\bibinfo {author} {\bibfnamefont {P.~D.}\ \bibnamefont
  {{Lasky}}}, \bibinfo {author} {\bibfnamefont {C.}~\bibnamefont {{Leris}}},
  \bibinfo {author} {\bibfnamefont {A.}~\bibnamefont {{Rowlinson}}}, \ and\
  \bibinfo {author} {\bibfnamefont {K.}~\bibnamefont {{Glampedakis}}},\ }\href
  {\doibase 10.3847/2041-8213/aa79a7} {\bibfield  {journal} {\bibinfo
  {journal} {ApJL}\ }\textbf {\bibinfo {volume} {843}},\ \bibinfo {eid} {L1}
  (\bibinfo {year} {2017})},\ \Eprint {http://arxiv.org/abs/1705.10005}
  {arXiv:1705.10005 [astro-ph.HE]} \BibitemShut {NoStop}%
\bibitem [{\citenamefont {Contopoulos}\ and\ \citenamefont
  {Spitkovsky}(2006)}]{Contopoulos:2005rs}%
  \BibitemOpen
  \bibfield  {author} {\bibinfo {author} {\bibfnamefont {I.}~\bibnamefont
  {Contopoulos}}\ and\ \bibinfo {author} {\bibfnamefont {A.}~\bibnamefont
  {Spitkovsky}},\ }\href {\doibase 10.1086/501161} {\bibfield  {journal}
  {\bibinfo  {journal} {Astrophys. J.}\ }\textbf {\bibinfo {volume} {643}},\
  \bibinfo {pages} {1139} (\bibinfo {year} {2006})},\ \Eprint
  {http://arxiv.org/abs/astro-ph/0512002} {arXiv:astro-ph/0512002} \BibitemShut
  {NoStop}%
\end{thebibliography}%

\end{document}